\documentclass[AER]{AEA}

\usepackage{lscape}
\usepackage{array}
\usepackage{xcolor}

\usepackage[round]{natbib}
\usepackage{graphicx}
\usepackage{booktabs}
\draftSpacing{1.5}
\usepackage{lscape}
\usepackage{gensymb}
\usepackage{newtxtext}
\usepackage{layout}
\usepackage{appendix}
\usepackage{comment}

\begin{document}
\pagestyle{plain}
\title{Air Pollution and Under-5 Child Mortality: Evidence from China's Coal Power Plant Phase-out Policy}
\shortTitle{Air Pollution and Under-5 Child Mortality}
\author{Xiaoying Liu and Heng Yu\thanks{Xiaoying Liu:  Affiliate researcher at Population Studies Center, University of Pennsylvania, Philadelphia, PA 19104, United States, xiaoyliu@sas.upenn.edu. Heng Yu: University of Chicago, 5757 S University Ave, Chicago, IL 60637, hyu8@uchicago.edu. Acknowledgements: we thank Jing Wei for sharing ChinaHighPM2.5 dataset. We also thank Jere Behrman, Emily Hannum, Shiqiu Zhang, Jo Van Biesebroeck and other participants at Development Economics seminar at KU Leuven for comments.}}
\date{\today}
\pubMonth{}
\pubYear{}
\pubVolume{}
\pubIssue{}

\begin{abstract}
This paper evaluates the impact of a mandatory shutdown policy of small-capacity coal power plants during China's $11^{th}$ 5-Year Plan period (2006--2010) on under-5 mortality. We collect capacity and location information on 2181 coal power plants that operated during 2000--2010 and compile a unique data set that combines coal power plants, county-level under-5 mortality and socioeconomic variables, high spatial resolution data of PM$_{2.5}$ and SO$_2$ and meteorological conditions. We model the impacts of air pollution on under-5 mortality using the IV-Lasso method, with distance-weighted sums of retired capacity and high-altitude wind status as instrumental variable candidates for air pollution. Our estimates imply that the phase-out policy saved around 46,000 lives during the $11^{th}$ 5-Year Plan period. We also find heterogeneity in the policy effects across regions. (\textit{JEL} I10, I18, P25, P28, Q52, Q53, Q58)
\end{abstract}

\maketitle

\begin{figure}[h]
    \centering
    \includegraphics[width=\linewidth]{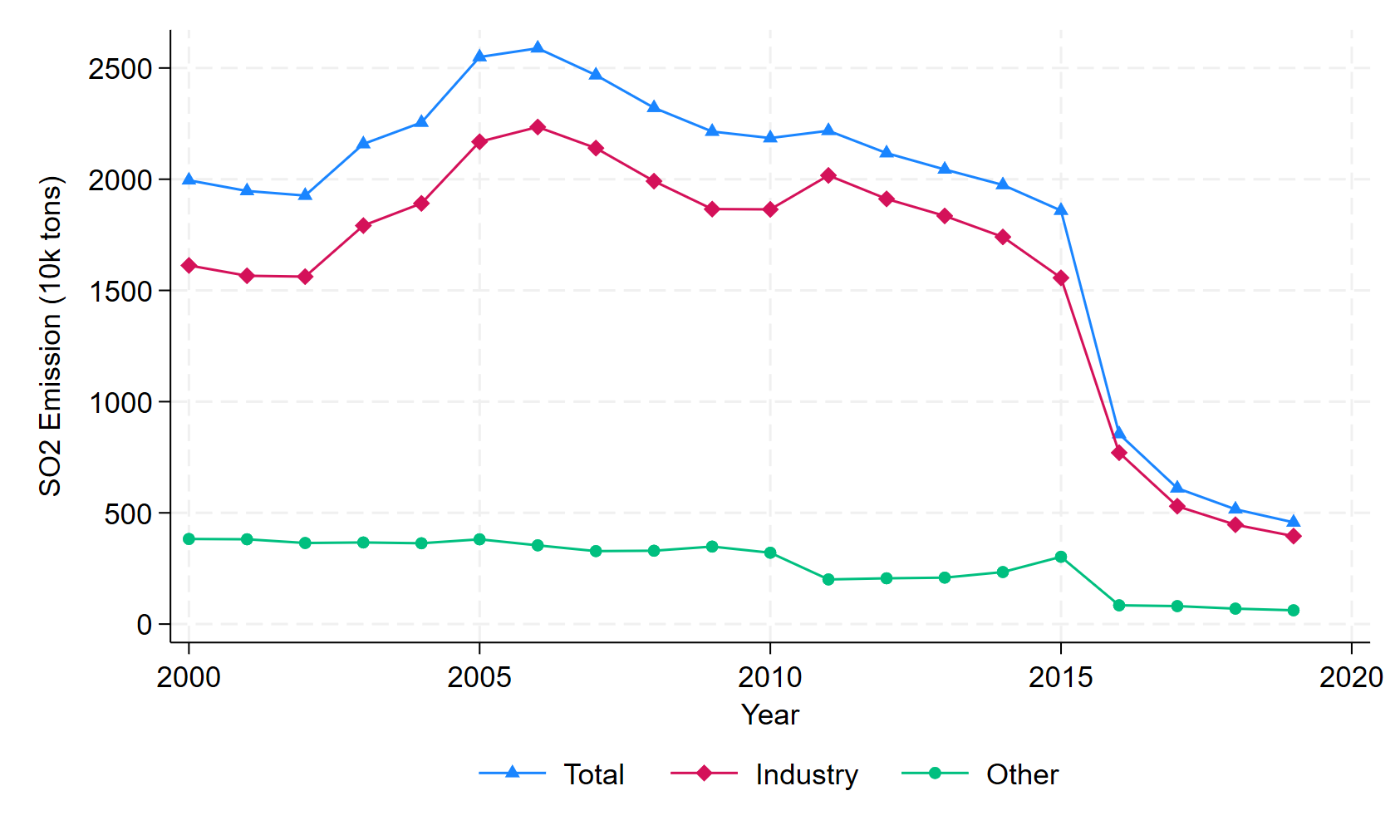}
    \caption{Yearly SO$_2$ Emission (2000 to 2019)}
    \begin{figurenotes}[Source]
        National Bureau of Statistics of China. 
    \end{figurenotes}
    \label{fig:so2emi}
\end{figure}

Coal-generated electricity in China accounts for 62\% of the country's total electricity supply \citep{IEAChina} and has been a key driver of economic growth over the past four decades. However, burning coal also generates air pollutants, such as $SO_2$ and $PM_{2.5}$. Over the three decades following 1990, the Chinese government implemented a series of regulations aimed at improving air quality, particularly by reducing $SO_2$ emissions, while still meeting coal demand and sustaining economic growth. As a result, total SO2 emissions peaked in 2006 and have steadily decreased since then (Figure \ref{fig:so2emi}). Understanding the contribution of these $SO_2$ emission control policies to public health, especially for children as the most vulnerable population, is of considerable interest, not only to China, but also to other fast-growing countries. 

Although the effects of $SO_2$ on human health have been studied extensively in the public health field, there are still areas that require further investigation. One such area is the impact of $SO_2$ on under-5 mortality, a topic that has rarely been explored specifically. Understanding this effect is particularly important for The United Nations' sustainable development goals, especially the target to end preventable deaths of newborns and children under 5 years of age by 2030, with the goal of reducing under-5 mortality to at least 25 per 1,000 live births \citep{UNECE}. Many developing and under-developed countries are still far from reaching this target, including India \citep{U5Mcountry}, which also relies heavily on coal for energy supply and is consistently increasing coal demand \citep{IEA}. Burning coal releases various toxic pollutants including mercury, lead, sulfur dioxide, nitrogen oxides, particulate matter and heavy metals into the air \citep{GASPAROTTO2021113}. These pollutants are known to increase susceptibility to respiratory infections by disrupting immune functions, imparing macrophage activity, altering cellular receptors used by pathogens, and disturbing the microbiome (see \cite{Monoson2023} for an extensive review of epidemiological evidence and laboratory mechanisms), particularly among vulnerable populations such as children and the elderly. According to a 2015 global study on the causes of under-5 mortality, pneumonia accounts for nearly one-sixth of deaths and is the leading cause of death among children aged 1-59 month period \citep{Liu2016}, surpassing both diarrhoea and malaria combined \citep{watkins_pneumonia_2018}. Although epidemiological studies over the past three decades have focused on the associations between $SO_2$ and public health, limited attention has been given to evaluating sulfur control policies and addressing bias from omitted variables that are linked to both air pollution and mortality. 

This paper examines the impacts of $SO_2$ and $PM_{2.5}$ on under-5 mortality in China, leveraging a policy that mandated the replacement of inefficient coal power plants with more efficient ones (also called "big-up, small-down" policy) and installation of flue gas desulfurization (FGD) system during the $11^{th}$ 5-Year Plan (FYP) period (2006–2010). The "big-up, small-down" policy and FGD installation policy, issued by the central government in 2006, required local governments to phase out low-efficiency power plants (particularly those with a capacity of less than 50 megawatts (MW)) and to ensure installation of FGD in other coal power plants over the course of the five-year period. Our identification strategy exploits the exogenous shocks (or improvements) to air pollution levels ($SO_2$ and $PM_{2.5}$) caused by this policy. We compile a unique national-scale dataset that links annual county-level under-5 mortality rates and socio-economic development, individual coal power plant operations and closures, high spatial and temporal resolution air pollution and meteorological conditions from remote sensing, all using geocoding. Specifically, we geocode all power plants that were operating or shut down between 2000 and 2010. We calculate the total capacity of power plant units that were shut down or required to install FGD each year within a certain radius from the geographic center of each county. Additionally, we obtain annual under-5 mortality rates at the county level and calculate monthly mean concentrations of $SO_2$ and $PM_{2.5}$ within each county boundary using high-resolution remote sensing data. We then estimate the impacts of $SO_2$ and $PM_{2.5}$ on under-5 mortality, using the closed-down capacities and capacities with installed FGD as instrumental variables to create exogenous variation in air pollution levels. 

This paper makes four contributions to the literature. Firstly, it contributes to the economic literature that examines the health effects of air pollution in quasi-experimental settings. Unlike studies that examine spontaneous natural events (e.g., a volcanic eruption\citep{halliday_vog_2019} or natural forest fire \citep{jayachandran_air_2009}) leading to short-term air quality deterioration, we evaluate a national policy that resulted in sustained improvements in air quality within the context of a highly polluted country. The period from 2006 to 2010 in China provides an ideal natural experiment for policy evaluation, offering two key advantages: 1) the national air quality policy during the $11^{th}$ FYP period focused solely on sulfur control, allowing us to disentangle the effects of $SO_2$ from other pollutants; 2) the policy was enforced in a top-down manner, which helps address potential endogeneity issues in estimation. Secondly, this study contributes to the literature on the health impacts of coal power plants specifically. While previous research has examined the health effects of coal power plant expansion or regulations in different countries \citep{Gupta2017,Luechinger2014,DeCicca2020}, this is the first study to focus on the impacts of coal power plants on child mortality in China. Thirdly, our study utilizes high spatial- and temporal-resolution satellite remote sensing data to measure air pollution during a period when ground-level monitoring data were unavailable, and it makes use of detailed information on individual coal power plants to measure policy exposure at county level. Our last contribution is methodological. This study adds to the few literature that implements the IV-Lasso method in selecting instrumental variables using machine learning. In addition to anthropogenic factors, air quality is affected by atmospheric conditions such as thermal inversions, planetary boundary layer height, altitude winds and altitude pressures \citep{deryugina2019, GODZINSKI2021102489} that are completely unrelated to local anthropological activities. Given the interdisciplinary nature of these studies involving high-frequency air quality observations, there may exist a large set of instrumental variable candidates available for use. Estimating with Lasso-IV method, we are able to efficiently select a smaller set of instrumental variables \citep{belloni2014}. 

More broadly, evaluating the health impacts of sulfur control policies in China has important implications for other developing countries conducting cost-benefit analyses of environmental policies. As coal remains the primary source of energy driving economic growth and the main contributor to $SO_2$ emissions, sulfur control policies may potentially slow economic growth. On the other hand, economic development can fund investments in public health, reducing mortality from malnutrition or lack of medical care. Given the ongoing rise in coal demand to support economic growth in India, Indonesia, Viet Nam and the Philippines \citep{IEA}, is it still feasible to reduce the under-5 mortality in these countries? What can be learned from China, the world's largest coal consumer and once the most polluted country, which has successfully reduced $SO_2$ significantly in the last two decades and achieved a very low under-5 mortality rate? 

The remainder of the paper is organized as follows. In Section I, we introduce the background of air quality control policies in the $11^{th}$ FYP in China; Section II introduces the data, which is followed by Section III that introduces the empirical model and identification strategy; Section IV presents our main results; Section V presents heterogeneity analysis results by subsamples; Section VI discusses the results and concludes. 

\vspace{0cm}

\section{Background and Policy Description}

\subsection{The historical background of air quality control policies in China}

Since China joined the World Trade Organization in 2001, the country has experienced an even more rapid phase of economic growth, with an average annual GDP growth rate of 10.55\% during the first decade of the 21st century, peaking at 14.2\% in 2007 \citep{GDPdata}. To meet the soaring demand for electricity, a large number of small-capacity coal power plants, which could be rapidly constructed with minimal investment, were brought into operation. As a result, total energy consumption in China increased by 70\% between 2000 and 2005, with coal consumption increasing by 75\% \citep{worldbank2007}. By 2005, the total capacity of power plants with units no larger than 100 MW reached 115 GW, accounting for over 27\% of the total capacity of coal-fired plants \citep{worldbank2009}. These small coal power plants are highly inefficient and emit significant amounts of pollutants. In fact, $SO_2$ emissions from the energy industry accounted for 53\% of the country's total $SO_2$ emissions\citep{Geng2016}. By 2005, 16 of the world's 20 most air-polluted cities were in China \citep{worldbank2007}, and 30\% of the China territory were polluted by acid rain \citep{lu_progress_2020}. 

China's air quality control policies in the 1990s and the first decade of 21st century primarily focused on controlling $SO_2$ emissions from coal combustion. In 1995, the Air Pollution Law was amended, designating "Two Control Zones" (TCZ) in areas with severe acid rain and sulfur dioxide emissions, respectively\footnote{Control zones for acid rain were designated for areas with high acidic precipitation (pH $\leq$ 4.5), high sulfur deposition exceeding the critical load, and elevated $SO_2$ emissions. The designation of control zones for $SO_2$ pollution was based on ambient $SO_2$ concentrations if they exceed National Ambient Air Quality Standard (NAAQS) limits.} \citep{lu_progress_2020}. During the $10^{th}$ FYP period (2001-2005), the Ministry of Environmental Protection issued a plan to prevent and control acid rain and sulfur dioxide pollution in these zones, aiming for a 10\% reduction in total sulfur dioxide emissions below the 2000 level. However, due to lack of specific enforcement plans, sulfur dioxide emissions did not decrease but instead increased by 27.8\% during the period \citep{hu2024synergistic}. By 2005, most sulfur control efforts had proven ineffective. 

The $11^{th}$ FYP (2006–2010) marked a turning point in air quality control in China, as the central government set concrete and mandatory quantitative environmental and energy targets for provincial and municipal governments for the first time \citep{Jin2016Air, hu2024synergistic}, with a goal to cap the total emissions of $SO_2$ by 2010 and to reduce total emissions by 10\% \citep{lu_progress_2020}. These policies included measures to limit the production and use of high-sulfur coal, mandate the installation of desulfurization units in newly built and renovated coal-fired power plants, and close small power plants as part of broader efforts to improve energy efficiency. Unlike the $10^{th}$ FYP, the $11^{th}$ FYP introduced several measures to ensure local compliance \citep{schreifels_sulfur_2012}. For example, local government officials were held politically accountable with clear evaluation metrics and consequences. Additionally, the mandatory installation of continuous emissions monitoring systems made real-time monitoring and verification technically feasible. Economic instruments, such as green pricing premiums for electricity generated from power plants with installed FGD systems and extra tax for electricity from small power plants, provided market incentives to enforce the mandates. 



\subsection{The specific sulfur control policies in the $11^{th}$ FYP in China}

Among the three measures of sulfur control issued by the central government (i.e., limiting production and use of high-sulfur coal, forced closure of small power plants and mandatory installation of FGD), we focus on Small Plants Closure Program and FGD Installation Program specifically. In 2004, the National Development and Reform Commission issued “a notice on the planning and construction of coal-fired power station projects” \citep{NEA2012}, recommending the replacement of small coal power plants with higher-efficiency plants\footnote{The high efficiency power plants utilize supercritical or ultra-supercritical technology that generates steam at pressures above the critical pressure (22.1 MPa) or even higher temperatures and pressures. The thermal efficiency of these plants can reach 42\%-45\%, compared to a maximum of 38\% for conventional pulverized coal combustion technology.}. The notice also specified that new power plant units should have capacities of at least 600 MW, with coal consumption for each unit of power generation not exceeding 286 grams of standard coal per 10,000 hours. It further mandated that newly constructed or expanded coal power plants install FGD facilities and encouraged existing power plants to do the same. 

In 2007, State Council issued a notice with several suggestions to accelerate the closure of small thermal plant units. The notice specifically targeted the shutdown of units meeting any of the following criteria: 1) units of capacity under 50 MW; 2) units of capacity under 100 MW with at least 20 years in operation; 3) units of capacity under 200 MW beyond their service life; 4) units of which coal consumption for power generation is 10\% higher than the provincial, regional or municipal average, or 15\% higher than the national average in 2005; 5) units failing to meet environmental protection standards. Furthermore, the notice emphasized that no new power plants should be constructed before closing small plant units in the region. New high-capacity units (over 300 MW), particularly those replacing retired plants, were prioritized for central government's financial support. The notice also provided guidelines for the proper relocation of employees following the closure of old power plants \citep{State2007}. 

As a result, the share of supercritical and ultra-supercritical capacity in China increased rapidly from the mid-2000s, increasing from less than 5\% in 2004 to approximately 28\% in 2010 \citep{IEA2012}. The total closed capacity between 2006 and 2008 amounted to 34,210 MW \citep{Price2011}, and by 2010, this figure had increased to approximately 70,000 to 80,000 MW. According to the Ministry of Environmental Protection, $SO_2$ emissions decreased by 14.3\% from 2005 to 2010 \citep{MEE2011}.  

Evidence from emission inventory, acid rain monitoring network, and the Ozone Monitoring Instrument (OMI) satellite sensor show a significant decrease in $SO_2$ emissions and concentrations during the $11^{th}$ FYP compared to $10^{th}$ FYP \citep{lu_progress_2020, Wang_2015, Xia2016}. 

\begin{figure}[h]
    \centering
    \includegraphics[width=\linewidth]{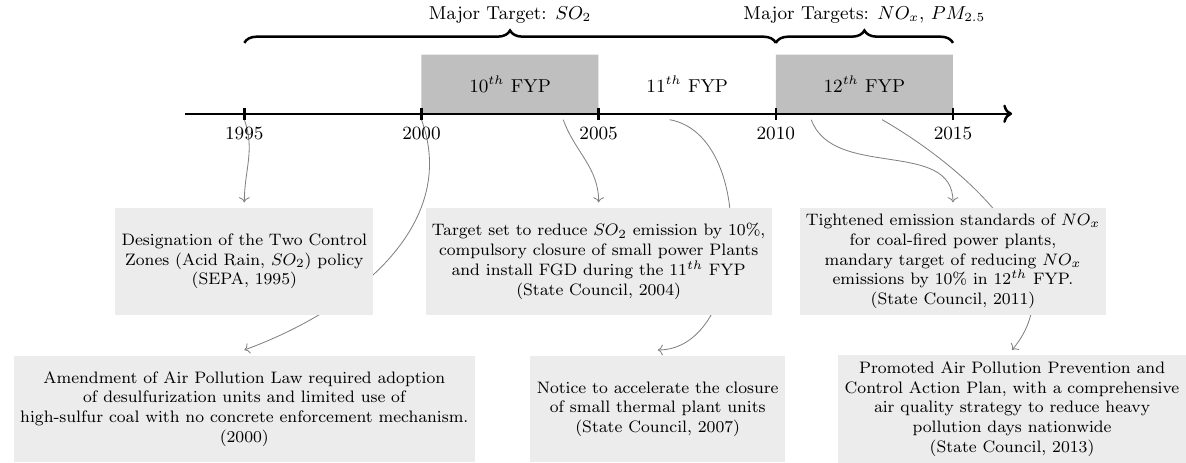}
    \caption{Major Stages of China's Air Pollution Control}
    \begin{figurenotes}
        SEPA = State Environmental Protection Administration. The figure is adapted from \citep{lu_progress_2020}.  
    \end{figurenotes}
    \label{fig:timeline_policy}
\end{figure}

It is important to note that during the $11^{th}$ FYP, the air quality control policy mainly focused on sulfur dioxide emission control. Although a plan to control nitrogen oxides ($NO_x$) emissions was also being developed, it was mainly enforced during the $12^{th}$ FYP (2011-2015) \citep{lu_progress_2020}. Furthermore, since 2013, China's environmental policy shifted from focusing on single pollutant emissions control to a more comprehensive air quality strategy, which required coordination across multiple sectors \citep{lu_progress_2020}. Therefore, focusing on the $11^{th}$ FYP allows for a clearer identification of $SO_2$ abatement effect on under-5 mortality. We document and summarize the three main stages of China's air pollution control actions and policies in Figure \ref{fig:timeline_policy}. The stages of pollution control are assigned based on targeted pollution, priority subjects and the five-year-plan periods. The government focused on acidic rain and sulphur dioxide controls in the late 1990s and early 2000s and on $NO_X$ and $PM_{2.5}$ since the $12^{th}$ FYP. 

\vspace{0cm}

\section{Data and Measurements}\label{section:data}
We compile a unique county-level panel dataset across China for the period 2001-2010, which consists of 1) under-5 mortality rates at county level for each year; 2) the total capacity of the coal power plant units within 25km, 50km or 100km radius from the geographic center of a county that were shut down or that were required to install FGD during the year, weighted by distance between power plant and county centroid and deviation from the dominant wind directions; 3) air pollution data including monthly averages of $SO_2$ and $PM_{2.5}$ concentration of a county; 4) monthly average of meteorological data, including temperature, rainfall and humidity of a county; 5) socio-economic development indicators including per capita GDP and the number of doctors per 10,000 people at county level for each year. This section outlines the sources of data and the construction of the variables used in the model estimation.

\subsection{Under-5 Mortality Rates}
The under-5 mortality rates used in our study is the estimated mortality at county level from 1996 to 2012, by the Institute for Health Metrics and Evaluation (IHME) at the University of Washington, USA. IHME synthesizes individual records, aggregated birth and death numbers at county level, and additional factors, including "maternal mortality, pregnancy management rates, visits for children under 7 years, pregnancy wellness checks, hospital deliveries, and newborn check-up rates" from the Annual Report System on Maternal and Child Health and employs a combination of a small area mortality estimation model, spatiotemporal smoothing, and Gaussian process regression to obtain the estimated mortality at county level for each year \citep{Wang2016}. 

\begin{figure}[htbp]
\centering
\begin{tabular}{c}
     \includegraphics[width=\textwidth]{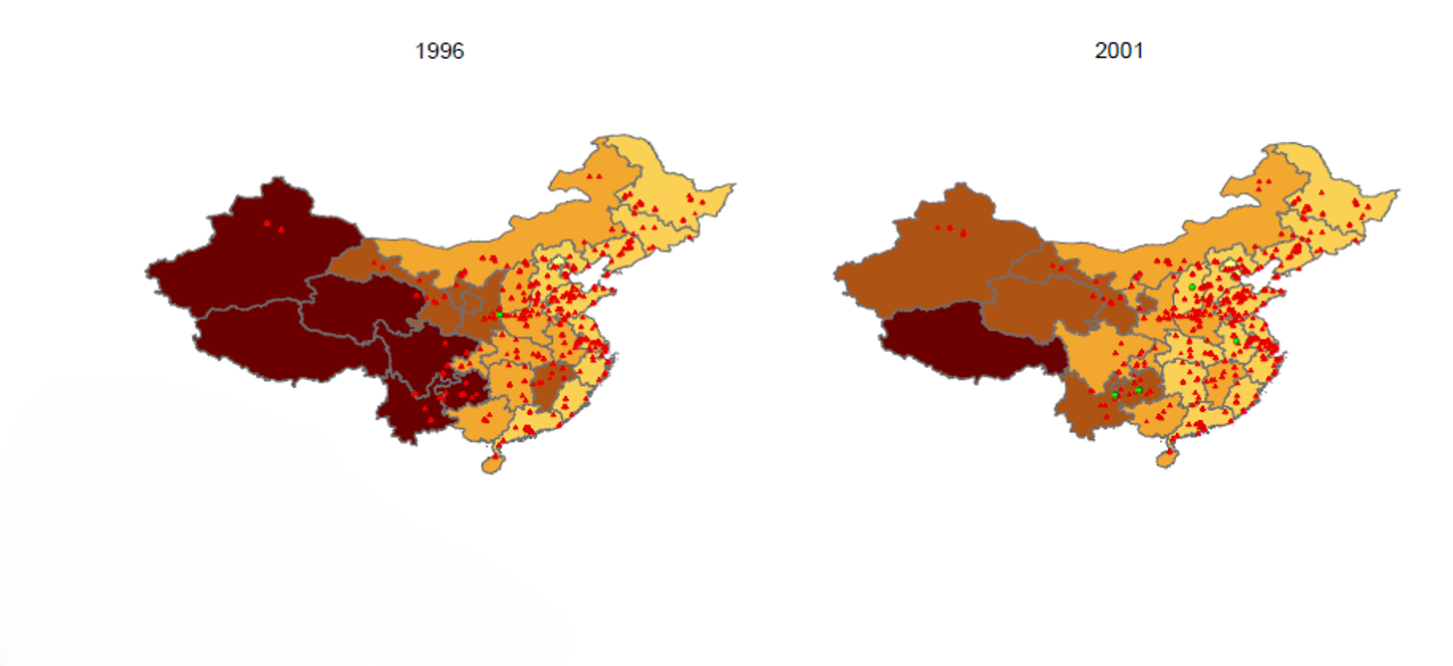}  \\
     \includegraphics[width=\textwidth]{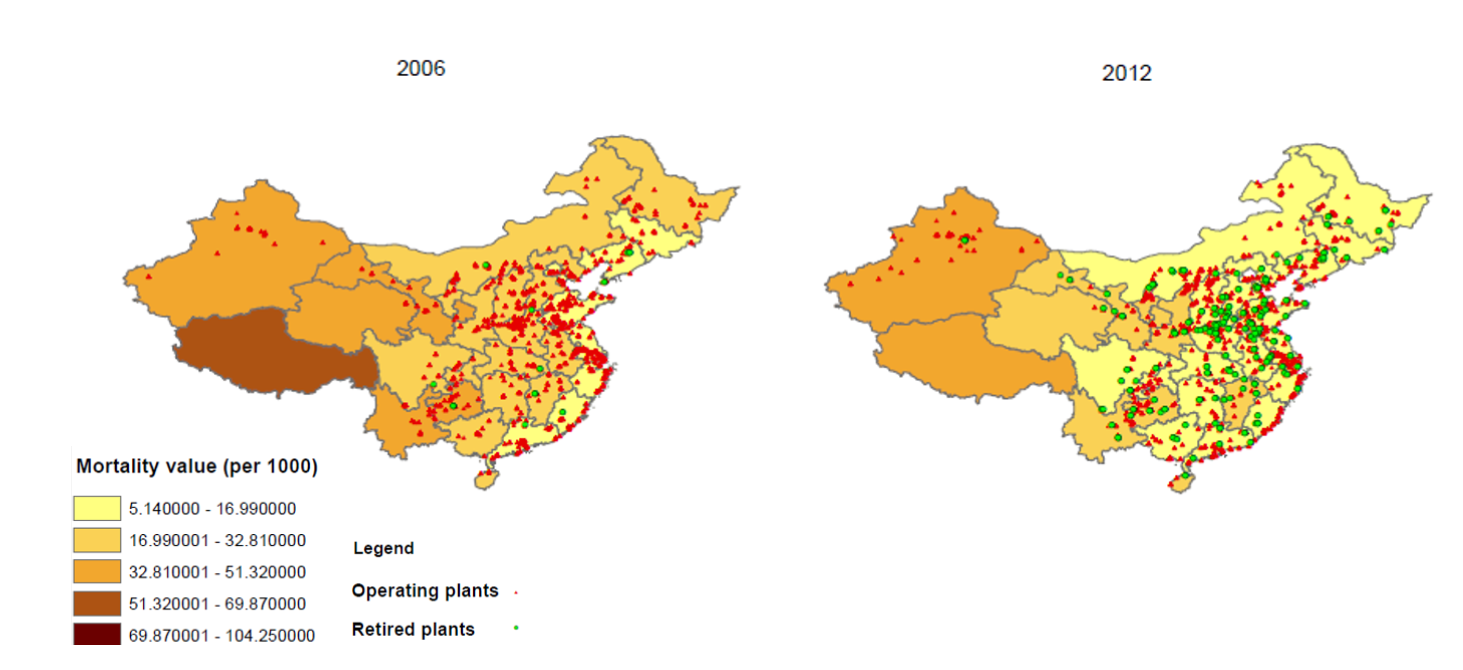} 
\end{tabular}
\caption{Under-5 Mortality and Coal-Fired Power Plants between 1996 and 2012}
\begin{figurenotes}
The figure displays the trend in province-level under-5 mortality from 1996 to 2012 in mainland China. It also shows the spatial distribution of operating coal power plants (represented by red triangles) and retired coal power plants (represented by green circles) from our compiled directory.
\end{figurenotes}
\label{fig:mortality1}
\end{figure}

Figure \ref{fig:mortality1} illustrates the change in under-5 mortality rates at the province level in 1996, 2001, 2006, and 2012. The data clearly show a nationwide decline in under-5 mortality over time. According to \cite{Wang2016}, the under-5 mortality rate in China dropped by 70\% from 46.0 per 1,000 live births in 1990 to approximately 13.7 per 1,000 live births in 2012. Our data reflect the same trend for the study period, with the national average under-5 mortality decreasing from 35.56 per 1,000 live births in 2001, to 18.19 per 1,000 in 2010. Furthermore, there is a significant regional disparity: in the eastern urban regions, under-5 mortality is as low as 3.3 per 1,000 live births, while in the western rural regions, it can be as high as 104.4 per 1,000 live births. Within provinces, there are also substantial variations across counties.

For the analysis in this paper, we use the under-5 mortality rate at county level from 2001 to 2010 as dependent variable. The summary statistics of the county-level mortality rates are also presented in Tables \ref{tab:summarystats1}. 

\vspace{0cm}
\vspace{0cm}

\subsection{China Coal Power Plants Directory}\label{section:coalplantsdata}
Global coal plant tracker (GCPT) by Global Energy Monitor\footnote{Website: https://globalenergymonitor.org/projects/global-coal-plant-tracker/} provides complete and comprehensive documentation of coal power plants in mainland China \citep{cui2019quantifying, cui2021plant, lu2022plant} and overseas \citep{gao2021challenges,springer2021empirical, montrone2023investment, wu2024characterizing}.\footnote{According to Global Energy Monitor, preliminary lists of plants in each country were gathered from public and private data sources including Global Energy Observatory, CARMA, BankTrack’s “Dirty Deals” list, Wikipedia, Enipedia, WRI’s “Global Coal Risk Assessment” report (2012), Platts World Energy Power Plant database, Industcards “Power Plants Around the World Photo Gallery”, national-level trackers developed by environmental organizations (Sierra Club (USA), Kara Atlas (Turkey), and Deutsche Umwelthilfe (Germany)), as well as various company and government sources. Coal plant data are validated and updated through five main sources: 1) government data on individual power plants, country energy and resource plans, and government websites tracking coal plant permits and applications; 2) reports by state-owned and private power companies; 3) news and media reports; 4) local non-governmental organizations tracking coal plants or permits; 5) On the ground contacts who can provide first-hand information about a project or plant. The data have also been reviewed by local experts where possible.} The GCPT catalogues every coal-fired generating unit in operation, as well as every new unit proposed since 2010 and those retired since 2000. For each power unit, it includes details such as the name, exact location and geocode, power generation capacity, operational status (operating, retired, under construction, or proposed), year of starting operation for active or retired units, year of retirement for retired units, combustion technology, coal source, heat rate, $CO_2$ emission factor, annual $CO_2$ emission, and more. We extracted the complete list for China up to the year of 2018. 


\par However, since the GCPT only tracks coal-fired power units with a capacity of 30 megawatts (MW) or above, it does not include information on units with capacities below 30 MW, which are the primary targets of the power plant closure policy during the $11^{th}$ FYP. To compile a complete list, we obtained data on a total of 2486 small thermal power plant units that were closed down in each year during 2006-2010. The lists are publicly available on the website of Ministry of Ecology and Environment of the People's Republic of China \citep{List2006, List2007, List2009, List2010}. This list of retired coal power units includes the name of each power plants and the units, the province and city where they are located, generating capacity, and the year and month of retirement. To obtain the precise distance measurements between power plants and nearby counties for our analysis, we geocode the power plants based on their exact locations - a task that turned out to be very challenging. During China's rapid economic growth and urbanization in the 2000s, cities expanded significantly, leading to major changes in urban planning. To overcome this challenge, we searched for the precise locations of these retired power plants through various sources, including local newspapers, government bulletins, and time-preserved information sources such as the names of bus stops and grocery stores. \footnote{In China, bus stops often take their names from nearby roads, villages, or notable landmarks such as hospitals and factories, especially in the 1990s and early 2000s. Despite the significant changes in urban planning over the past 30 years, these bus stop names have largely been preserved over time.} 

\par In the end, we manually geocode these power plants using Google Maps Geocoding Web Service. For some retired small-unit power plants that we were unable to find the exact locations, we assume it was situated at the geographical center of the county. We also made efforts to document the year each plant unit started operating through online searches whenever possible. For those plants where the exact start year could not be determined, we assumed they started operation before 2000, which marks the beginning of our analysis period. The summary statistics of the closed capacity variables are presented in Table \ref{tab:summarystats1}.

\begin{figure}[h]
    \centering
    \includegraphics[width=\linewidth]{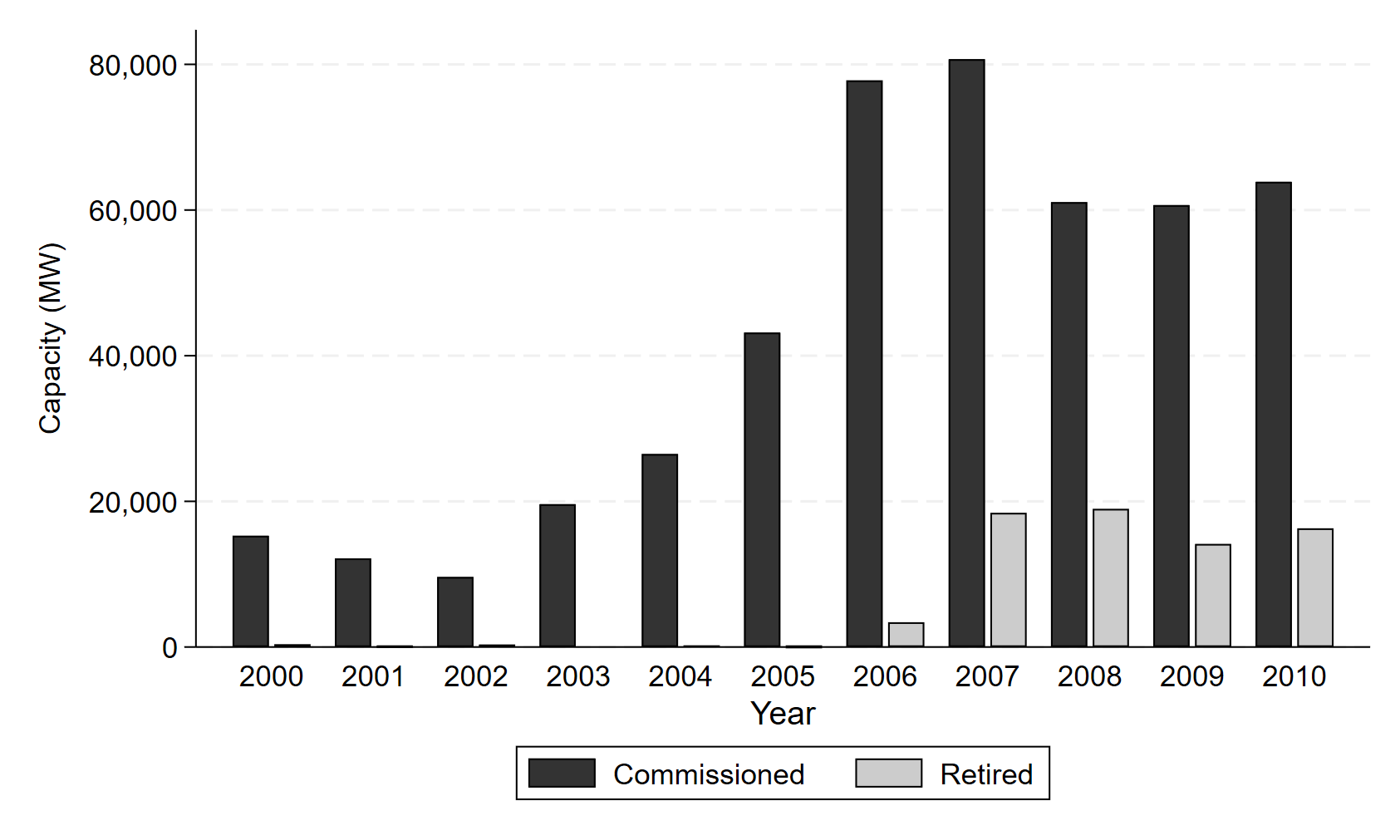}
    \caption{Total Capacities of the Commissioned and Retired Coal-Fired Power Plants from 2000 to 2010}
    \label{fig:pollutantdensity}
\end{figure}
\vspace{0cm}

\par We calculate the total capacities of the coal power plants closed and commissioned each year from 2000 to 2010, as shown in Figure \ref{fig:pollutantdensity}. As mentioned in the background section, although the coal power plants phase-out plan was initially proposed in the early 2000s, its implementation lagged behind. As a result, most of the closures occurred after 2005, with a significant acceleration in 2007 and 2008, particularly after the notice was issued in 2007 by the central government. According to the data we collected, the total capacity shut down during this period is 71,315 MW, which closely aligns with the National Energy Administration's reported figure of 72,100 MW \citep{govreport}. We also calculate the total capacity of retired coal power plant units for those with capacity under 50 MW only, for the same period. In the first half of $11^{th}$ FYP, the majority of the retired plants had capacities under 50MW, as these were the least efficient plants and prioritized for closure. The peak of closures for coal power units with capacities under 50MW occurred in 2007 and 2008, accounting for approximately two thirds of those closed under-50 MW capacity during the $11^{th}$ FYP. 

\par To measure the exposure to the policy regulating $SO_2$ emission and the retirement of coal power plants, we calculate the sum of capacities of all coal power plants within a given radius (25 km, 50 km, and 100 km) from the county's geographic centroid that were retired in a given year.\footnote{For reference, we calculate the distance of each county centroid to the closest county border. The mean distance is 12.98 km. The 25th, medium and 75th percentiles are 5.87, 10.64 and 16.08 km, respectively.} Additionally, we calculate the sum of capacities only for plants with a capacity of 50 MW or less within each of these radii from the county's center.

\par Considering the impact of distance and wind direction on the transmission of air pollutants, we calculate the weighted sum of the capacities of closed power plants or those required to install FGD within a specified radius from a county center. We assign a weight to each power plant's capacity based on the inverse of the distance to the county center and the alignment of its direction to the county with the annually averaged wind direction for the county, as illustrated in Figure \ref{fig:projection} in Appendix. \footnote{We do not account for the wind direction at the individual plant locations or the complete air pollutants transmission process along the route as discussed in literature on dynamic transmission in complex terrains \citep{SHI2023116268}} 

\par Figure \ref{fig:mortality1} also shows the geographic distribution of operating and retired coal-powered plant units in our database for a specific year. The red triangles represent plants that were operating in each of the selected years and the green circles represent plants that had retired by the year of observation. We observe the following patterns: 1) most power plants were concentrated in eastern and central China, particularly in the Yangtze River Delta region (YRD), North China plain (NCP), and Pearl River Delta (PRD) where industry and economy were most developed \footnote{For reference, we show the distribution of county-level nominal GDP in 2000 in appendix Figure \ref{fig:gdp2000}}; 2) the retired power plant units from 2006 to 2012 were also concentrated in eastern and central China, especially around NCP and YRD regions. 

\subsection{Air Quality Data}\label{densitydata}
\begin{figure}[htbp]
\centering
\begin{tabular}{cc}
\centering
    (a) $SO_2$ 2005 & (b) $SO_2$ 2010 \\
     \includegraphics[width=.5\textwidth]{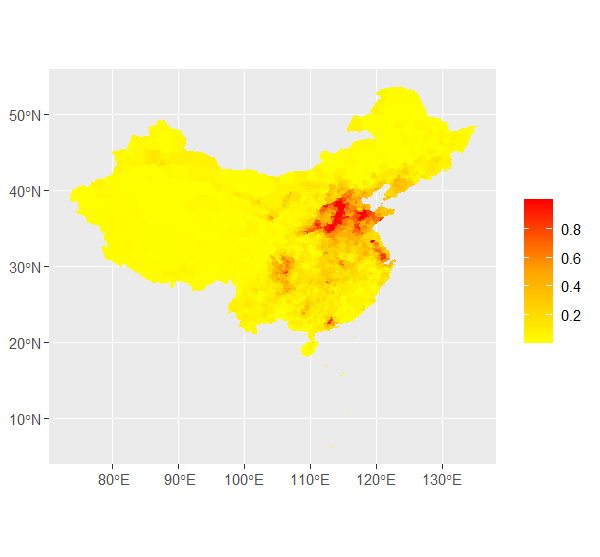}  &
     \includegraphics[width=.5\textwidth]{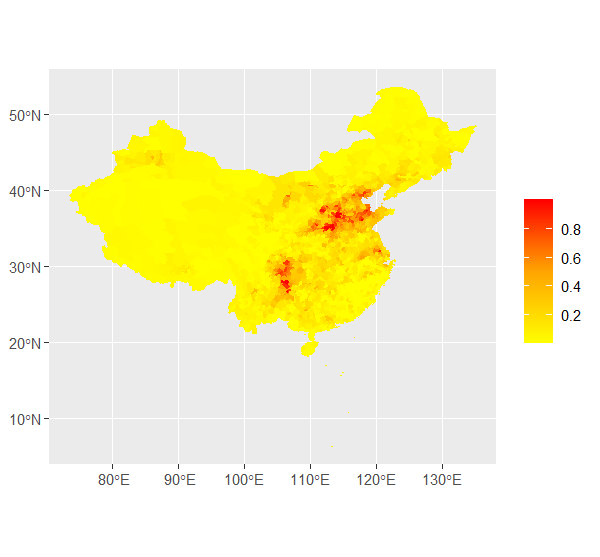} \\
     (c) $PM_{2.5}$ 2000 & (d)  $PM_{2.5}$ 2005 \\
     \includegraphics[width=.5\textwidth]{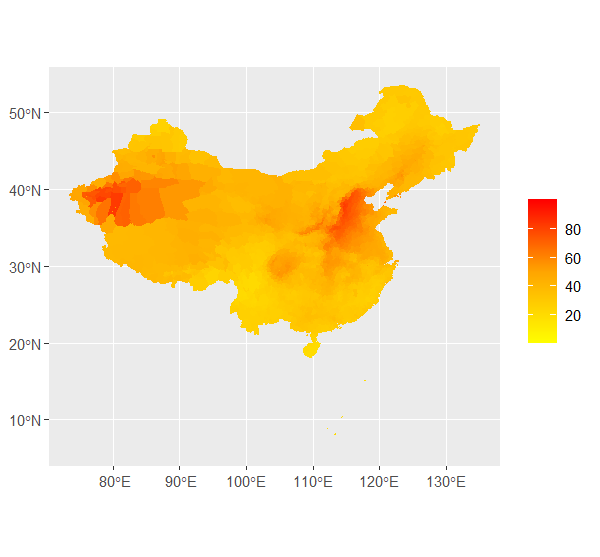}  &
     \includegraphics[width=.5\textwidth]{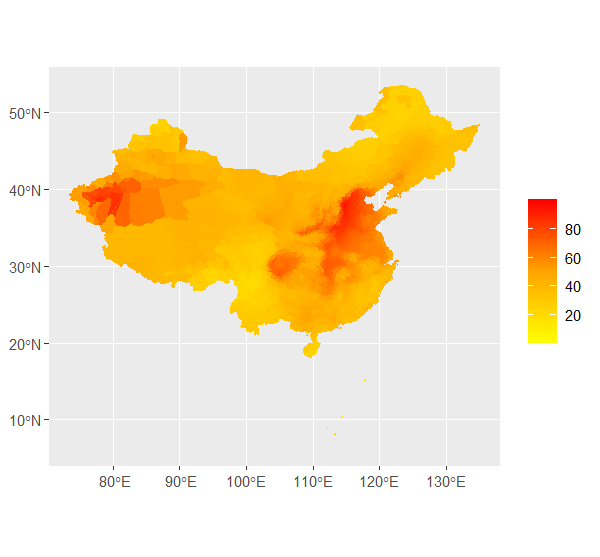} \\
     \multicolumn{2}{c}{(e) $PM_{2.5}$ 2010}\\
     \multicolumn{2}{c}{\includegraphics[width=.5\textwidth]{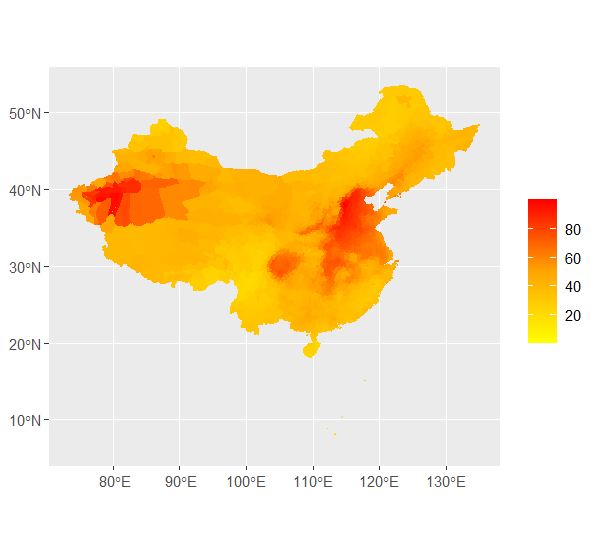}}
     
\end{tabular}
\caption{$SO_2$ and $PM_{2.5}$ Density Heat Maps}
\begin{figurenotes}
The figure shows the density of $SO_2$ VCDs (in DU) and $PM_{2.5}$ (in $\mu\text{g/m}^{3}$) throughout China in 2000, 2005 and 2010. Since $SO_2$ VCDs are only available from October 2004, $SO_2$ density heatmaps are only shown for 2005 and 2010.
\end{figurenotes}
\label{fig:wholesample_density}
\end{figure}

Following \cite{Xia2016}, we obtain a gridded data of daily observation of $SO_2$ vertical column densities (VCDs) in the Planetary Boundary Layer from NASA's Ozone Monitoring Instrument (OMI) on the Aura Satellite for the period between 2005 and 2010, at a 0.25 degree by 0.25 degree spatial resolution \citep{Li2020}\footnote{Since Aura satellite was put into use in October 2004, data on $SO_2$ VCDs is only available since then. Ground monitoring data for $SO_2$ concentration for that time period were only available for several metropolitan cities.}. VCDs measure total number of molecules of $SO_2$ in a vertical column through
the Earth’s atmosphere in Dobson Unit (DU). Although $SO_2$ VCDs data can not completely reflect ground level human exposure, its spatial and temporal variation has been found to be consistent with ground emission data \citep{Xia2016}. We calculate the arithmetic mean for each county using daily observations from all the grids within the county, then calculate the monthly and annual averages for each county. Heat maps of the annual mean of $SO_2$ VCDs for 2005 and 2010 are displayed in Figure \ref{fig:wholesample_density}. The spatial and temporal trends align with the findings in \cite{Xia2016}, namely: 1) the most severe $SO_2$ pollution was observed in NP, YRD, PRD and the Sichuan basin (SB); 2) the average concentration of $SO_2$ VCDs across China decreased from 2005 to 2010, initially rising from 0.241 DU in 2005 to 0.287 DU in 2007, before declining to 0.184 DU in 2010.\footnote{One DU corresponds to $2.69$x$10^{26}$ molecules per square kilometer at standard temperature and pressure \citep{Nasa2018}.} This trend was especially evident in the NP, YRD and PRD regions, with the exception of SB, where $SO_2$ concentration increased throughout the entire $11^{th}$ FYP period.
\vspace{0cm}
\vspace{0cm}

\par In addition to $SO_2$ VCDs data, we also obtain high spatial and temporal resolution density data for $PM_{2.5}$ from 2000 to 2010. Since China only classified $PM_{2.5}$ as a critical pollutant in 2012 \citep{Jiang2015}, ground-level observations of $PM_{2.5}$ were not available during our study period (2001-2010). The ChinaHigh$PM_{2.5}$ dataset provides daily $PM_{2.5}$ concentration at a 1 km resolution, estimated using long-term, high-spatial-resolution aerosol optical depths (AOD) generated by the Moderate Resolution Imaging Spectroradiometer (MODIS) Multi-Angle Implementation of Atmospheric Correction algorithm \citep{Wei2021}. Similar to $SO_2$, we calculate the monthly and annual mean concentrations of $PM_{2.5}$ for each county. Figure \ref{fig:wholesample_density} also displays the annual mean of county-level $PM_{2.5}$ concentrations for 2000, 2005 and 2010 respectively. $PM_{2.5}$ concentrations did not show a significant change from 2005 to 2010, relative to the change from 2000 to 2005. In fact, the national average of $PM_{2.5}$ consistently increased from 47.63 ${\mu}g/m^3$ in 2001 to 56.86 ${\mu}g/m^3$ in 2007, and then gradually declined to 52.96 ${\mu}g/m^3$ in 2010. 



\subsection{Meteorological data}
We obtain meteorological data from ERA5 provided by the European Centre for Medium-Range Weather Forecasts (ECMWF) \citep{Hersbach2023}. ERA5 reanalyzes observational meteorological data and provides hourly estimates of atmospheric, ocean-wave, and land-surface variables at $0.25^\circ \times 0.25^\circ$ resolution. We extract monthly-averaged reanalysis data for key meteorological variables, including 2-meter above-ground temperature, dew point temperature, total precipitation, 10-meter and 100-meter above-ground wind speed and wind direction\footnote{Wind speed and wind directions are calculated from the u and v components of wind with a positive u component coming from the west and a positive v component coming from the south.}, covering the period from 1950 to 2012. Additionally, we calculate relative humidity using 2-meter above-ground temperature and dew point temperature. 

We standardize the monthly 2-meter above-ground temperature and relative humidity during the period 2000-2010 against the historical trend defined by the 50-year period (1950-1999) for each month. The standardized temperature and relative humidity are included in the mortality function as control variables. 

\begin{table}[p]
  \centering
  \caption{Summary Statistics of Key Variables}
  \resizebox{\textwidth}{!}{\begin{tabular}{lcccc}
    \toprule
    \toprule
      & (1) & (2) & (3) & (4) \\
    Year  & 2001 & 2004 & 2007 & 2010 \\
    \midrule
    Under-5 Mortality (per 1,000) & 33.38 & 26.03 & 20.49 & 16.28 \\
      & (15.17) & (11.55) & (8.401) & (6.570) \\
    $SO_2$ (DU) & 0 & 0 & 0.318 & 0.202 \\
      & (0) & (0) & (0.308) & (0.224) \\
    PM$_{2.5}$ ($\mu\text{g/m}^{3}$) & 48.26 & 54.10 & 59.33 & 54.75 \\
      & (13.93) & (15.50) & (18.94) & (17.04) \\
          Prim. GDP (CNY per capita) & 1478.7 & 1923.7 & 2518.7 & 3556.6 \\
      & (696.9) & (880.1) & (1168.7) & (1740.7) \\
    Sec. GDP (CNY per capita) & 2407.0 & 4208.2 & 7378.0 & 12450.6 \\
      & (2415.8) & (4341.7) & (8577.8) & (13496.8) \\
    Hospital Bed (per 10,000 Population) & 24.24 & 24.02 & 26.12 & 31.90 \\
      & (9.293) & (9.247) & (9.680) & (10.28) \\
    Unweighted Sum of Retired Capacity (0 to 100 km) & 0.485 & 2.199 & 170.1 & 138.9 \\
      & (6.950) & (15.37) & (243.5) & (262.5) \\
    Unweighted Sum of Retired Capacity (0 to 50 km) & 0.0808 & 0.788 & 42.20 & 33.55 \\
      & (2.843) & (9.279) & (110.3) & (98.35) \\
    Unweighted Sum of Retired Capacity (0 to 25 km) & 0.0808 & 0.332 & 9.978 & 7.229 \\
      & (2.843) & (6.188) & (50.95) & (43.48) \\
    Unweighted Sum of Retired Capacity (50 to 100 km) & 0.404 & 1.411 & 127.9 & 105.4 \\
      & (6.347) & (12.34) & (207.5) & (209.3) \\
    Unweighted Sum of Retired Capacity (25 to 100 km) & 0.404 & 1.868 & 160.1 & 131.7 \\
      & (6.347) & (14.11) & (236.9) & (251.0) \\
    Unweighted Sum of Retired Capacity Under 50 MW (50 to 100 km) & 0 & 0 & 10.17 & 21.15 \\
      & (0) & (0) & (23.13) & (44.38) \\
    Unweighted Sum of Retired Capacity Under 50 MW (25 to 100 km) & 0 & 0 & 12.69 & 26.54 \\
      & (0) & (0) & (26.95) & (52.40) \\
    Unweighted Sum of Retired Capacity Under 50 MW (0 to 100 km) & 0 & 0 & 13.64 & 28.40 \\
      & (0) & (0) & (28.52) & (55.61) \\
    Unweighted Sum of Retired Capacity Under 50 MW (0 to 50 km) & 0 & 0 & 3.618 & 7.322 \\
      & (0) & (0) & (12.79) & (22.52) \\
    Unweighted Sum of Retired Capacity Under 50 MW (0 to 25 km) & 0 & 0 & 0.989 & 1.876 \\
      & (0) & (0) & (6.189) & (9.892) \\
    Weighted Sum of Retired Capacity (0 to 100 km) & 0.0138 & 0.0521 & 3.437 & 2.674 \\
      & (0.297) & (0.511) & (6.720) & (5.873) \\
    Weighted Sum of Retired Capacity (0 to 50 km) & 0.00804 & 0.0344 & 1.719 & 1.269 \\
      & (0.283) & (0.488) & (5.796) & (4.416) \\
    Weighted Sum of Retired Capacity (0 to 25 km) & 0.00804 & 0.0237 & 0.858 & 0.561 \\
      & (0.283) & (0.460) & (5.082) & (3.637) \\
    Weighted Sum of Retired Capacity (50 to 100 km) & 0.00572 & 0.0176 & 1.718 & 1.405 \\
      & (0.0914) & (0.157) & (2.847) & (2.831) \\
    Weighted Sum of Retired Capacity (25 to 100 km) & 0.00572 & 0.0284 & 2.579 & 2.114 \\
      & (0.0914) & (0.226) & (4.039) & (4.210) \\
    Weighted Sum of Retired Capacity Under 50 MW (50 to 100 km) & 0 & 0 & 0.136 & 0.280 \\
      & (0) & (0) & (0.313) & (0.590) \\
    Weighted Sum of Retired Capacity Under 50 MW (25 to 100 km) & 0 & 0 & 0.203 & 0.422 \\
      & (0) & (0) & (0.454) & (0.862) \\
    Weighted Sum of Retired Capacity Under 50 MW (0 to 100 km) & 0 & 0 & 0.302 & 0.567 \\
      & (0) & (0) & (1.140) & (1.495) \\
    Weighted Sum of Retired Capacity Under 50 MW (0 to 50 km) & 0 & 0 & 0.169 & 0.287 \\
      & (0) & (0) & (1.065) & (1.243) \\
    Weighted Sum of Retired Capacity Under 50 MW (0 to 25 km) & 0 & 0 & 0.1000 & 0.145 \\
      & (0) & (0) & (1.000) & (1.058) \\
    Yearly Desulphurization Capacity (10k ton) & 0 & 0 & 0.509 & 0.202 \\
      & (0) & (0) & (1.316) & (0.730) \\
    Sum of Operating Capacity Under Desulphurization & 0 & 0 & 121.8 & 74.01 \\
      & (0) & (0) & (307.1) & (245.8) \\
    \bottomrule
    \bottomrule
Standard Errors are in Parenthesis. &   &   &   &    \\
    \end{tabular}}%
  \label{tab:summarystats1}%
  \\
  \begin{tablenotes}
      This table documents the mean and standard deviation of key variables of interest in 2001, 2004, 2007 and 2010 at county level.
  \end{tablenotes}
\end{table}%
\break
\clearpage
\subsection{Socioeconomic Data}
Socio-economic development data are from county-level Statistical Yearbooks covering the period 2000 to 2010. Since the statistics collected in the Yearbooks varied between years, we were able to obtain only a small set of consistently available variables that may influence public health. These include GDP per capita (or primary and secondary industrial outputs from the County Statistical Yearbooks) and the number of hospital beds per 10,000 people. 




\section{Empirical Method}

We model under-5 mortality as an outcome of both economic development and air pollution. As introduced earlier, omitted variables that are determinants to both air pollution and public health may bias the estimates of air pollution to under-5 mortality. Our empirical method outlined as follows tries to address this concern by exploiting an exogenous change in air quality brought by the small coal power plants closure policy and installation of FGD. As introduced in the Background section, these are compulsory top-down policies and cause exogeneous change in air quality. We show this change is not related to economic growth, and argue to a great extent the estimated impacts on mortality can be deemed as causal. 

In this section, we first present an empirical model that evaluates the effect of policy-induced change in air pollutants levels ($SO_2$ and $PM_{2.5}$) on under-5 mortality in China in section \ref{empiricalframework}, and then introduce our empirical estimation strategy in section \ref{section:lassoiv}. We then use the weighted and unweighted total capacities, along with high-altitude wind direction and speed, as potential instrument candidates for the concentrations of air pollutants. In estimation, we apply the Lasso-IV method, a machine learning process, to select most efficient instrumental variables from a list of the candidates, which are then used in the post-Lasso mortality estimation.


\subsection{Empirical Model}\label{empiricalframework}

The empirical model includes two stages. The first stage estimates the effectiveness of power plant closure and required installation of FGD on air quality. 

\begin{singlespace}\vspace*{-\baselineskip}

\begin{equation} 
P_{pcy}  =  \beta_X X_{pcy} + \Sigma_{m=1}^{12}\gamma_{m}{W_{cym}} + \Sigma_{t=y-3}^{y}\sigma_t\Sigma_{r\geq0}^{R}w_rC_{tr} + u_p + v_y + e_{cy} \label{eq:firststage}
\end{equation}
\end{singlespace}\noindent\ignorespaces

Equation \eqref{eq:firststage} is the first-stage estimation of air pollutant concentration $P_{pcy}$ in county $c$ of province $p$ in year $y$. Air quality $P_{pcy}$ is a function of the county socio-economic development that is described by a vector $X_{pcy}$ (for example, a county's GDP), a set of policy related variables ($C_{tr}$) and high-altitude wind conditions ($W_{cym}$), as well as province fixed effects $u_p$, year fixed effects $v_y$, and the time-varying county-level unobserved characteristics $e_{cy}$. 

Policy-related variables ($C_{tr}$) include variables measuring exposure to coal power plants closure policy and variables measuring exposure to power plant desulphurization (installation of FGD) policy. The plants closure policy-related instruments include sums of closed capacities within a certain radius $R$ ($R=$ 25km, 50km, 100km) from a county's geographical center for the current year or with a lag of up to 3 years. The capacity of each coal power plant may be unweighted or weighted by the inverse of the distance to the center of the county and the direction relative to the dominant wind at the county center (see \ref{section:coalplantsdata} for details about weighting). Similarly, the desulphurization policy-related instrument variables include the unweighted or weighted sum of capacities of coal power plant units that were required to install FGD, as well as the total reduction in sulfate emission from the plants after installing FGD. $W_{cym}$ includes a set of high-altitude wind variables - u and v components of wind at 100 meters above ground that have direct impacts on air pollutants diffusion but do not have direct impacts on human health, especially after controlling for near-ground meteorological conditions. \footnote{A positive u vector denotes wind coming from west and a positive v vector denotes wind coming from south. From the u and v vectors of NASA's daily averaged data we are able to calculate the monthly mean wind direction and wind speed for each county.} 

These policy-related variables ($C_{tr}$) and high-altitude wind variables ($W_{cym}$) will serve as instrumental variables in the estimation of the second stage function, which models the impacts of the induced change in air pollution from Eq.\ref{eq:firststage} on under-5 mortality. 

\begin{singlespace}\vspace*{-\baselineskip}
\begin{equation} 
Mort_{pcy} =  \alpha_Z Z_{pcy} + \Sigma_{m=1}^{12}\varsigma_{m}\ Me_{cym} + \theta P_{pcy} + p*y + U_{cp} + \epsilon_{pcy}\label{eq:main}
\end{equation}
\end{singlespace}\noindent\ignorespaces

\par Under-5 mortality at county level $Mort_{pcy}$ is a function of socio-economic indicators $Z_{pcy}$, meteorological conditions characterised by a vector $Me_{cym}$ that includes standardized monthly mean temperature and humidity for county $c$ in year $y$ at month $m$, air pollutant concentrations $P_{cpy}$, as well as province and year fixed effects $p*y$, county fixed effects $U_{cp}$, and time-varying county effects $\epsilon_{pcy}$. $\theta$ is the coefficient of interest, which measures the effect of reduced air pollution induced by coal power plants closure on under-5 mortality during 2001-2010.

\subsection{Lasso-IV Optimal Instrument Estimation Method}\label{section:lassoiv}
Given a big set of instrumental variables candidates (120 in total), we apply Lasso Method \citep{belloni2014} to select the statistically significant IVs, instead of implementing the traditional 2-stage IV estimation with predetermined instrumental variables. The key assumption of the Lasso-based IV estimation, i.e., Approximately Sparse Optimal Instrument, requires that, given known support of the instrumental variables available, there exist at most $s$ instrument variables in a setting of high-dimensional metrics, such that given the sample size $n$, the approximation error of the expectations of endogenous variables is bounded above by the estimation error, $\sqrt{s/n}$, of the oracle estimator (the estimator on the true support without penalization we would apply in the Lasso approach)\citep{belloni2014}. Different instruments could be selected for different endogenous variables, and the total number of instruments selected for each specification is unknown. 

Lasso method has been applied in several recent studies estimating the impacts of air pollution on health. For example, \citet{deryugina2019} applies Cox-Lasso machine learning model to select excluded instruments from a large set of weather variables for $PM_{2.5}$ in the first-stage estimation. \citet{GODZINSKI2021102489} apply Lasso-IV model to select optimal instruments from a large set of altitude-weather instrumental variables for five air pollutants. 


Our estimation method is similar to \citet{GODZINSKI2021102489}. We apply the R package \textbf{hdm} developed by \citet{chernozhukov2016hdm} for Lasso estimation. We then use the selected instrument variables for the post-Lasso estimation. 

\section{Results}


\subsection{Instruments Selected by Lasso Method}
\par Using the iteration-based IV-Lasso process described above, we conduct the Lasso-based IV selection analysis for a single-pollutant model first. We document the first stage results for the single-pollutant model ($SO_2$ and $PM_{2.5}$ separately) in Table \ref{tab:firststage_projiv} using the whole country data based on equation \ref{eq:firststage}.  For each air pollutant we list the instrumental variables selected, the sample size, the F-statistics of the first stage regression and the adjusted $R^2$ from the first stage regression. Only coefficients of the instrumental variables selected are presented in the tables. 

\par Lasso method has successfully selected a set of instrumental variables that are statistically significantly related to both air pollutants (as shown by F-statistics). Although the variables constructed that calculate the unweighted sum of capacities are included as instrumental variable candidates, only the ones that calculate the weighted sum of retired capacities are selected. For $SO_2$, the weighted sums of retired capacities of power plants within 100km from the county center that are closed in the previous two years and those between 25-100km that are closed in the same year are negatively correlated with $SO_2$ VCDs. In other words, more/bigger power plants were closed within 100 km from a county center in the past 2 years, or more/bigger power plants were closed between 25 and 100 km from a county center in the same year, the average $SO_2$ VCDs of the counties are lower. Although the finding that the weighted sum of capacities of retired coal plants within 100 km in the same year were positively correlated with $SO_2$ VCDs seems inconsistent with our policy assumption, this may be due to imprecise timing of plants closure in a year, since we aggregate the capacities of plants closed from January through December in the same year. We also find the desulphurization policy has lagging effectiveness in $SO_2$ control. The reduced emission of $SO_2$ of the coal plants that were required to install FGD in the previous year is also significantly negatively correlated with the $SO_2$ VCDs.
As we have only considered upwind plants in calculating the weighted sum of closed capacities, wind directions are mostly insignificant, and wind speed 100m above ground in several seasons are found positively correlated with $SO_2$ VCDs. 


Interestingly, a different set of instrumental variables was selected for $PM_{2.5}$, although some IVs overlapped between $PM_{2.5}$ and $SO_2$. Among the variables that calculate the sum of retired capacities, only the weighted sum of projected retired capacities located between 50 and 100km 3 years ago is significantly negatively correlated with concentration of $PM_{2.5}$. On the other hand, the sum of the operating capacities of plants required to install FGD in the same year and 2 years ago were both negatively correlated with $PM_{2.5}$. In addition, the reduced emission of $SO_2$ due to FGD 1 year ago is also negatively correlated with $PM_{2.5}$. However, the reduced $SO_2$ emission in the current year and 2 years ago was positively correlated with $PM_{2.5}$. Different from $SO_2$ which increases with the wind speed, 100 meter above-ground wind helps the diffusion of $PM_{2.5}$, especially in late spring (May and June) and winter months (November and December). Wind direction also has an impact on $PM_{2.5}$. \footnote{we use continuous monthly averaged wind direction variables, with a minimum value of 0 denoting north wind (wind coming from north), 90 denotes east wind, 180 denotes south wind and 270 denotes west wind. In the next version of the paper, we will provide results with categorical wind direction variables}. The different patterns of the effects of wind on $SO_2$ and $PM_{2.5}$ reflect the differences in the mechanism between the two air pollutants. As $SO_2$ was mainly generated from coal-fired power plants, $PM_{2.5}$ is the secondary air pollutant that is formed from the chemical reactions of gases, including $SO_2$ and $NO_2$. Although $SO_2$ VCDs were reduced following the closure of coal power plants and installation if FGDs during the $11^{th}$ FYP, $NO_2$ was not regulated and instead increased significantly due to the surge in automobile use in the same period. As a result, the variation in  $PM_{2.5}$ concentration observed in the first stage was induced by the sulfur control policies and the diffusion effects of high-altitude wind conditions.  

\clearpage

\clearpage
\setlength{\textwidth}{350pt}

\begin{table}[htb]
\tiny
  \centering
  \caption{First Stage Results by Lasso Method (County Level, Projected IV)}
  \setlength\extrarowheight{-3pt}
    \resizebox{\textwidth}{!}{\begin{tabular}{lcccc}
    \toprule
    \toprule
      & \multicolumn{1}{c}{(1)} & \multicolumn{1}{c}{(2)} \\
    VARIABLES & \multicolumn{1}{c}{$SO_2$} & \multicolumn{1}{c}{PM$_{2.5}$} \\
    \midrule
      & \multicolumn{1}{c}{} & \multicolumn{1}{c}{} \\
    Weighted Sum of Projected Retired Capacity (25 to 100km) & \multicolumn{1}{c}{-0.00385**} & \multicolumn{1}{c}{0.0279} \\
      & \multicolumn{1}{c}{(0.00186)} & \multicolumn{1}{c}{(0.0329)} \\
    Weighted Sum of Projected Retired Capacity (25 to 100km, 1-year lag) & \multicolumn{1}{c}{-0.00240*} & \multicolumn{1}{c}{0.0331} \\
      & \multicolumn{1}{c}{(0.00130)} & \multicolumn{1}{c}{(0.0326)} \\
    Weighted Sum of Projected Retired Capacity (25 to 100km, 2-year lag) & \multicolumn{1}{c}{0.000646} & \multicolumn{1}{c}{0.000865} \\
      & \multicolumn{1}{c}{(0.00330)} & \multicolumn{1}{c}{(0.0590)} \\
    Weighted Sum of Projected Retired Capacity (50 to 100km) & \multicolumn{1}{c}{0.00367*} & \multicolumn{1}{c}{0.0634} \\
      & \multicolumn{1}{c}{(0.00219)} & \multicolumn{1}{c}{(0.0386)} \\
    Weighted Sum of Projected Retired Capacity (50 to 100km, 1-year lag) & \multicolumn{1}{c}{} & \multicolumn{1}{c}{-0.0626} \\
      & \multicolumn{1}{c}{} & \multicolumn{1}{c}{(0.0469)} \\
    Weighted Sum of Projected Retired Capacity (50 to 100km, 2-year lag) & \multicolumn{1}{c}{0.00124} & \multicolumn{1}{c}{0.00909} \\
      & \multicolumn{1}{c}{(0.00346)} & \multicolumn{1}{c}{(0.0608)} \\
    Weighted Sum of Projected Retired Capacity (50 to 100km, 3-year lag) & \multicolumn{1}{c}{} & \multicolumn{1}{c}{-0.225***} \\
      & \multicolumn{1}{c}{} & \multicolumn{1}{c}{(0.0818)} \\
    Weighted Sum of Projected Retired Capacity (0 to 100km) & \multicolumn{1}{c}{0.00387***} & \multicolumn{1}{c}{0.00477} \\
      & \multicolumn{1}{c}{(0.00116)} & \multicolumn{1}{c}{(0.0254)} \\
    Weighted Sum of Projected Retired Capacity Under 50 MW (0 to 100km) & \multicolumn{1}{c}{} & \multicolumn{1}{c}{0.00305} \\
      & \multicolumn{1}{c}{} & \multicolumn{1}{c}{(0.0325)} \\
    Weighted Sum of Projected Retired Capacity (0 to 100km, 2-year lag) & \multicolumn{1}{c}{-0.00930***} & \multicolumn{1}{c}{-0.0283} \\
      & \multicolumn{1}{c}{(0.00235)} & \multicolumn{1}{c}{(0.0419)} \\
    Weighted Sum of Projected Retired Capacity (0 to 100km, 3-year lag) & \multicolumn{1}{c}{-0.00956***} & \multicolumn{1}{c}{-0.0398} \\
      & \multicolumn{1}{c}{(0.00196)} & \multicolumn{1}{c}{(0.0478)} \\
    Sum of Operating Capacity Under Desulphurization & \multicolumn{1}{c}{} & \multicolumn{1}{c}{-0.000349**} \\
      & \multicolumn{1}{c}{} & \multicolumn{1}{c}{(0.000173)} \\
    Sum of Operating Capacity Under Desulphurization (1-year lag) & \multicolumn{1}{c}{} & \multicolumn{1}{c}{0.000518***} \\
      & \multicolumn{1}{c}{} & \multicolumn{1}{c}{(0.000187)} \\
    Sum of Operating Capacity Under Desulphurization (2-year lag) & \multicolumn{1}{c}{} & \multicolumn{1}{c}{-0.00118***} \\
      & \multicolumn{1}{c}{} & \multicolumn{1}{c}{(0.000210)} \\
    Yearly Desulphurization Capacity (10k ton) & \multicolumn{1}{c}{0.00231*} & \multicolumn{1}{c}{0.200***} \\
      & \multicolumn{1}{c}{(0.00129)} & \multicolumn{1}{c}{(0.0396)} \\
    Yearly Desulphurization Capacity (10k ton, 1-year lag) & \multicolumn{1}{c}{-0.00407***} & \multicolumn{1}{c}{-0.154***} \\
      & \multicolumn{1}{c}{(0.00129)} & \multicolumn{1}{c}{(0.0419)} \\
    Yearly Desulphurization Capacity (10k ton, 2-year lag) & \multicolumn{1}{c}{} & \multicolumn{1}{c}{0.142***} \\
      & \multicolumn{1}{c}{} & \multicolumn{1}{c}{(0.0458)} \\
    Yearly Desulphurization Capacity (10k ton, 3-year lag) & \multicolumn{1}{c}{} & \multicolumn{1}{c}{-0.0153} \\
      & \multicolumn{1}{c}{} & \multicolumn{1}{c}{(0.0314)} \\
    100m Wind Speed (Jan) & \multicolumn{1}{c}{0.00773**} & \multicolumn{1}{c}{} \\
      & \multicolumn{1}{c}{(0.00301)} & \multicolumn{1}{c}{} \\
    100m Wind Speed (Feb) & \multicolumn{1}{c}{0.00256} & \multicolumn{1}{c}{} \\
      & \multicolumn{1}{c}{(0.00300)} & \multicolumn{1}{c}{} \\
    100m Wind Speed (Mar) & \multicolumn{1}{c}{-0.00147} & \multicolumn{1}{c}{} \\
      & \multicolumn{1}{c}{(0.00435)} & \multicolumn{1}{c}{} \\
    100m Wind Speed (May) & \multicolumn{1}{c}{0.00845**} & \multicolumn{1}{c}{-0.459***} \\
      & \multicolumn{1}{c}{(0.00330)} & \multicolumn{1}{c}{(0.0430)} \\
    100m Wind Speed (Jun) & \multicolumn{1}{c}{} & \multicolumn{1}{c}{-0.449***} \\
      & \multicolumn{1}{c}{} & \multicolumn{1}{c}{(0.0515)} \\
    100m Wind Speed (Jul) & \multicolumn{1}{c}{} & \multicolumn{1}{c}{1.167***} \\
      & \multicolumn{1}{c}{} & \multicolumn{1}{c}{(0.0411)} \\
    100m Wind Speed (Sep) & \multicolumn{1}{c}{0.0243***} & \multicolumn{1}{c}{} \\
      & \multicolumn{1}{c}{(0.00383)} & \multicolumn{1}{c}{} \\
    100m Wind Speed (Oct) & \multicolumn{1}{c}{0.0367***} & \multicolumn{1}{c}{} \\
      & \multicolumn{1}{c}{(0.00372)} & \multicolumn{1}{c}{} \\
    100m Wind Speed (Nov) & \multicolumn{1}{c}{} & \multicolumn{1}{c}{-0.617***} \\
      & \multicolumn{1}{c}{} & \multicolumn{1}{c}{(0.0443)} \\
    100m Wind Speed (Dec) & \multicolumn{1}{c}{0.0145***} & \multicolumn{1}{c}{-0.778***} \\
      & \multicolumn{1}{c}{(0.00336)} & \multicolumn{1}{c}{(0.0401)} \\
    100m Wind Direction (Feb) & \multicolumn{1}{c}{} & \multicolumn{1}{c}{0.00332***} \\
      & \multicolumn{1}{c}{} & \multicolumn{1}{c}{(0.000286)} \\
    100m Wind Direction (Mar) & \multicolumn{1}{c}{} & \multicolumn{1}{c}{-0.00410***} \\
      & \multicolumn{1}{c}{} & \multicolumn{1}{c}{(0.000290)} \\
    100m Wind Direction (Apr) & \multicolumn{1}{c}{} & \multicolumn{1}{c}{-0.000826**} \\
      & \multicolumn{1}{c}{} & \multicolumn{1}{c}{(0.000321)} \\
    100m Wind Direction (Aug) & \multicolumn{1}{c}{-2.24e-05} & \multicolumn{1}{c}{-0.00185***} \\
      & \multicolumn{1}{c}{(1.81e-05)} & \multicolumn{1}{c}{(0.000244)} \\
    100m Wind Direction (Oct) & \multicolumn{1}{c}{5.81e-05**} & \multicolumn{1}{c}{} \\
      & \multicolumn{1}{c}{(2.69e-05)} & \multicolumn{1}{c}{} \\
    100m Wind Direction (Nov) & \multicolumn{1}{c}{1.93e-05} & \multicolumn{1}{c}{-0.00215***} \\
      & \multicolumn{1}{c}{(3.14e-05)} & \multicolumn{1}{c}{(0.000425)} \\
    100m Wind Direction (Dec) & \multicolumn{1}{c}{-7.80e-05*} & \multicolumn{1}{c}{0.00380***} \\
      & \multicolumn{1}{c}{(4.16e-05)} & \multicolumn{1}{c}{(0.000579)} \\
    \midrule
    \# Instruments Selected & \multicolumn{1}{c}{21} & \multicolumn{1}{c}{29} \\
    Sample Size & \multicolumn{1}{c}{7,275} & \multicolumn{1}{c}{10,723} \\
    Time periods & \multicolumn{1}{c}{2005-2010} & \multicolumn{1}{c}{2001-2010} \\
    Aggregation Level & \multicolumn{1}{c}{County} & \multicolumn{1}{c}{County} \\
    Fixed Effect & \multicolumn{1}{c}{Year \& County} & \multicolumn{1}{c}{Year \& County} \\
    R-squared & \multicolumn{1}{c}{0.861} & \multicolumn{1}{c}{0.987} \\
    F-stat & \multicolumn{1}{c}{26.89} & \multicolumn{1}{c}{451.89} \\
    \bottomrule
    \bottomrule
    \multicolumn{2}{l}{Standard Errors in Parenthesis} &   &   &  \\
    \multicolumn{2}{l}{*** p$<$0.01, ** p$<$0.05, * p$<$0.1} &   &   &  \\
    \end{tabular}}%
  \label{tab:firststage_projiv}%
  \\
  \begin{tablenotes}
      Only selected instrumental variables by Lasso are included. Column (1) and (2) are the results of instruments selected for $SO_2$ and $PM_{2.5}$, respectively. We control for year and county fixed effects and the same set of monthly standardized meteorological variables (2-meter temperature and humidity) in each model.
  \end{tablenotes}
\end{table}%
\clearpage
\pagebreak

\subsection{County Level Whole Sample Results}
Table \ref{tab:singleresult_whole_city_proj} presents the county-level post-Lasso estimation results for the multi-pollutant model with the instrumental variables selected by Lasso method for the whole country. We present the bi-pollutant IV-Lasso results in column (1) and its baseline fixed-effects model results without including IVs in column (2) for comparison. For both IV-Lasso and baseline models, we control for year fixed effects and county fixed effects, as well as a set of socio-economic covariates, meteorological variables as introduced in \ref{section:data}. Standard errors are clustered at province level. 

As expected, per capita GDP growth of a county decreases under-5 mortality. In the baseline fixed effects model, only primary industry output decreases under-5 mortality. Although secondary industry output also has negative coefficients in Column (2), it is not statistically significant. This is likely due to collinearity between secondary industry output and the concentrations of air pollutants. As a result, the coefficient on $SO_2$ is insignificant in the baseline model, while the coefficient on $PM_{2.5}$ is positive but weakly significant. In the Lasso-IV model, both primary and secondary industry outputs decrease under-5 mortality (column (1)). A 10,000 RMB increase in a county's primary and secondary industry output per capita population decreases under-5 mortality by 2.58\textperthousand \space and 1.32\textperthousand \space separately.

Different from the baseline fixed effects model, Lasso-IV estimation model gives positive and significant effects of both $SO_2$ and $PM_{2.5}$ on under-5 mortality at county level in column (1). More specifically, one standard deviation increase in $SO_2$ VCDs (around 0.3 DU) would lead to an increase in under-5 mortality by 3 per 1,000,000 while one standard deviation increase in concentration of $PM_{2.5}$ (around 18 $\mu$g$^{-3}$) would result in an increase of under-5 mortality by 1.8\textperthousand, assuming other control variables remain unchanged. As the units between column $SO_2$ and $PM_{2.5}$ concentration are different, it is hard to compare the scales of the effects between these two pollutants. Moreover, total column $SO_2$ is a measurement of vertical column densities of the sulfur dioxide in the atmosphere. Although it reflects emission by volcanic eruptions and anthropogenic pollution especially by coal power plants, it can not be directly translated into ambient $SO_2$ concentration. When interpreting the results, it is cautioned that the estimated coefficient can not be directly interpreted as the dose-response relationship between ambient $SO_2$ and under-5 mortality.

Constrained by the data availability, we are not able to examine age-specific and cause-specific mortality in analysis.  

    \begin{table}[htb]
\scriptsize
  \centering
   \caption{ Post-Lasso Estimation Results (County Level)}
   \begin{tabular}{lcc}
    \toprule
    \toprule
      & \multicolumn{1}{c}{IV-Lasso}  & \multicolumn{1}{c}{Baseline Fixed Effects}\\
    Variables & (1) & (2) \\
    \midrule
    $SO_2$ (DU)   & 0.00134***  & -0.405  \\
         & (0.00027)  & (0.728)\\
    PM$_{2.5}$ ($\mu$g$^{-3}$)  & 0.176*** & 0.108* \\
       &  (0.0381) & (0.0654)\\
    Hospital Bed (per 10,000 population) & 0.00863&  -0.00407**\\
       & (0.0150)  & (0.00176) \\
    Prim GDP (10,000 CNY per capita)   & -2.58** & -6.82*** \\
        & ((1.26) & (1.95)\\
    Sec GDP  (10,000 CNY per capita)   & -1.32*** &  -0.0001 \\
        & (0.328)  & (0.179)\\
    \midrule
    Sample Size  & 7275 & 7275\\
    Aggregation Level  & County & County\\
    Fixed Effect  & Year \& County & Year \& County\\
    \bottomrule
    \bottomrule
    \multicolumn{1}{l}{Standard Errors in Parenthesis} &   & \\
    \multicolumn{1}{l}{*** p$<$0.01, ** p$<$0.05, * p$<$0.1} &    &  \\
    \end{tabular}
  \label{tab:singleresult_whole_city_proj}%
  \begin{tablenotes}
We document the IV-Lasso results (post-Lasso estimation) in column (1). The baseline Fixed Effect estimates for the multi-pollutant model is documented in column (2). We control for year and county fixed effects and the same set of monthly standardized meteorological variables (2-meter temperature and humidity) in all the documented results above. Both variables Hospital bed is measured as per capita term. Primary and Secondary GDP are per capita measures in 10,000 CNY. 
\end{tablenotes}
\end{table}%

\subsection{Validity of Instruments}\label{balancetest}
To interpret the effect of air pollution on under-5 mortality identified through the mandated closure of small power plants and installation of FGDs, we have to make assumption that power plants closure and installation of FGDs only affects under-5 mortality through improved air quality. However, one might be concerned that counties with small power plants being closed could have experienced higher unemployment rate and slowed economic growth, which might, in turn, increase under-5 mortality. If this is the case, our model may underestimate the impacts of the air pollution policy on under-5 mortality.  

\begin{figure}[htp]
\centering
\includegraphics[width=\textwidth]{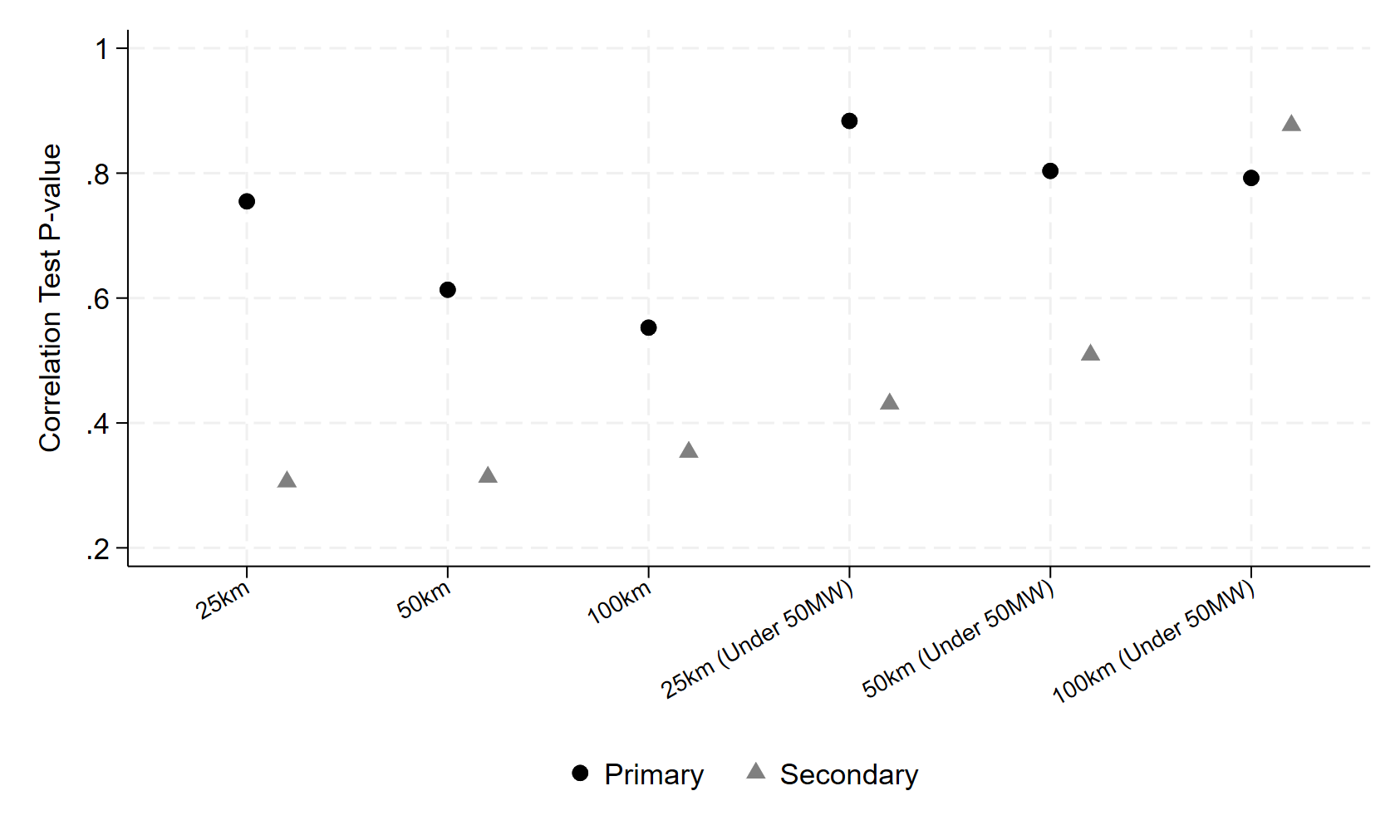}
\caption{Correlation: Weighted Sum of Retired Capacity and Per Capita Primary \& Secondary GDP Growth Rate}
\begin{figurenotes}
This figure presents the p-values of correlation tests between policy-related instrumental variables (distance-weighted sums of retired capacities within 25km, 50km and 100 km from a county centroid) and annual economic growth rates of primary and secondary per capita GDP from 2001 to 2010 with the null hypothesis that the correlation equals 0.
\end{figurenotes}
\label{fig:balancetest_corr_proj}
\end{figure}

Given the data constraint, no unemployment rate data at county level were available during the study period for us to test if unemployment increases after small power plants closed. Instead, we conduct a balance test on the correlations between the policy instrumental variables and county-level economic growth. Figure \ref{fig:balancetest_corr_proj} presents the p-values of the correlations between county-level distance-weighted sums of retired capacities within specific radius (25km, 50km, 100km) from a county centroid and annual economic growth rates (primary and secondary GDP per capita separately). The figure shows all p-values are above 0.1, which implies that the capacities of closed coal power plants within given radius from a county centroid are not correlated with annual growth in primary and secondary GDP per capita. This result addresses part of the concern that policy-related instrumental variables may influence under-5 mortality not only through air pollution but also through other channels, which could undermine the validity of the policy-related instrumental variables. 

\section{Heterogeneity Analysis}\label{section:sensitivity}

In the heterogeneity analysis, we examine whether there exist displacement effects and spillover effects of the health benefits of coal power plants policies across regions. Through displacement effects, we hypothesize that coal power plants may increase production and air pollution in regions with less stringent environmental regulations, potentially diminishing or even negating the health benefits in the displaced areas. Regarding spillover effects, we test the hypothesis that the health benefits of sulfur control policies in one region may extend to neighboring regions, even in the absence of sulfur control policies there. In this section, we first test for displacement effects, focusing on the impact of the West-to-East Power Transfer Program on air quality, as well as the varying health impacts of power plant policies across regions, categorized as either exporters or importers of electricity under the program. Next, we examine spillover effects, specifically analyzing the heterogeneous health impacts of power plant policies across regions, categorized according to sulfur control zoning policy. 

\subsection{West-to-east power transfer program and displacement effects}\label{section:regionsubsample}
\subsubsection{West-to-east power transfer program}
In 1999, China Central Government initiated China Western Development strategy. The goal is to address the regional disparity in economic development between the more developed eastern China and less developed western China, by increasing investment in infrastructure including transportation, hydropower, fossil fuel, telecommunications, and other policies. In this context, the West-to-East Power transfer Program was launched to transfer electricity generated in western China to the eastern regions. The purposes of the West-to-East Power Transfer Program are two folds: improving economic development of the West China and reducing air pollutants in the already polluted East. Three major power grid networks were formed: the northern network that transmits electricity generated from hydropower on the upper and middle reaches of the Yellow River and from coal-fired power plants in the mid-West China (Shanxi, Shaanxi and Inner Mongolia) to the North China (Beijing, Tianjin, Hebei, Shandong); the central network that transports hydropower from the Three Gorges Dam and the Jinsha River and its tributaries to the eastern China region (Shanghai, Zhejiang, Jiangsu etc); and the southern network that develops and transmits electricity from hydropower and coal-fired power plants in the Southwest (Sichuan, Yunnan, Guizhou), to Guangdong and Guangxi. For the southern transfer network alone, transfer capacity and electricity amount increased by 10 times from 2000 to 2010. By 2010, the maximum power capacity of electricity to Guangdong transmitted from the west reached 18.55 gigawatt (GW), accounting for 27\% of the maximum load in Guangdong \citep{Zeng2013}. 

In the later heterogeneity analysis, we divide counties into 3 subsamples according to West-to-East Power Transfer Program: East, Southwest and Northwest subsamples. The East China subsample includes all the provinces and municipalities in East and Central China (including Beijing, Tianjin, Hebei, Henan, Shanxi, Shandong, Jiangsu, Anhui, Shanghai, Hubei, Hunan, Jiangxi, and Zhejiang). These regions are mainly the recipients of the west-east power transfer program, characterized by high population and industry densities which create the high demand for electricity. The Northwest China subsample includes four northwestern provinces/autonomous regions that transfer electricity through the north power grid network: Shanxi\footnote{Shanxi appears in both the northwest subsample and the east subsample because of its high power generation capacities and high energy usage.}, Shaanxi, Ningxia, and Inner Mongolia. These provinces/regions are characterized by relatively low population density (especially in Ningxia and Inner Mongolia) and rich coal deposits, as well as densely distributed thermal power plants (see Figure \ref{fig:mortality1}). The third regional subsample, the Southwest China subsample, consists of 5 provinces (Sichuan, Yunnan, Guangxi, Hunan and Guizhou) and 1 municipality city (Chongqing) in the southwest part of China with both rich coal deposits and hydropower deposits that transfer electricity through the south power grid network. After many power plants were closed during the $11^{th}$ FYP, provinces in the East and Central China imported electricity from the West China to meet the increasing demand for energy. Given the less stringent environmental regulations and less efficient governance in the western provinces, we hypothesize that counties in the exporting regions - Southwest and Northwest China - experienced less abatement in air pollution during the $11^{th}$ FYP and that under-5 mortality were less affected by the coal plants closure policy.

\par
\begin{figure}[htbp]
\centering
\begin{tabular}{c}
     \includegraphics[width=.5\textwidth]{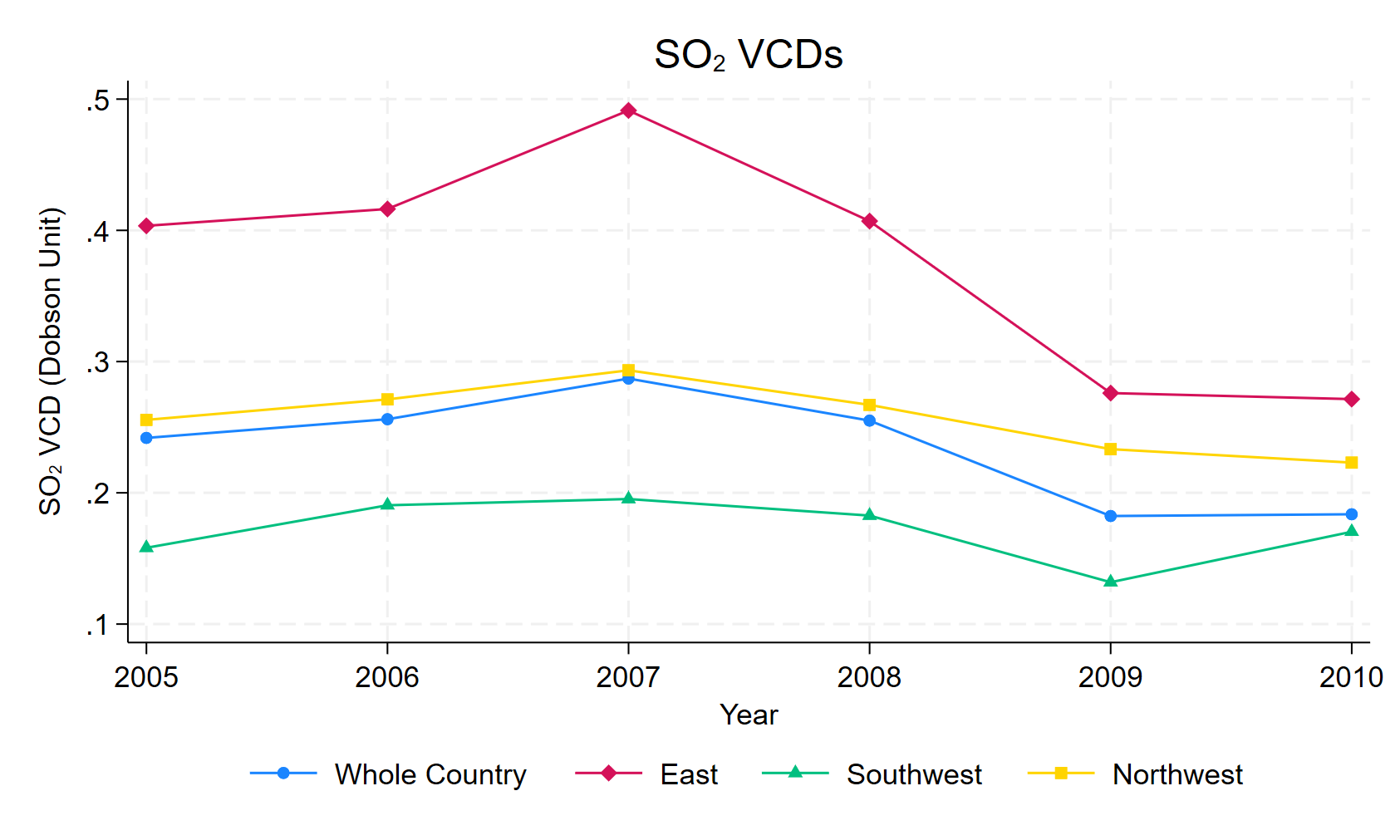} 
     \includegraphics[width=.5\textwidth]{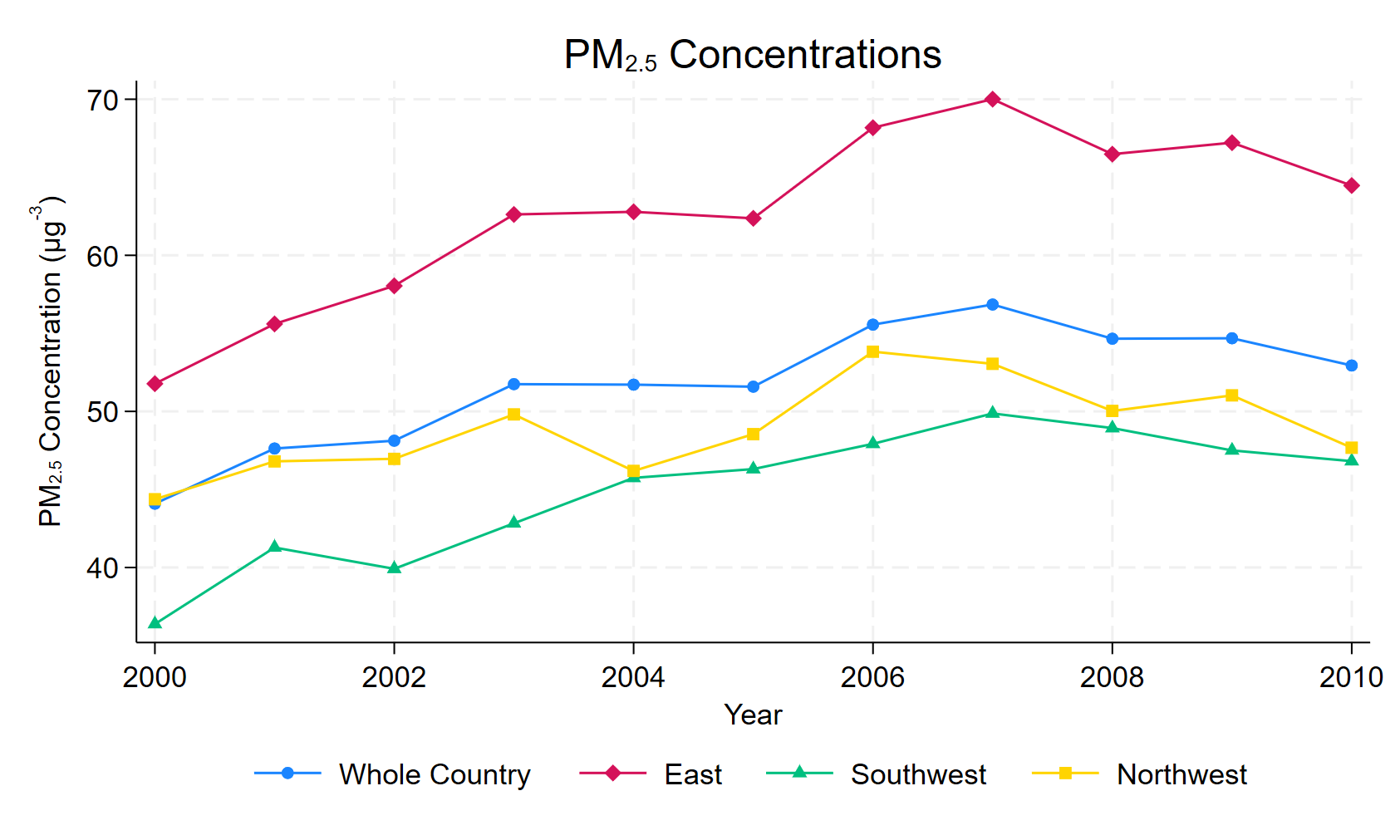} 
\end{tabular}
\caption{Yearly Trends of $SO_2$ and $PM_{2.5}$}
\begin{figurenotes} 
This figure documents and compares the time trends of annual average of (a) $SO_2$ and (b) $PM_{2.5}$ in the whole sample and three subsamples defined by geographic regions (East, Southwest and Northwest). 
\end{figurenotes}
\label{fig:subsampletrend_geo}
\end{figure}

 Figure \ref{fig:subsampletrend_geo} illustrates the annual trends in average $SO_2$ VCDs and $PM_{2.5}$ densities for the entire country sample as well as for regional subsamples. For both $SO_2$ and $PM_{2.5}$, it is evident that East subsample has significantly higher average pollution levels compared to the Southwest and Northwest regions in all years. In terms of the time trend, all regions show an increase in $SO_2$ levels until 2007, followed by a consistent decline thereafter, except in the Southwest, where $SO_2$ VCD rose from 2009 to 2010. Due to the Southwest's abundant coal reserves with higher sulfur content and the region's weaker government enforcement, the increased demand for electricity driven by the West-to-East Energy Transfer Program has led to a rise in $SO_2$ emissions in the area. This is consistent with the finding that $SO_2$ emissions increased during $11^{th}$ FYP in Sichuan basin \citep{Xia2016}. However, the decline in the East subsample after 2007 is much more pronounced than in the other two regions. A similar general trend is observed for $PM_{2.5}$, which increased until 2007 (or 2006 for the Northwest region) and then declined, except in the Northwest, where $PM_{2.5}$ rose from 2008 to 2009. Furthermore, the declines in $PM_{2.5}$ after 2007 across all regions are much less pronounced than those in $SO_2$, which aligns with the observations for the national average. 


\vspace{0cm}
\vspace{0cm}

\par  We document the capacity commissioned and retired from 2000 to 2010 in Figure \ref{fig:pollutantdensity_ea}. One interesting finding from the capacity distribution figures is that, after increasing investments in thermal power plants in 2006 and 2007, we observe a huge cascade in the thermal capacity commissioned in 2008 for the southwest subsample in comparison with our other subsamples and the preceding year within the same subsample. From the breakdown statistics of total capacity closed, we also observe a concentration of under-50 capacity retired in 2007 for the southwest subsample that greatly distinguished from the other years during the policy period accounting for nearly half of total under-50 capacity closed from 2001 to 2010. The Northwest China subsample and the East China subsample show similar patterns in terms of retired and commissioned capacity distributions. 

\subsubsection{Displacement Effects}
\par Table \ref{tab:geosubsample_proj} compares the bi-pollutant Lasso-IV model results across the three geographical subsamples using the projected policy instruments (Columns (1)-(3)). For comparison, we also present the baseline fixed effect estimates for the three subsamples in Columns (4)-(6). 
While baseline fixed effects show insignificant effects of $SO_2$ across all subsamples and adverse effects of $PM_{2.5}$ for East and Northwest subsamples, the Lasso-IV results show heterogeneous adverse effects across geographic regions for both $SO_2$ and $PM_{2.5}$. With Lasso-IV results, the partial effect estimates of both $SO_2$ and $PM_{2.5}$ are highest in the Northwest subsample, and are lowest but significant in the East subsample, except for Southwest where $SO_2$ shows insignificant impact on under-5 mortality. The insignificant effect does not imply $SO_2$ has no adverse impacts on under-5 mortality in the Southwest. Rather, it is most likely due to the weak instrumental variables. Only one power plant closure policy-related variable was selected to predict $SO_2$ in the first-stage estimation for the Southwest subsample, significantly fewer than the ten variables in the East subsample and three in the Northwest subsample. The poor performance of the first-stage estimation in the Southwest region suggests that $SO_2$ levels in the Southwest show little response to coal plant closures and FGD installation policies, as indicated by the increase in $SO_2$ VCDs from 2009 to 2010 in the Southwest. As a result, the variation in $SO_2$ induced by these policies is too minimal to predict under-5 mortality effectively in this region. 

The marginal impact of $PM_{2.5}$ on under-5 mortality is at the similar magnitude across three subsamples with Lasso-IV method, with the Northwest having the highest effect, followed by Southwest and East subsample. The findings that the coefficients on air pollutants on under-5 mortality are higher in the Northwest and Southwest (for $PM_{2.5}$ only) regions than in the East China imply that the marginal effects of air pollution on under-5 mortality are higher in the less developed West China than in the more developed East China. This finding is consistent with the evidence of socioeconomic modification of air pollution health effects at individual level \citep{ONeill2003}. At the regional level, less developed areas have fewer public health care resources to mitigate the adverse effects of air pollution on under-5 mortality.

\begin{landscape}
    \begin{table}[htb]
\scriptsize
  \centering
  \caption{Post-Lasso Estimation Results (County Level, Geographic Subsamples)}
    \begin{tabular}{lcccccc}
    \toprule
    \toprule
      & \multicolumn{3}{c}{IV-Lasso}  & \multicolumn{3}{c}{Baseline Fixed Effects}\\
    Variables & (1) & (2) & (3) & (4) & (5) & (6)\\
    \midrule
    $SO_2$ (DU) & 0.00148*** & 0.00247*** & 0.00157&-0.704 & -2.228 & 1.304 \\
      & (0.00028) & (0.00012) & (0.00153) & (0.694) & (2.028) & (1.405)\\
    $PM_{2.5}$ ($\mu$g$^{-3}$) & 0.178*** & 0.405*** & 0.291** & 0.161** & 0.278** & -0.333  \\
      & (0.0355) & (0.0194) & (0.103)& (0.0715) & (0.0695) & (0.239) \\
    Hospital Bed (per 10,000 population) & -0.0327** & -0.0272 & -0.167*** & -0.0793*** & -0.000259 & -0.000739*\\
      & (0.0140) & (0.0250) & (0.0500) & (0.0211) & (0.0413) & (0.000489) \\
    Prim GDP (10,000 CNY per capita) & -3.13* & 16.0*** & -33.7*** & -4.13 & -11.6** & 12.6\\
      & ((2.44) & (2.21) & (10.90) & (3.68) & (3.21) & (7.90) \\
    Sec GDP (10,000 CNY per capita) & -1.58*** & -0.0269 & -2.15***& 0.310 & -0.096 & 1.91 \\
      & (0.381) & (0.212) & (0.670) & (0.311) & (0.089) & (1.790)\\
    \midrule
    Subsample & East & Northwest & Southwest & East & Northwest & Southwest\\
    \# Policy Instruments Selected ($SO_2$) & 10 & 3 & 1 &  &  & \\
    First Stage F-stat ($SO_2$) & 0.856 & 0.899 & 0.767 &  &  & \\
    Sample Size & 4653 & 1258 & 1274 & 4653 & 1258 & 1274\\
    Aggregation Level & County & County & County & County & County & County\\
    Fixed Effect & Year \& County & Year \& County & Year \& County & Year \& County & Year \& County & Year \& County\\
    \bottomrule
    \bottomrule
    \multicolumn{2}{l}{Standard Errors in Parenthesis} &   &   &  \\
    \multicolumn{2}{l}{*** p$<$0.01, ** p$<$0.05, * p$<$0.1} &   &   &  \\
    \end{tabular}%
  \label{tab:geosubsample_proj}%
  \\
   \begin{tablenotes}
We present the multi-pollutant model IV-Lasso results (post-Lasso estimation) in columns (1), (2) and (3) for our three geographical subsamples: the East China subsample, the Northwest China subsample and the Southwest China subsample, respectively. The baseline Fixed Effect multi-pollutant estimates for the three subsamples are documented in columns (4), (5) and (6), respectively. We control for year and county fixed effects and the same set of monthly standardized meteorological variables (2-meter temperature and relative humidity) in all the documented results above. Only multi-pollutant post-Lasso estimation results are documented in the main text for comparison and heterogeneity analysis purposes. Hospital bed is measured as per capita term. Primary and Secondary GDP are per capita measures in 10,000 CNY. 
\end{tablenotes}
\end{table}%
\end{landscape}


\subsection{Spill Over Effects? 
- Heterogeneity analysis by Sulfur Control Policy-Specific Subsamples}\label{section:policysubsample}

\par In the second exercise, we define the subsamples according to sulfur control zoning policy, which designates the counties into Acid Rain Control Zones (ARCZ) and the Sulphur Dioxide Pollution Control Zones (SO2CZ). The two types of control zones were delimited in 1998 by Ministry to Ecology and Environment, with the targets to control the total amount of discharged sulphur dioxide and the density of sulphur dioxide in cities as well as to significantly decrease the area of the acid rain control zones with precipitation pH value below 4.5 \citep{controlzones1998}. SO2CZ contains areas in the north and northeastern of China, which was the traditional center of heavy industry in China centered on coal and steel. Although SO2CZ has high emission of $SO_2$, acid rain is not prominent because wind-blown, alkaline soil dust commonly existing in the northern China can neutralize rain acidity \citep{Larssen1999}. 192 counties and cities located in the south part of China were included into the Acid Rain Control Zone based on the pH value and frequencies of precipitation while excluding the poorest counties for which economic development took priority. Most high-sulfur coal (with more than 3\% sulfur) in China was produced in this area in the time period, with 6.4\% of China's coal output while accounting for over 20\% of $SO_2$ emission in 1998 \citep{china1998coal}. For comparison, we also define a Non-sulfur policy subsample (NSP), i.e., the counties that are out of both control zones. NPS is interwoven closely with ARCZ and SO2CZ, especially in the east and central China. It consists of mainly poor counties for which economic development took priority, and therefore was not subject to sulfur control regulations at that time. In this section, we are examining whether $SO_2$ is reduced in this subsample and whether the benefits of sulfur control policies have been spilled over to these counties. 

\par
\begin{figure}[htbp]
\centering
\begin{tabular}{c}
     \includegraphics[width=.5\textwidth]{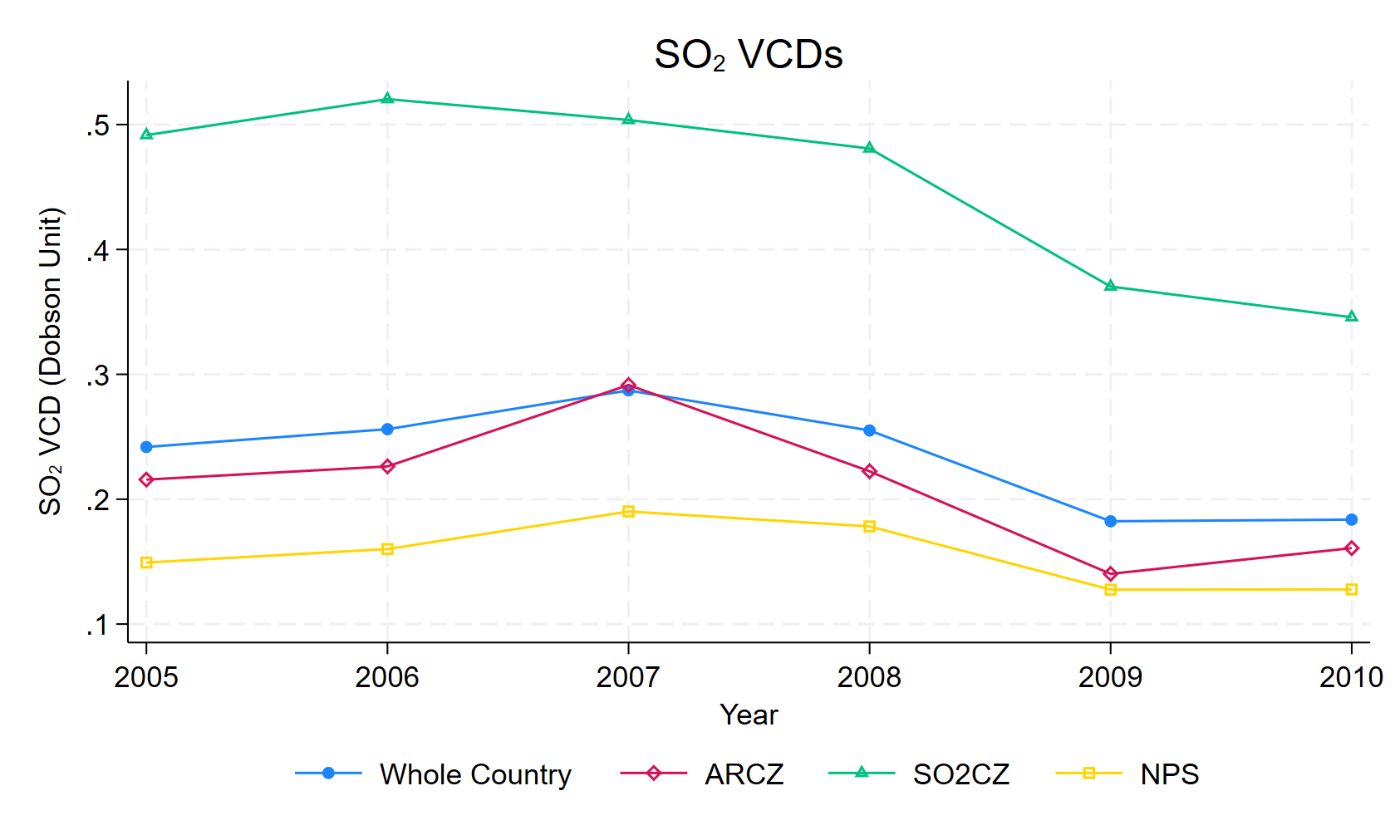} 
     \includegraphics[width=.5\textwidth]{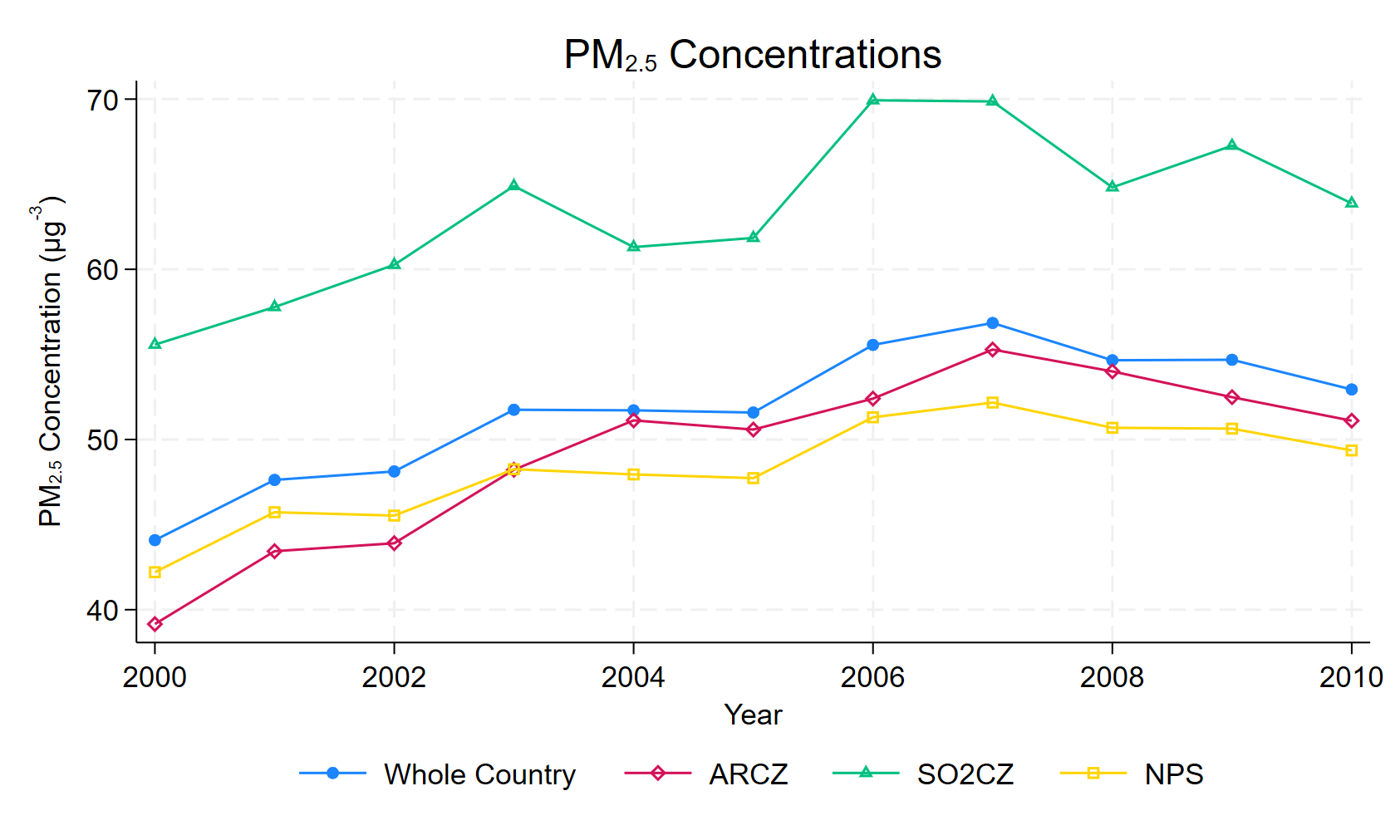} 
\end{tabular}
\caption{Yearly Trends of $SO_2$ and $PM_{2.5}$}
\begin{figurenotes} ARCZ = Acid Rain Control Zones; SO2CZ = Sulfur Dioxide Control Zones; NPS = Areas not belonged to ARCA or SO2CZ. 
This figure documents and compares the time trends of annual average of (a) $SO_2$ and (b) $PM_{2.5}$ in the whole sample and three policy subsamples defined by sulfur control zoning policy (ARCZ, SO2CZ and NPS). 
\end{figurenotes}
\label{fig:subsampletrend_pol}
\end{figure}

 Figure \ref{fig:subsampletrend_pol} illustrates the average $SO_2$ VCDs across three regions during the period from 2005 to 2010.\footnote{The density heat maps of $SO_2$ and $PM_{2.5}$ for each subsample are presented in Appendix Figures \ref{fig:ARCZ_density}, \ref{fig:sca_density}, \ref{fig:none_density}} SO2CZ experienced the highest average $SO_2$ VCDs, followed by ARCZ and NPS. $SO_2$ VCDs peaked in 2006 for SO2CZ, in 2007 for both ARCZ and NPS, before subsequently declining in all three regions. Similarly, SO2CZ experienced the highest average $PM_{2.5}$, followed by ARCZ and NPS. ARCA surpassed NPS in 2004 with a higher average $PM_{2.5}$. Across all regions, $PM_{2.5}$ concentrations have increased significantly since 2000, peaking around 2007 before experiencing a moderate decline. 

\par Coal power plants closure policy was rigorously enforced during $11^{th}$ FYP in ARCZ and SO2CZ. Comparing Figure \ref{fig:pollutantdensity_ARCA} for ARCZ with Figure \ref{fig:pollutantdensity} for the entire country, we find that the under-50MW capacity closed at the start of the policy period (2006) in ARCZ accounted for more than half of the total under-50MW capacity closed nationwide in 2006. This share remained high for the ARCZ subsample until 2008, during which the under-50MW share remained much higher than the country average.  
\par Although no substantial policy-related closures took place in 2006, SO2CZ experienced substantial plants closure in 2007, 2008 and 2010. Again, in the earlier years, under-50MW capacity units were closed with priority. (Figure \ref{fig:pollutantdensity_sca}). With very different meteorological conditions (much lower humidity and precipitation levels in the north), we expect different health impacts of $SO_2$ between these two control zones. 
\par Although NPS is not subject to the sulfur control zoning policy, a significant amount of capacity was still closed down in this area (Figures \ref{fig:pollutantdensity_none}). Large scale of closure started in 2007 and showed similar trend as in SO2CZ.

\vspace{0cm}
\vspace{0cm}

\par The results of bi-pollutant estimation for the three policy-specific subsamples are presented in Table \ref{tab:policysub_proj}. Columns (1)-(3) show results for the ARCZ, SO2CZ and NPS, respectively, and columns (4)-(6) show results of baseline fixed effect estimates for the three regions in the same order. While the effects of $SO_2$ are insignificant and negative in the baseline fixed effects model, the IV-Lasso model obtains the significant and positive effects on under-5 mortality, with the effects being highest in NPS, and similar between SO2CZ and ARCZ. For $PM_{2.5}$, NPS also shows the highest marginal effect on under-5 mortality, while SO2CZ shows the lowest marginal effect. As part of the heterogeneity analysis, we performed similar analysis on the NPS subsample by removing counties from the 4 western provinces (Tibet, Xinjiang, Gansu and Qinghai) with lower economic development and industrial activities, the results are robust to the sample change in both magnitude and direction of partial effects.

\begin{landscape}
    \begin{table}[htb]
\scriptsize
  \centering
  \caption{Post-Lasso Estimation Results (County Level, Policy Subsamples)}
    \begin{tabular}{lcccccc}
    \toprule
    \toprule
      & \multicolumn{3}{c}{IV-Lasso}  & \multicolumn{3}{c}{Baseline Fixed Effects}\\
    Variables & (1) & (2) & (3)& (4) & (5) & (6) \\
    \midrule
    $SO_2$ (DU) & 0.00116*** & 0.00116*** & 0.00120***& -0.713 & -0.612 & -0.732  \\
      & (0.00006) & (0.00009) & (0.00043) & (0.780) & (0.473) & (0.818) \\
    PM$_{2.5}$ ($\mu$g$^{-3}$) & 0.182*** & 0.122*** & 0.192*** & 0.0946 & 0.104 & 0.141*  \\
      & (0.012) & (0.00945) & (0.0689) & (0.155) & (0.0861) & (0.0774)\\
    Hospital Bed (per 10,000 population) & 0.00642 & -0.0612*** & 0.108*** & -0.114*** & -0.0367** & -0.00268***\\
      & (0.020) & (0.019) & (0.024) & (0.0328) & (0.0138) & (0.000518) \\
    Prim GDP (10,000 CNY per capita) & -5.01*** & 0.655 & -9.63*** & -3.97* & -5.61*** & -7.53** \\
      & (1.090) & (0.776) & (3.690) & (2.34) & (1.21) & (2.75) \\
    Sec GDP (10,000 CNY per capita) & -1.17*** & -0.17*** & {-1.48***}& 0.955*** & -0.0442 & -0.516*** \\
      & (0.284) & (0.234) & (0.463) & (0.264) & (0.277) & (0.175)  \\
    \midrule
    Subsample & ARCZ & SO2CZ & Non-policy & ARCZ & SO2CZ & Non-policy \\
    \# Policy Instruments Selected ($SO_2$) & 4 & 5 & 7 &  &  & \\
    First Stage F-stat ($SO_2$) & 0.838 & 0.870 & 0.846 &  &  & \\    Sample Size & 2559 & 1884 & 3417 & 2559 & 1884 & 3417\\
    Aggregation Level & County & County & County & County & County & County\\
    Fixed Effect & Year \& County & Year \& County & Year \& County & Year \& County & Year \& County & Year \& County\\
    \bottomrule
    \bottomrule
    \multicolumn{2}{l}{Standard Errors in Parenthesis} &   &   &  \\
    \multicolumn{2}{l}{*** p$<$0.01, ** p$<$0.05, * p$<$0.1} &   &   &  \\
    \end{tabular}%
  \label{tab:policysub_proj}%
  \\
   \begin{tablenotes}
We document the multi-pollutant model IV-Lasso results (post-Lasso estimation) in columns (1), (2) and (3) for our three policy-specific subsamples: the Acid Rain Control Zones (ARCZ) Policy subsample, the Sulphur Dioxide Pollution Control Zones (SO2CZ) Policy subsample and the Non-policy subsample, respectively. The baseline Fixed Effect multi-pollutant estimates for the three subsamples are documented in columns (4), (5) and (6), respectively. We control for year and county fixed effects and the same set of monthly standardized meteorological variables (2-meter temperature and humidity) in all the documented results above. Only multi-pollutant post-Lasso estimation results are documented in the main text for comparison and heterogeneity analysis purposes. Hospital bed is measured as per capita term. Primary and Secondary GDP are per capita measures in 10,000 CNY. 
\end{tablenotes}
\end{table}%
\end{landscape}

\section{Discussion and Conclusion}
To establish a causal relationship between decreased air pollution and improved public health statistics, this paper examines the role of mandatory coal power plant phase-out policy and desulphurization policy during 2005-2010 in reducing under-5 child mortality. We calculate the weighted capacities of coal power plants closed or required to install FGDs within a certain radius from a county centroid and use it to serve as instrumental variable candidates, along with high-altitude wind information. We use Lasso-IV estimation to select a small set of efficient instrumental variables and estimate the causal impacts on under-5 mortality using data from 1314 counties throughout China. Our analysis shows that the coal power plant phase-out policy the Chinese government carried out during the Eleventh Five-Year Plan period significantly reduced densities of two major air pollutants -- SO$_2$ and PM$_{2.5}$ -- which contributed to the reduction in under-5 mortality. Since the average decrease in $SO_2$ VCDs and $PM_{2.5}$ density at the end of the policy period compared with peak values was 0.1 DU for $SO_2$ and 3.9 $\mu$g$^{-3}$ for PM$_{2.5}$, and the under-5 population in China was 68,978,374 according to the 2000 census   \citep{chinacensus2000}, the back-of-envelope calculation based on our estimates suggests the under-5 lives saved during the period of 2006-2010 was 46,012. This calculation provides a lower bound of the estimate for evaluating the health benefit of the sulfur control policies during $11^{th}$ FYP. Since $PM_{2.5}$ is a secondary pollutant formed in the atmosphere through chemical reactions of gases such as $SO_2$ and $NO_X$ and organic compounds, and given that $NO_X$ increased instead of decreased during $11^{th}$ FYP, the reduction in $PM_{2.5}$ resulting from $SO_2$ reduction could have been greater. 

The heterogeneous effects observed across regions suggest that the marginal impact of air pollution can vary based on a region's socio-economic development. In areas with lower economic development and weaker healthcare systems, air pollution may be more harmful than in more developed regions, even when pollution levels are lower. This implies a potential non-linearity in the relationship between air pollution and mortality, driven not only by pollution levels but also by the interplay between air pollution and socio-economic factors. 

China’s economic developments and efforts in air pollution control since its economic reform in 1978 serve as a valuable lesson to developing countries with large energy consumption and needs for economic development. Environmental concerns such as pollution usually arise as the by-product of fast economic development. China witnessed significant improvements in public health statistics rate since the 1990s with a series of healthcare and environmental policy reforms. Our study is not only relevant for the comprehensive evaluation of environmental regulations during $11^{th}$ and $12^{th}$ FYP in China, but also provides valuable insights for cost-benefit analyses of air pollution policies or the construction of coal-fired power plants in other developing countries. 

\par 
\bibliography{ref}

\clearpage

\appendix
\begin{appendices}

\counterwithin{figure}{section}

\section{Appendices}

\subsection{Construction of Instrumental Variables}
In this section, we demonstrate in detail the construction of the weighted sum of projected capacities that are used as policy instruments in the estimations. 
\begin{figure}[h]
    \centering
    \includegraphics[width=0.8\linewidth]{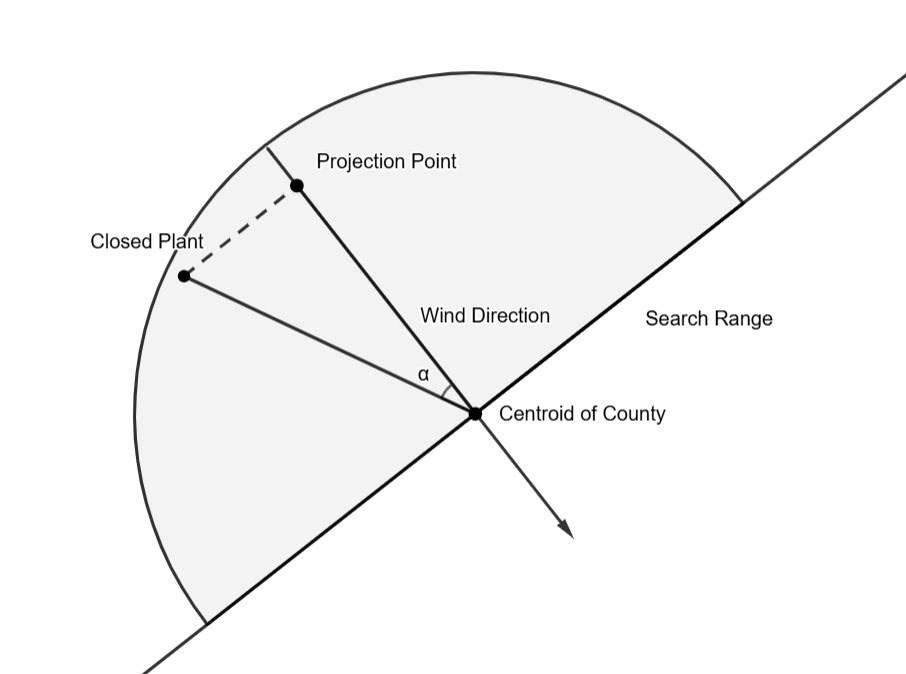}
     \caption{Showcase of Projection}   
    \begin{figurenotes}
        We multiply $\cos{\alpha}$ with the capacity of each plant in the semicircle within the specified radius (25, 50 or 100km) from a county center to calculate the weighted sums of retired capacities. We draw the semicircle based on the centroid of each county and the annually averaged wind direction which serves as the central axis. Only plants in the upwind semicircle are included in the calculation.
    \end{figurenotes}
    \label{fig:projection}
\end{figure}

To measure the exposure of a county to the power plant closure policy, we calculate the weighted sum of all the closed plants' capacities within a specific radius (25km, 50km and 100km) from a county centroid. We only consider the coal-fired power plants located within the upwind fan of the county. The upwind fan is defined as the semicircle located within Quadrants 3 and 4 of a coordinate system, which is centered on the country's centroid and oriented according to its yearly-averaged wind direction vector (see Figure \ref{fig:projection}). We apply two weights on each closed plant's capacity based on its location relative to the county centroid: 1) inverse distance weighting (IDW) to reduce the influence of closed plants in more remote areas; and 2) a projection of the vector connecting the coal plant and the county centroid, aligned with the county's average wind direction. We do not consider the effect of terrain on the dispersion of air pollutants, which is a limitation of the current paper. For simplicity, we also ignore the plants located on the downwind fan of the county centroid.

\subsection{Density Maps and Summary Statistics of Subsamples}\label{appx:maps}
In this section, we present the density heat maps of pollutants and the summary statistics of key variables by subsample. Before proceeding on to the subsections, we present the overview of China's county-level nominal GDP in 2000 which is the year before our policy period of interest (2001 to 2010). Compared with the density heatmaps shown in the following subsections, we find, as expected, a high degree of overlap between counties with higher nominal GDP and counties with high pollutant density.\\
\begin{figure}[h]
    \centering
    \includegraphics[width=0.8\linewidth]{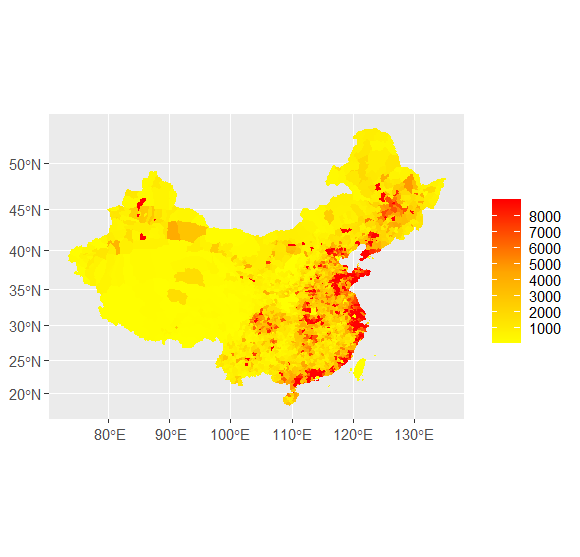}
     \caption{County-level GDP in 2000}   
    \begin{figurenotes}
        We map the nominal yearly county-level GDP (in million CNY) in all mainland provinces and municipalities in 2000 in the figure above. 
    \end{figurenotes}
    \begin{figurenotes}[Source]
        2001 China Statistical Yearbook for Regional Economy
    \end{figurenotes}
    \label{fig:gdp2000}
\end{figure}
\break
\subsection{East China subsample}\label{eastfig}
\begin{figure}[htbp]
    \centering
    \includegraphics[width=0.7\linewidth]{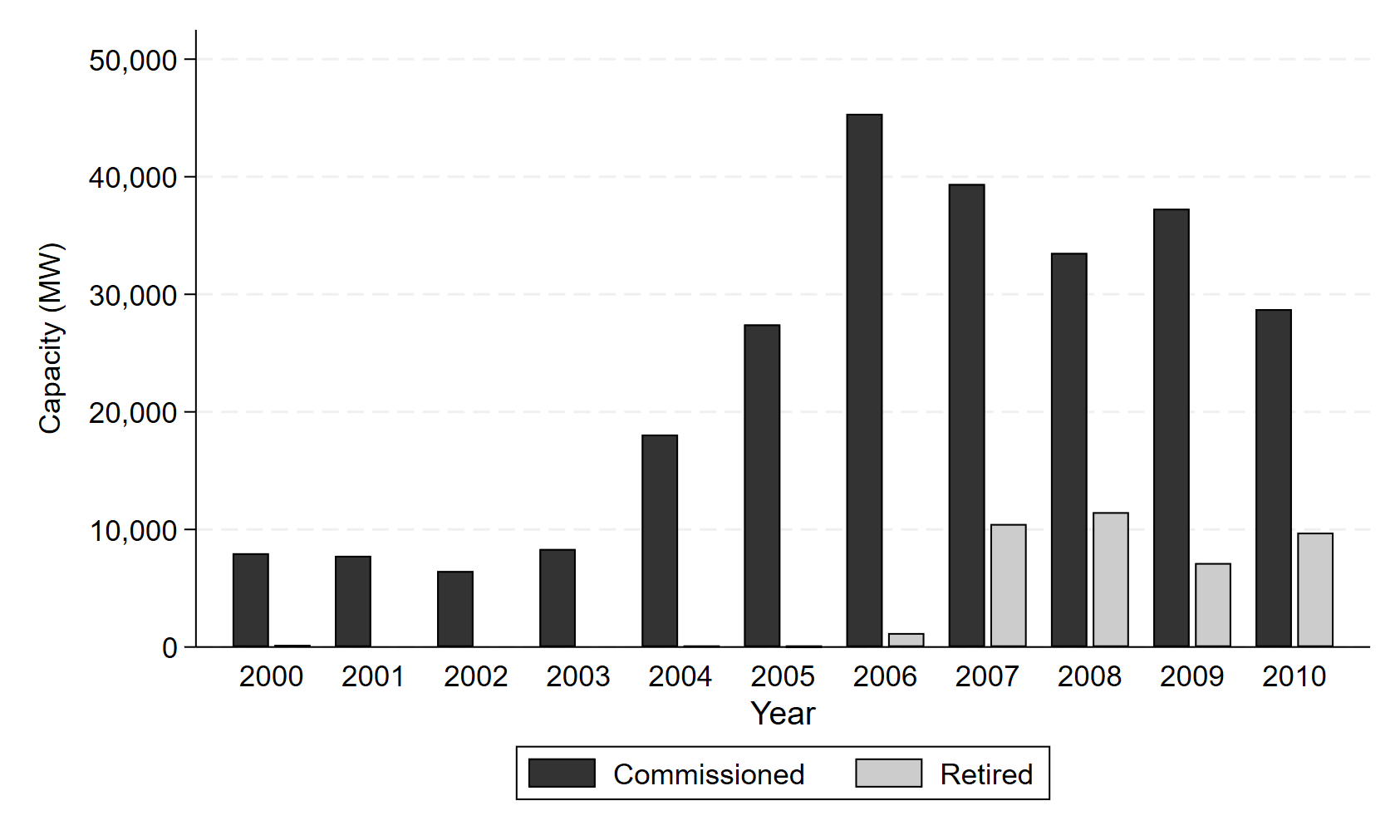}
    \caption{Commissioned and Retired Thermal Capacity Count from 2000 to 2010 (East China Subsample)}
    \label{fig:pollutantdensity_ea}
\end{figure}

\begin{figure}[htb]
\centering
\begin{tabular}{cc}
\centering
    (a) $SO_2$ 2005 & (b) $SO_2$ 2010 \\
     \includegraphics[width=.5\textwidth]{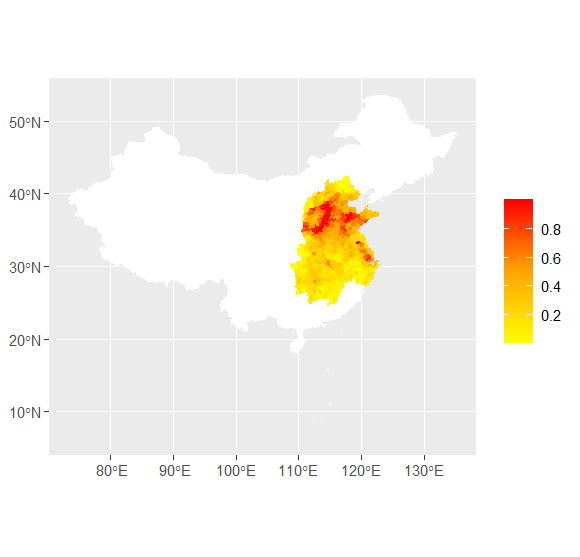}  &
     \includegraphics[width=.5\textwidth]{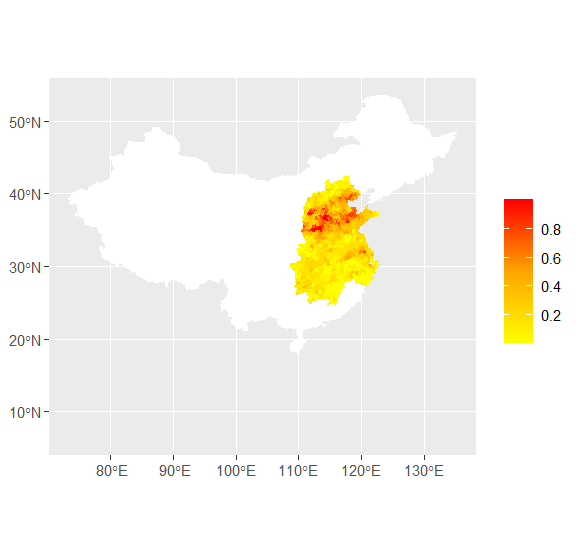} \\
     (c)$PM_{2.5}$ 2000 & (d)  $PM_{2.5}$ 2005 \\
     \includegraphics[width=.5\textwidth]{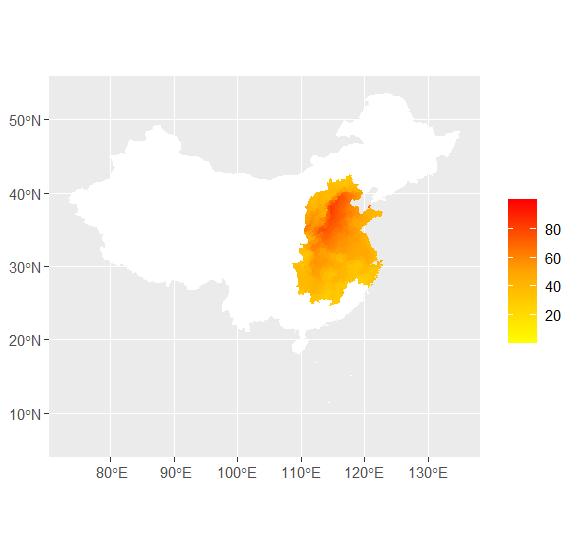}  &
     \includegraphics[width=.5\textwidth]{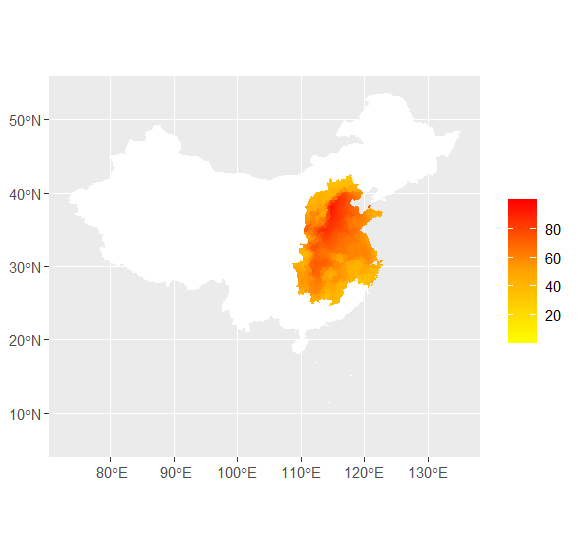} \\
     \multicolumn{2}{c}{(e) $PM_{2.5}$ 2010}\\
     \multicolumn{2}{c}{\includegraphics[width=.5\textwidth]{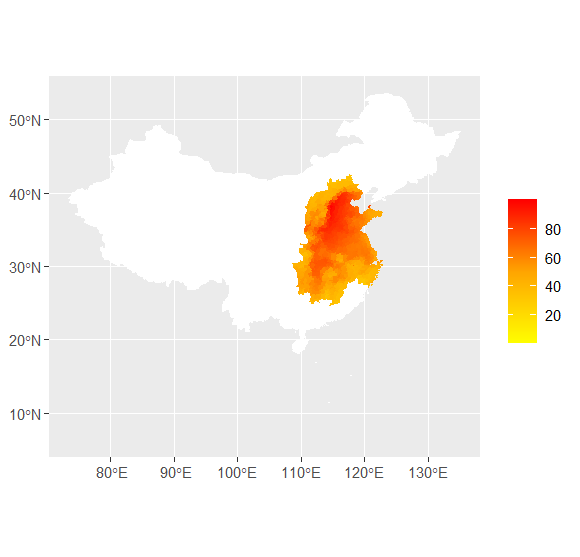}}
     
\end{tabular}
\caption{$SO_2$ and PM$_{2.5}$ Density Heat Maps (East China Subsample)}
\begin{figurenotes}
The figure shows the density of $SO_2$ (in DU) and $PM_{2.5}$ (in $\mu\text{g/m}^{3}$) in East China Subsample. Due to data availability, we document $SO_2$ density heatmaps in 2005 and 2010 while documenting $PM_{2.5}$ density heatmaps in 2000, 2005 and 2010.
\end{figurenotes}
\label{fig:ea_density}
\end{figure}

In this subsection we present the summary figures and the summary stats of key variables of the East China subsample. The bar chart in Figure \ref{fig:pollutantdensity_ea} presents the sum of yearly retired and commissioned capacities while breaking down the total yearly capacity retired into capacity closed under 50 MW (the smaller-unit plant units forced to shut down by the phase-out policy). We present the density heat maps of the East China subsample in Figure \ref{fig:ea_density}. Table \ref{tab:summary_ea1} documents the mean and standard deviations of our key variables of interest in 2001, 2004, 2007, and 2010, respectively for the East China subsample. 
\vspace{0cm}


\begin{table}[htbp]
  \centering
  \caption{Summary Stats of Key Variables (East China subsample)}
    \resizebox{\textwidth}{!}{\begin{tabular}{lrrrrr}
    \toprule
    \toprule
      & (1) & (2) & (3) & (4) \\
    Year  & 2001 & 2004 & 2007 & 2010 \\
    \midrule
    Under-5 Mortality (per 1000) & 31.95 & 24.40 & 18.83 & 14.90 \\
      & (11.78) & (9.111) & (6.661) & (5.201) \\
    $SO_2$ (DU) & 0 & 0 & 0.464 & 0.270 \\
      & (0) & (0) & (0.322) & (0.246) \\
    PM$_{2.5}$ ($\mu\text{g/m}^{3}$) & 54.06 & 61.89 & 68.91 & 63.17 \\
      & (14.18) & (14.79) & (18.37) & (16.49) \\
    Unweighted Sum of Retired Capacity (0 to 100 km) & 0 & 2.532 & 229.5 & 202.0 \\
      & (0) & (15.72) & (287.2) & (323.7) \\
    Unweighted Sum of Retired Capacity (0 to 50 km) & 0 & 0.886 & 56.54 & 47.94 \\
      & (0) & (9.377) & (130.1) & (118.3) \\
    Unweighted Sum of Retired Capacity (0 to 25 km) & 0 & 0.253 & 13.46 & 10.03 \\
      & (0) & (5.028) & (62.36) & (51.35) \\
    Unweighted Sum of Retired Capacity (50 to 100 km) & 0 & 1.646 & 173.0 & 154.1 \\
      & (0) & (12.73) & (246.2) & (257.3) \\
    Unweighted Sum of Retired Capacity (25 to 100 km) & 0 & 2.278 & 216.1 & 192.0 \\
      & (0) & (14.93) & (280.4) & (309.2) \\
    Unweighted Sum of Retired Capacity Under 50 MW (50 to 100 km) & 0 & 0 & 12.21 & 29.91 \\
      & (0) & (0) & (25.79) & (52.63) \\
    Unweighted Sum of Retired Capacity Under 50 MW (25 to 100 km) & 0 & 0 & 15.22 & 37.67 \\
      & (0) & (0) & (30.07) & (62.40) \\
    Unweighted Sum of Retired Capacity Under 50 MW (0 to 100 km) & 0 & 0 & 16.36 & 40.48 \\
      & (0) & (0) & (31.80) & (66.60) \\
    Unweighted Sum of Retired Capacity Under 50 MW (0 to 50 km) & 0 & 0 & 4.271 & 10.71 \\
      & (0) & (0) & (13.90) & (27.46) \\
    Unweighted Sum of Retired Capacity Under 50 MW (0 to 25 km) & 0 & 0 & 1.168 & 2.837 \\
      & (0) & (0) & (6.993) & (12.30) \\
    Weighted Sum of Retired Capacity (0 to 100 km) & 0 & 0.0620 & 4.634 & 3.768 \\
      & (0) & (0.579) & (8.115) & (6.780) \\
    Weighted Sum of Retired Capacity (0 to 50 km) & 0 & 0.0409 & 2.311 & 1.719 \\
      & (0) & (0.556) & (7.038) & (4.740) \\
    Weighted Sum of Retired Capacity (0 to 25 km) & 0 & 0.0256 & 1.168 & 0.694 \\
      & (0) & (0.521) & (6.288) & (3.614) \\
    Weighted Sum of Retired Capacity (50 to 100 km) & 0 & 0.0210 & 2.322 & 2.049 \\
      & (0) & (0.167) & (3.384) & (3.481) \\
    Weighted Sum of Retired Capacity (25 to 100 km) & 0 & 0.0363 & 3.465 & 3.074 \\
      & (0) & (0.255) & (4.740) & (5.183) \\
    Weighted Sum of Retired Capacity Under 50 MW (50 to 100 km) & 0 & 0 & 0.164 & 0.395 \\
      & (0) & (0) & (0.352) & (0.694) \\
    Weighted Sum of Retired Capacity Under 50 MW (25 to 100 km) & 0 & 0 & 0.244 & 0.600 \\
      & (0) & (0) & (0.502) & (1.029) \\
    Weighted Sum of Retired Capacity Under 50 MW (0 to 100 km) & 0 & 0 & 0.373 & 0.827 \\
      & (0) & (0) & (1.413) & (1.886) \\
    Weighted Sum of Retired Capacity Under 50 MW (0 to 50 km) & 0 & 0 & 0.212 & 0.434 \\
      & (0) & (0) & (1.338) & (1.602) \\
    Weighted Sum of Retired Capacity Under 50 MW (0 to 25 km) & 0 & 0 & 0.130 & 0.227 \\
      & (0) & (0) & (1.267) & (1.384) \\
    Yearly Desulphurization Capacity (10k ton) & 0 & 0 & 0.666 & 0.239 \\
      & (0) & (0) & (1.366) & (0.839) \\
    Sum of Operating Capacity Under Desulphurization & 0 & 0 & 173.7 & 92.77 \\
      & (0) & (0) & (372.3) & (286.1) \\
    Prim. GDP (CNY per capita) & 1398.1 & 1858.8 & 2390.8 & 3323.3 \\
      & (614.7) & (780.8) & (995.2) & (1373.5) \\
    Sec. GDP (CNY per capita) & 2613.1 & 4793.2 & 8626.7 & 13577.6 \\
      & (2538.5) & (4650.5) & (9085.2) & (13259.0) \\
    Hospital Bed (per 10,000 Population) & 22.53 & 22.71 & 25.33 & 31.32 \\
      & (7.266) & (7.927) & (8.806) & (9.196) \\
    \bottomrule
    \bottomrule
    Standard Errors in Parenthesis &   &     &   &  \\
    \end{tabular}}%
  \label{tab:summary_ea1}%
  \\
  \begin{tablenotes}
      This table documents the mean and standard deviation of key variables of interest from 2001 to 2005 for the East China sample at county level. Variables documented are: under-5 mortality (outcome variable), $SO_2$ and PM$_{2.5}$ density (endogenous variable of interest), weighted and unweighted sums of capacities (coal plant phase-out policy instruments), and desulphurization capacities (desulphurization policy instruments).
  \end{tablenotes}
\end{table}%
\clearpage
\subsection{Northwest China subsample}\label{nwfig}
\begin{figure}[h]
    \centering
    \includegraphics[width=0.67\linewidth]{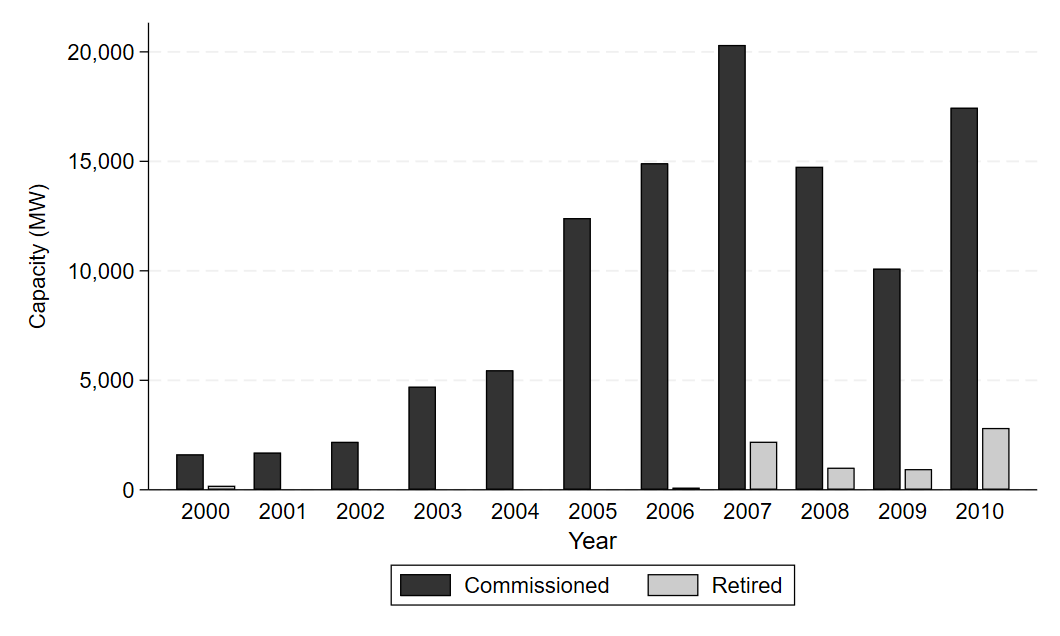}
    \caption{Commissioned and Retired Thermal Capacity Count from 2000 to 2010  (Northwest China Subsample)}
    \label{fig:pollutantdensity_nw}
\end{figure}

\begin{figure}[htbp]
\centering
\begin{tabular}{cc}
\centering
    (a) $SO_2$ 2005 & (b) $SO_2$ 2010 \\
     \includegraphics[width=.5\textwidth]{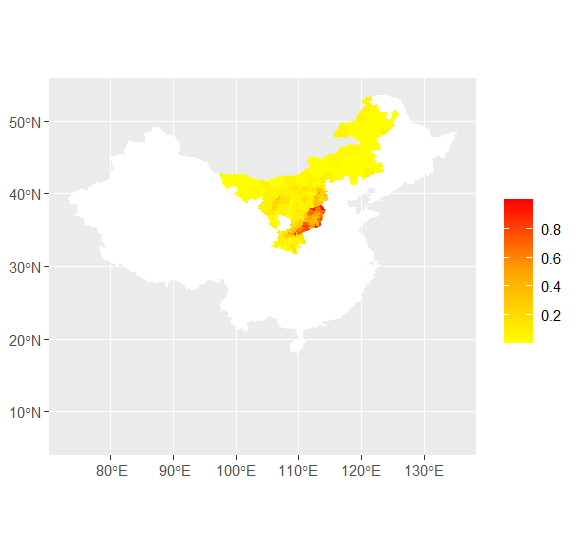}  &
     \includegraphics[width=.5\textwidth]{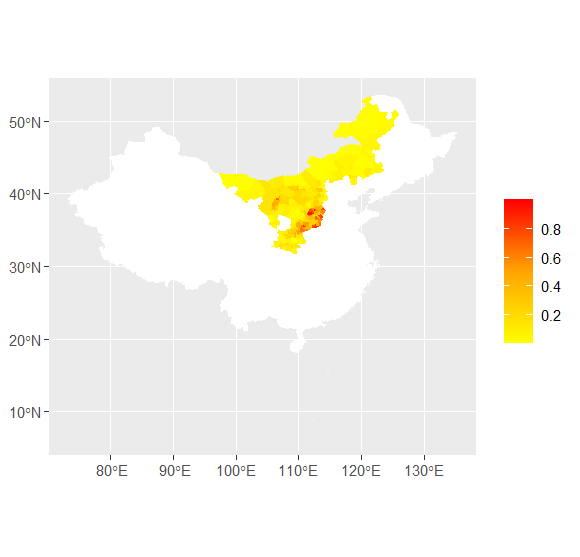} \\
     (c)$PM_{2.5}$ 2000 & (d)  $PM_{2.5}$ 2005 \\
     \includegraphics[width=.5\textwidth]{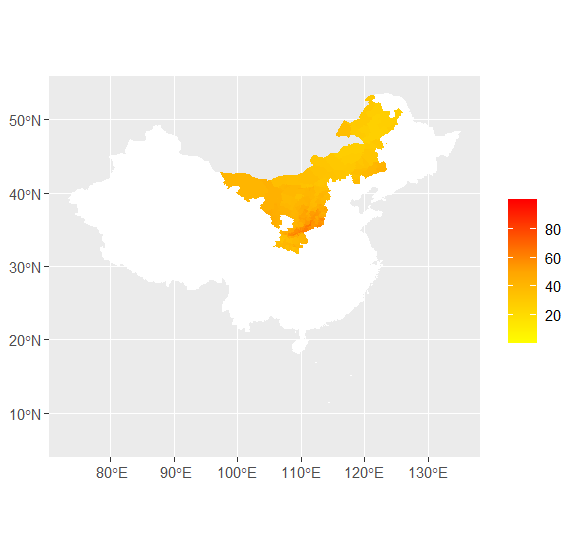}  &
     \includegraphics[width=.5\textwidth]{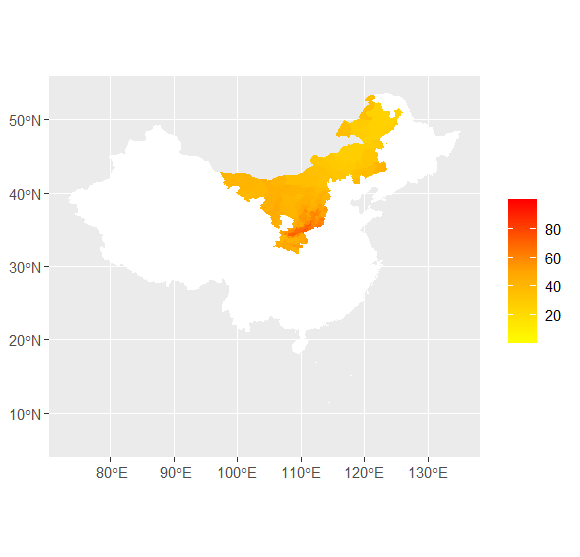} \\
     \multicolumn{2}{c}{(e) $PM_{2.5}$ 2010}\\
     \multicolumn{2}{c}{\includegraphics[width=.5\textwidth]{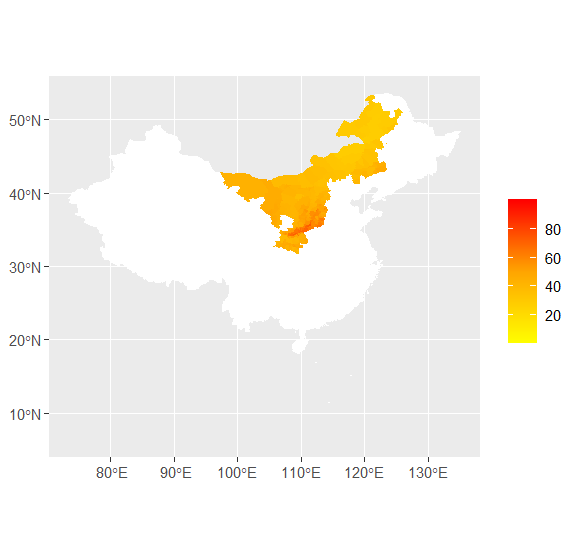}}
     
\end{tabular}
\caption{$SO_2$ and PM$_{2.5}$ Density Heat Maps (Northwest China Subsample)}
\begin{figurenotes}
The figure shows the density of $SO_2$ (in DU) and PM$_{2.5}$ (in $\mu\text{g/m}^{3}$) in Northwest China Subsample. Due to data availability, we document $SO_2$ density heatmaps in 2005 and 2010 while documenting PM$_{2.5}$ density heatmaps in 2000, 2005 and 2010.
\end{figurenotes}
\label{fig:nw_density}
\end{figure}

In this subsection we present the summary figures and the summary stats of key variables of the Northwest China subsample. The bar chart in Figure \ref{fig:pollutantdensity_nw} presents the sum of yearly retired and commissioned capacities while breaking down the total yearly capacity retired into capacity closed under 50 MW (the smaller-unit plant units forced to shut down by the phase-out policy). We present the density heat maps of the Northwest China subsample in Figure \ref{fig:nw_density}. Table \ref{tab:summary_nw1} documents the mean and standard deviations of our key variables of interest in 2001, 2004, 2007, and 2010, respectively for the Northwest China subsample.
\vspace{0cm}

\begin{table}[htbp]
  \centering
  \caption{Summary Stats of Key Variables (Northwest China subsample)}
    \resizebox{\textwidth}{!}{\begin{tabular}{lcccc}
    \toprule
    \toprule
      & (1) & (2) & (3) & (4) \\
    Year  & 2001 & 2004 & 2007 & 2010 \\
    \midrule
    Under-5 Mortality (per 1000) & 40.50 & 31.13 & 23.91 & 18.68 \\
      & (13.61) & (11.44) & (8.332) & (6.590) \\
    $SO_2$ (DU) & 0 & 0 & 0.313 & 0.232 \\
      & (0) & (0) & (0.297) & (0.227) \\
    PM$_{2.5}$ ($\mu\text{g/m}^{3}$) & 49.55 & 49.25 & 55.15 & 49.48 \\
      & (7.895) & (9.208) & (11.63) & (9.918) \\
    Unweighted Sum of Retired Capacity (0 to 100 km) & 0 & 3.429 & 111.6 & 181.1 \\
      & (0) & (18.25) & (164.7) & (271.4) \\
    Unweighted Sum of Retired Capacity (0 to 50 km) & 0 & 1.143 & 25.38 & 47.79 \\
      & (0) & (10.66) & (66.50) & (94.21) \\
    Unweighted Sum of Retired Capacity (0 to 25 km) & 0 & 0 & 9.465 & 13.35 \\
      & (0) & (0) & (44.31) & (47.41) \\
    Unweighted Sum of Retired Capacity (50 to 100 km) & 0 & 2.286 & 86.17 & 133.3 \\
      & (0) & (14.99) & (150.5) & (223.2) \\
    Unweighted Sum of Retired Capacity (25 to 100 km) & 0 & 3.429 & 102.1 & 167.7 \\
      & (0) & (18.25) & (162.3) & (262.7) \\
    Unweighted Sum of Retired Capacity Under 50 MW (50 to 100 km) & 0 & 0 & 10.57 & 47.13 \\
      & (0) & (0) & (22.61) & (63.50) \\
    Unweighted Sum of Retired Capacity Under 50 MW (25 to 100 km) & 0 & 0 & 13.38 & 59.75 \\
      & (0) & (0) & (25.01) & (79.00) \\
    Unweighted Sum of Retired Capacity Under 50 MW (0 to 100 km) & 0 & 0 & 13.64 & 65.00 \\
      & (0) & (0) & (25.13) & (84.96) \\
    Unweighted Sum of Retired Capacity Under 50 MW (0 to 50 km) & 0 & 0 & 3.066 & 17.87 \\
      & (0) & (0) & (13.62) & (38.29) \\
    Unweighted Sum of Retired Capacity Under 50 MW (0 to 25 km) & 0 & 0 & 0.253 & 5.250 \\
      & (0) & (0) & (3.553) & (16.56) \\
    Weighted Sum of Retired Capacity (0 to 100 km) & 0 & 0.0543 & 2.444 & 3.676 \\
      & (0) & (0.298) & (5.741) & (5.766) \\
    Weighted Sum of Retired Capacity (0 to 50 km) & 0 & 0.0231 & 1.333 & 1.908 \\
      & (0) & (0.216) & (5.553) & (4.297) \\
    Weighted Sum of Retired Capacity (0 to 25 km) & 0 & 0 & 0.925 & 0.988 \\
      & (0) & (0) & (5.444) & (3.595) \\
    Weighted Sum of Retired Capacity (50 to 100 km) & 0 & 0.0312 & 1.111 & 1.768 \\
      & (0) & (0.209) & (1.906) & (2.948) \\
    Weighted Sum of Retired Capacity (25 to 100 km) & 0 & 0.0543 & 1.519 & 2.688 \\
      & (0) & (0.298) & (2.354) & (4.125) \\
    Weighted Sum of Retired Capacity Under 50 MW (50 to 100 km) & 0 & 0 & 0.141 & 0.630 \\
      & (0) & (0) & (0.309) & (0.846) \\
    Weighted Sum of Retired Capacity Under 50 MW (25 to 100 km) & 0 & 0 & 0.216 & 0.969 \\
      & (0) & (0) & (0.456) & (1.366) \\
    Weighted Sum of Retired Capacity Under 50 MW (0 to 100 km) & 0 & 0 & 0.227 & 1.359 \\
      & (0) & (0) & (0.478) & (2.361) \\
    Weighted Sum of Retired Capacity Under 50 MW (0 to 50 km) & 0 & 0 & 0.0855 & 0.729 \\
      & (0) & (0) & (0.397) & (1.989) \\
    Weighted Sum of Retired Capacity Under 50 MW (0 to 25 km) & 0 & 0 & 0.0113 & 0.390 \\
      & (0) & (0) & (0.159) & (1.573) \\
    Yearly Desulphurization Capacity (10k ton) & 0 & 0 & 0.316 & 0.196 \\
      & (0) & (0) & (1.075) & (0.725) \\
    Sum of Operating Capacity Under Desulphurization & 0 & 0 & 97.78 & 44.10 \\
      & (0) & (0) & (309.0) & (156.3) \\
    Prim. GDP (CNY per capita) & 817.1 & 1195.9 & 1805.4 & 3027.2 \\
      & (457.5) & (598.8) & (1140.1) & (2085.2) \\
    Sec. GDP (CNY per capita) & 1628.5 & 3474.6 & 8034.2 & 14170.6 \\
      & (1441.7) & (3696.8) & (10205.7) & (17697.9) \\
    Hospital Bed (per 10,000 Population) & 28.81 & 28.26 & 29.90 & 36.31 \\
      & (9.359) & (8.123) & (8.219) & (7.859) \\
    \bottomrule
    \bottomrule
    Standard Errors in Parenthesis &     &   &   &  \\
    \end{tabular}}%
  \label{tab:summary_nw1}%
  \\
  \begin{tablenotes}
      This table documents the mean and standard deviation of key variables of interest from 2001 to 2005 for the Northwest China subsample at county level. Variables documented are: under-5 mortality (outcome variable), $SO_2$ and PM$_{2.5}$ density (endogenous variable of interest), weighted and unweighted sums of capacities (coal plant phase-out policy instruments), and desulphurization capacities (desulphurization policy instruments).
  \end{tablenotes}
\end{table}%

\clearpage
\subsection{Southwest China subsample}\label{swfig}
\begin{figure}[h]
    \centering
    \includegraphics[width=0.7\linewidth]{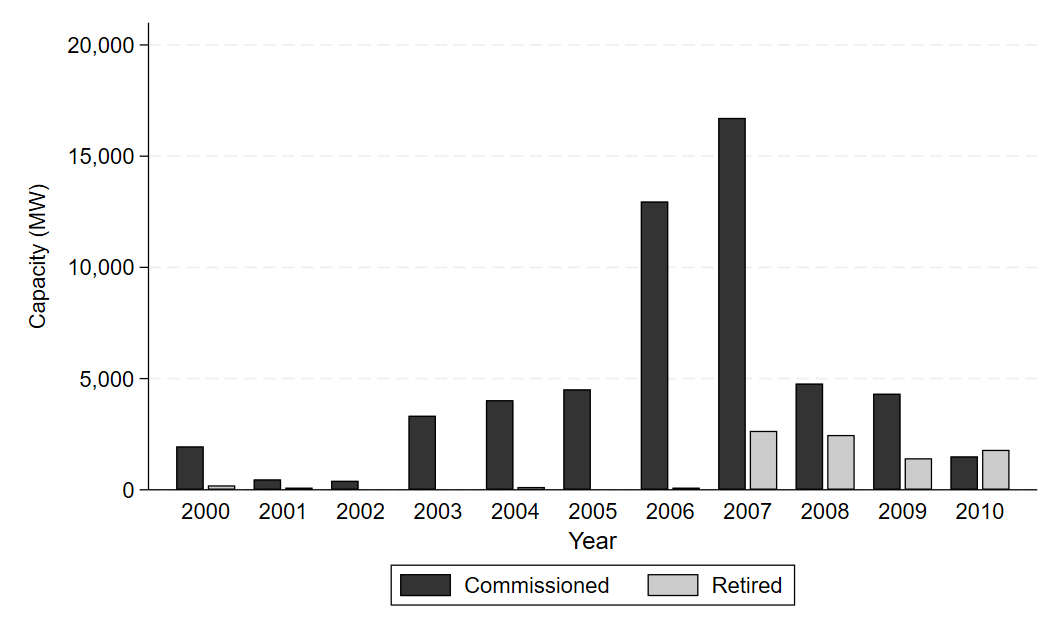}
    \caption{Commissioned and Retired Thermal Capacity Count from 2000 to 2010 (Southwest China Subsample)}
    \label{fig:pollutantdensity_sw}
\end{figure}

\begin{figure}[htbp]
\centering
\begin{tabular}{cc}
\centering
    (a) $SO_2$ 2005 & (b) $SO_2$ 2010 \\
     \includegraphics[width=.5\textwidth]{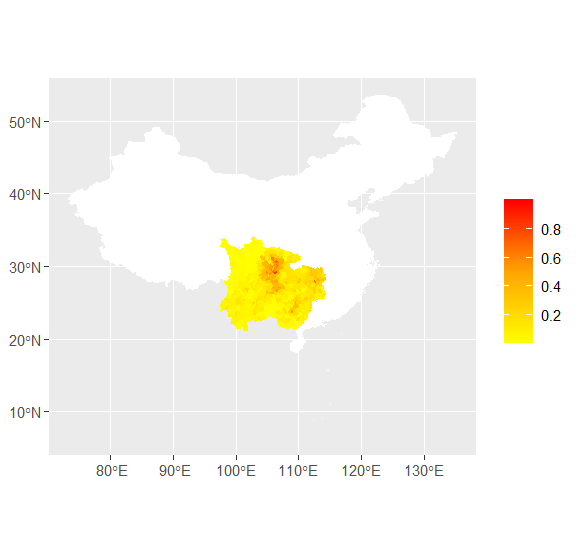}  &
     \includegraphics[width=.5\textwidth]{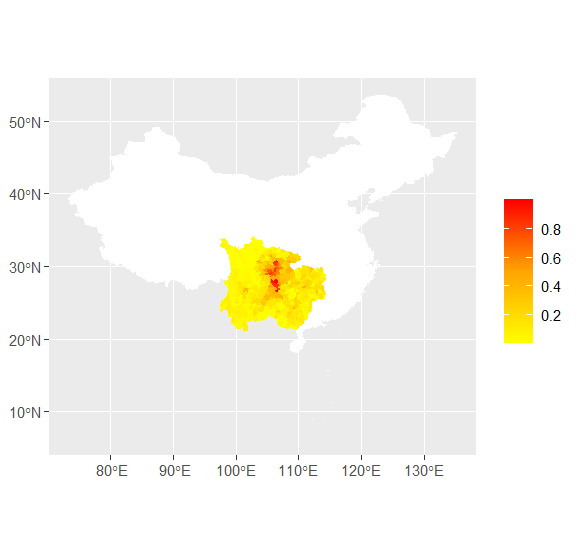} \\
     (c) PM$_{2.5}$ 2000 & (d)  $SO_2$ 2005 \\
     \includegraphics[width=.5\textwidth]{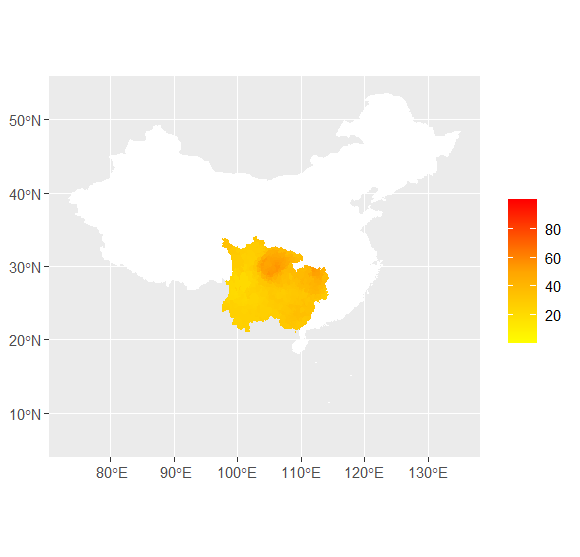}  &
     \includegraphics[width=.5\textwidth]{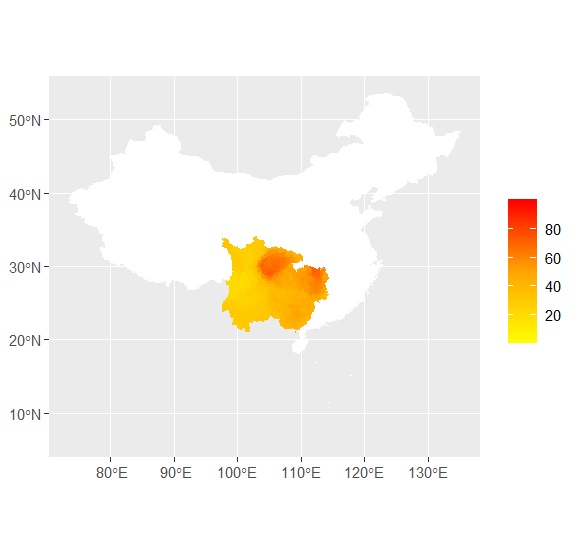} \\
     \multicolumn{2}{c}{(e) PM$_{2.5}$ 2010}\\
     \multicolumn{2}{c}{\includegraphics[width=.5\textwidth]{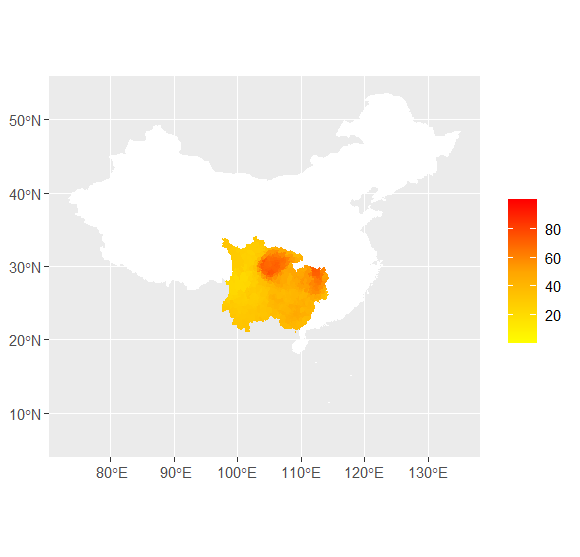}}
     
\end{tabular}
\caption{$SO_2$ and PM$_{2.5}$ Density Heat Maps (Southwest China Subsample)}
\begin{figurenotes}
The figure shows the density of $SO_2$ (in DU) and PM$_{2.5}$ (in $\mu\text{g/m}^{3}$) in Southwest China Subsample. Due to data availability, we document $SO_2$ density heatmaps in 2005 and 2010 while documenting PM$_{2.5}$ density heatmaps in 2000, 2005 and 2010.
\end{figurenotes}
\label{fig:sw_density}
\end{figure}

In this subsection we present the summary figures and the summary stats of key variables of the Southwest China subsample. The bar chart in Figure \ref{fig:pollutantdensity_sw} presents the sum of yearly retired and commissioned capacities while breaking down the total yearly capacity retired into capacity closed under 50 MW (the smaller-unit plant units forced to shut down by the phase-out policy). We present the density heat maps of the Southwest China subsample in Figure \ref{fig:sw_density}. Table \ref{tab:summary_sw1} documents the mean and standard deviations of our key variables of interest in 2001, 2004, 2007, and 2010, respectively for the Southwest China subsample. 
\vspace{0cm}
\begin{table}[htbp]
  \centering
  \caption{Summary Stats of Key Variables (Southwest China subsample)}
    \resizebox{\textwidth}{!}{\begin{tabular}{lcccc}
    \toprule
    \toprule
      & (1) & (2) & (3) & (4) \\
    Year  & 2001 & 2004 & 2007 & 2010 \\
    \midrule
    Under-5 Mortality (per 1000) & 42.33 & 32.77 & 23.83 & 18.42 \\
      & (20.04) & (14.68) & (9.113) & (5.995) \\
    $SO_2$ (DU) & 0 & 0 & 0.227 & 0.183 \\
      & (0) & (0) & (0.164) & (0.187) \\
    PM$_{2.5}$ ($\mu\text{g/m}^{3}$) & 44.01 & 49.19 & 54.11 & 49.93 \\
      & (9.132) & (11.65) & (13.65) & (13.32) \\
    Unweighted Sum of Retired Capacity (0 to 100 km) & 2.564 & 4.031 & 94.55 & 41.76 \\
      & (15.84) & (22.40) & (156.9) & (89.65) \\
    Unweighted Sum of Retired Capacity (0 to 50 km) & 0.427 & 1.512 & 28.66 & 11.24 \\
      & (6.537) & (13.86) & (92.93) & (48.73) \\
    Unweighted Sum of Retired Capacity (0 to 25 km) & 0.427 & 1.008 & 6.791 & 2.645 \\
      & (6.537) & (11.34) & (47.56) & (21.41) \\
    Unweighted Sum of Retired Capacity (50 to 100 km) & 2.137 & 2.520 & 65.89 & 30.52 \\
      & (14.49) & (17.82) & (135.5) & (75.61) \\
    Unweighted Sum of Retired Capacity (25 to 100 km) & 2.137 & 3.024 & 87.75 & 39.12 \\
      & (14.49) & (19.48) & (151.8) & (87.63) \\
    Unweighted Sum of Retired Capacity Under 50 MW (50 to 100 km) & 0 & 0 & 3.530 & 5.567 \\
      & (0) & (0) & (11.11) & (12.98) \\
    Unweighted Sum of Retired Capacity Under 50 MW (25 to 100 km) & 0 & 0 & 4.532 & 7.224 \\
      & (0) & (0) & (13.01) & (15.94) \\
    Unweighted Sum of Retired Capacity Under 50 MW (0 to 100 km) & 0 & 0 & 5.202 & 7.959 \\
      & (0) & (0) & (14.60) & (17.58) \\
    Unweighted Sum of Retired Capacity Under 50 MW (0 to 50 km) & 0 & 0 & 1.832 & 2.391 \\
      & (0) & (0) & (8.810) & (8.614) \\
    Unweighted Sum of Retired Capacity Under 50 MW (0 to 25 km) & 0 & 0 & 0.702 & 0.734 \\
      & (0) & (0) & (4.572) & (4.710) \\
    Weighted Sum of Retired Capacity (0 to 100 km) & 0.0727 & 0.0892 & 2.046 & 1.115 \\
      & (0.681) & (0.607) & (5.705) & (4.800) \\
    Weighted Sum of Retired Capacity (0 to 50 km) & 0.0425 & 0.0591 & 1.172 & 0.699 \\
      & (0.650) & (0.571) & (5.507) & (4.716) \\
    Weighted Sum of Retired Capacity (0 to 25 km) & 0.0425 & 0.0486 & 0.554 & 0.457 \\
      & (0.650) & (0.547) & (4.999) & (4.555) \\
    Weighted Sum of Retired Capacity (50 to 100 km) & 0.0302 & 0.0301 & 0.873 & 0.416 \\
      & (0.209) & (0.214) & (1.828) & (1.058) \\
    Weighted Sum of Retired Capacity (25 to 100 km) & 0.0302 & 0.0406 & 1.492 & 0.658 \\
      & (0.209) & (0.271) & (2.904) & (1.643) \\
    Weighted Sum of Retired Capacity Under 50 MW (50 to 100 km) & 0 & 0 & 0.0481 & 0.0765 \\
      & (0) & (0) & (0.154) & (0.184) \\
    Weighted Sum of Retired Capacity Under 50 MW (25 to 100 km) & 0 & 0 & 0.0750 & 0.122 \\
      & (0) & (0) & (0.229) & (0.299) \\
    Weighted Sum of Retired Capacity Under 50 MW (0 to 100 km) & 0 & 0 & 0.142 & 0.175 \\
      & (0) & (0) & (0.675) & (0.510) \\
    Weighted Sum of Retired Capacity Under 50 MW (0 to 50 km) & 0 & 0 & 0.0972 & 0.0987 \\
      & (0) & (0) & (0.648) & (0.415) \\
    Weighted Sum of Retired Capacity Under 50 MW (0 to 25 km) & 0 & 0 & 0.0678 & 0.0532 \\
      & (0) & (0) & (0.610) & (0.353) \\
    Yearly Desulphurization Capacity (10k ton) & 0 & 0 & 0.768 & 0.125 \\
      & (0) & (0) & (1.790) & (0.477) \\
    Sum of Operating Capacity Under Desulphurization & 0 & 0 & 124.1 & 38.48 \\
      & (0) & (0) & (229.1) & (138.6) \\
    Prim. GDP (CNY per capita) & 1377.1 & 1684.7 & 2387.6 & 2983.7 \\
      & (471.2) & (581.5) & (830.9) & (1133.3) \\
    Sec. GDP (CNY per capita) & 1566.7 & 2428.0 & 4170.9 & 7381.0 \\
      & (1463.6) & (2370.7) & (3850.2) & (6694.8) \\
    Hospital Bed (per 10,000 Population) & 22.61 & 22.04 & 23.97 & 30.55 \\
      & (9.012) & (9.234) & (9.605) & (10.92) \\
    \bottomrule
    \bottomrule
    Standard Errors in Parenthesis   &   &   &   &  \\
    \end{tabular}}%
  \label{tab:summary_sw1}%
  \\
  \begin{tablenotes}
      This table documents the mean and standard deviation of key variables of interest from 2001 to 2005 for the Southwest China subsample at county level. Variables documented are: under-5 mortality (outcome variable), $SO_2$ and PM$_{2.5}$ density (endogenous variable of interest), weighted and unweighted sums of capacities (coal plant phase-out policy instruments), and desulphurization capacities (desulphurization policy instruments).
  \end{tablenotes}
\end{table}%
\clearpage
\subsection{Acid Rain Control Zones Subsample}\label{ARCZfig}
\begin{figure}[h]
    \centering
    \includegraphics[width=0.7\linewidth]{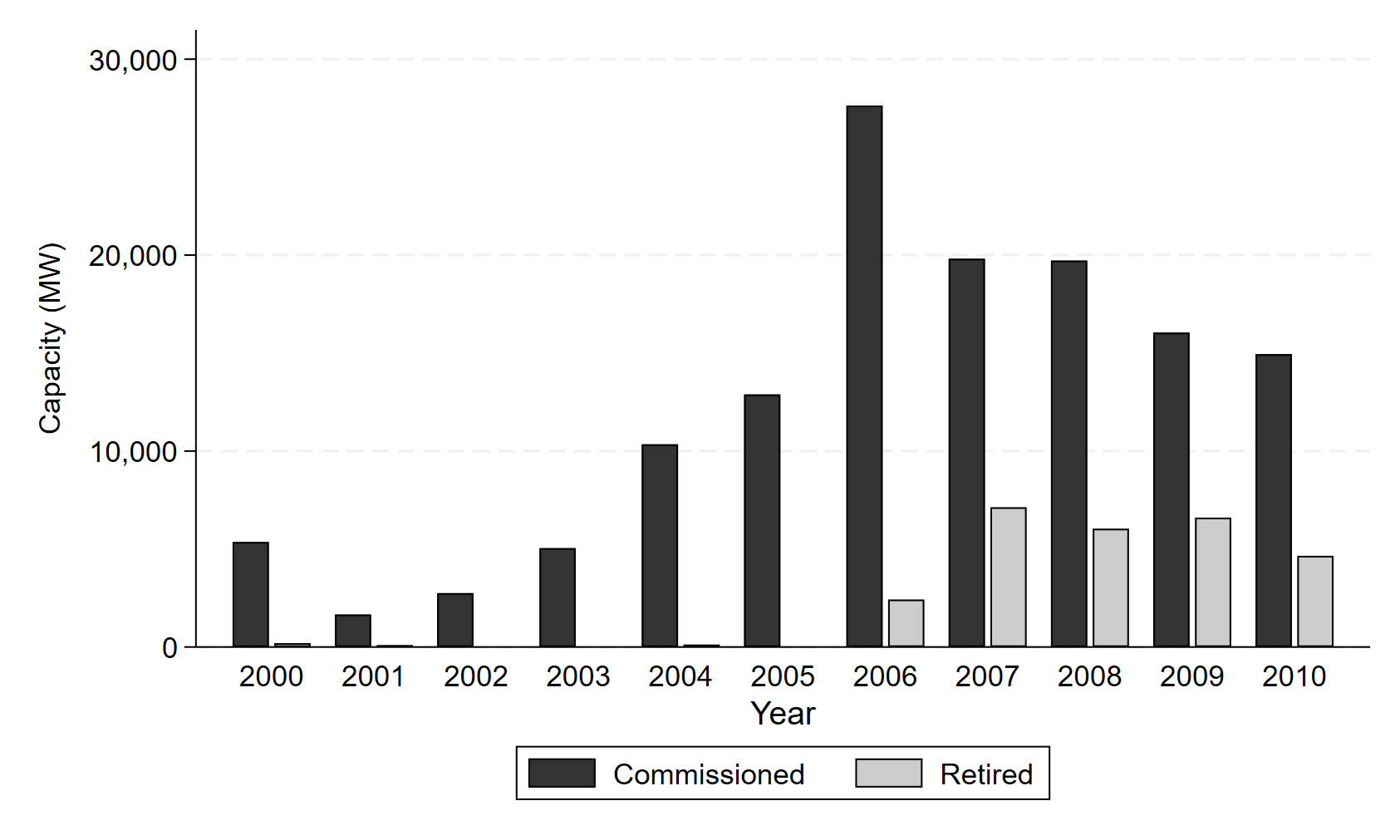}
      \caption{Commissioned and Retired Thermal Capacity Count from 2000 to 2010 (ARCZ Subsample)}
  \label{fig:pollutantdensity_ARCA}
\end{figure}

\begin{figure}[htbp]
\centering
\begin{tabular}{cc}
\centering
    (a) $SO_2$ 2005 & (b) $SO_2$ 2010 \\
     \includegraphics[width=.5\textwidth]{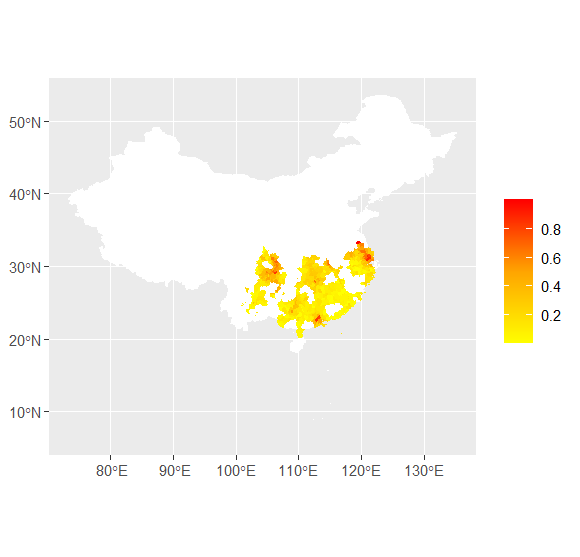}  &
     \includegraphics[width=.5\textwidth]{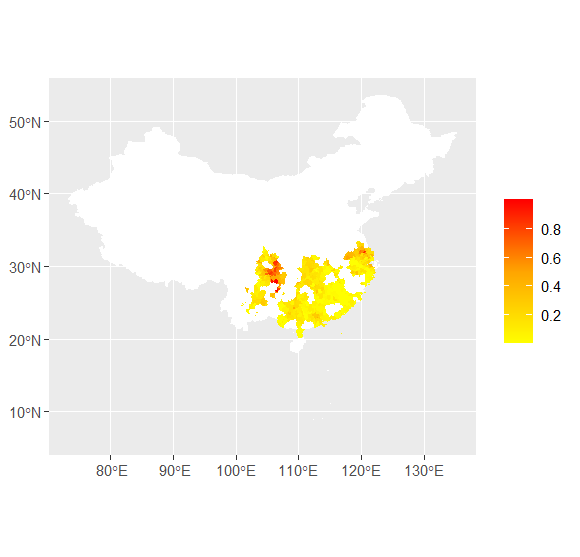} \\
     (c) $PM_{2.5}$ 2000 & (d)  $PM_{2.5}$ 2005 \\
     \includegraphics[width=.5\textwidth]{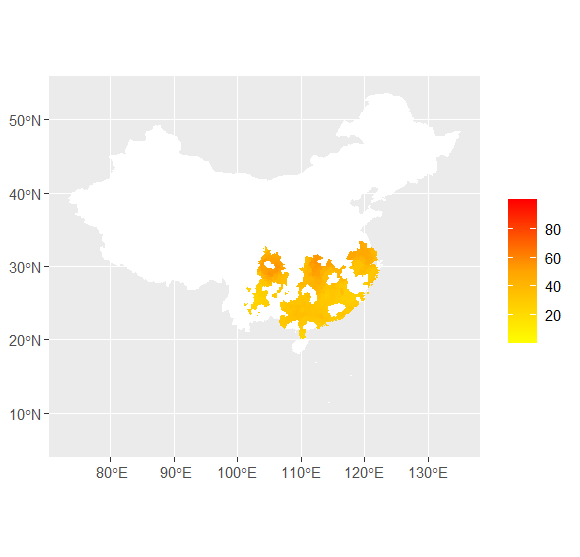}  &
     \includegraphics[width=.5\textwidth]{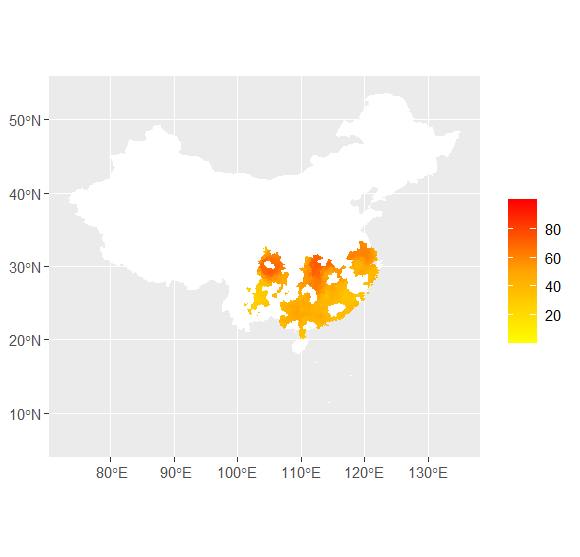} \\
     \multicolumn{2}{c}{(e) $PM_{2.5}$ 2010}\\
     \multicolumn{2}{c}{\includegraphics[width=.5\textwidth]{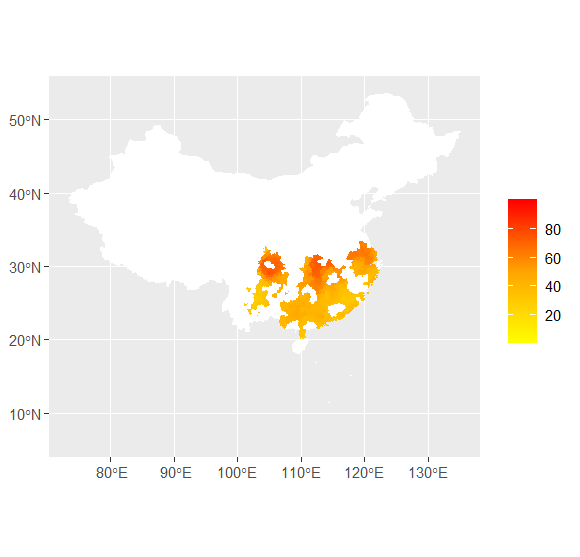}}
     
\end{tabular}
\caption{$SO_2$ and PM$_{2.5}$ Density Heat Maps (ARCZ Subsample)}
\begin{figurenotes}
The figure shows the density of $SO_2$ (in DU) and PM$_{2.5}$ (in $\mu\text{g/m}^{3}$) in ARCZ Subsample. Due to data availability, we document $SO_2$ density heatmaps in 2005 and 2010 while documenting PM$_{2.5}$ density heatmaps in 2000, 2005 and 2010.
\end{figurenotes}
\label{fig:ARCZ_density}
\end{figure}

In this subsection we present the summary figures and the summary stats of key variables of the Acid Rain Control Zones policy subsample. The bar chart in Figure \ref{fig:pollutantdensity_ARCA} presents the sum of yearly retired and commissioned capacities while breaking down the total yearly capacity retired into capacity closed under 50 MW (the smaller-unit plant units forced to shut down by the phase-out policy). We present the density heat maps of the ARCZ Policy subsample in Figure \ref{fig:ARCZ_density}. Table \ref{tab:summary_ARCA1} documents the mean and standard deviations of our key variables of interest in 2001, 2004, 2007, and 2010, respectively for the ARCZ subsample.
\begin{table}[htbp]
  \centering
  \caption{Summary Stats of Key Variables (ARCZ Subsample)}
    \resizebox{\textwidth}{!}{\begin{tabular}{lcccc}
    \toprule
    \toprule
      & (1) & (2) & (3) & (4) \\
    Year  & 2001 & 2004 & 2007 & 2010 \\
    \midrule
    Under-5 Mortality (per 1000) & 33.66 & 26.21 & 19.86 & 15.41 \\
      & (18.09) & (13.60) & (8.959) & (6.256) \\
    $SO_2$ (DU) & 0 & 0 & 0.245 & 0.141 \\
      & (0) & (0) & (0.214) & (0.175) \\
    PM$_{2.5}$ ($\mu\text{g/m}^{3}$) & 41.63 & 48.37 & 52.44 & 48.26 \\
      & (9.233) & (10.82) & (12.50) & (12.70) \\
    Unweighted Sum of Retired Capacity (0 to 100 km) & 0.871 & 2.111 & 155.6 & 76.75 \\
      & (9.305) & (16.32) & (240.2) & (165.5) \\
    Unweighted Sum of Retired Capacity (0 to 50 km) & 0.218 & 0.792 & 41.13 & 16.39 \\
      & (4.668) & (10.05) & (113.9) & (55.55) \\
    Unweighted Sum of Retired Capacity (0 to 25 km) & 0.218 & 0.528 & 10.43 & 2.549 \\
      & (4.668) & (8.211) & (52.23) & (20.38) \\
    Unweighted Sum of Retired Capacity (50 to 100 km) & 0.654 & 1.320 & 114.5 & 60.36 \\
      & (8.067) & (12.94) & (206.9) & (145.4) \\
    Unweighted Sum of Retired Capacity (25 to 100 km) & 0.654 & 1.584 & 145.2 & 74.20 \\
      & (8.067) & (14.16) & (235.1) & (164.6) \\
    Unweighted Sum of Retired Capacity Under 50 MW (50 to 100 km) & 0 & 0 & 8.046 & 9.344 \\
      & (0) & (0) & (18.50) & (25.94) \\
    Unweighted Sum of Retired Capacity Under 50 MW (25 to 100 km) & 0 & 0 & 9.780 & 11.74 \\
      & (0) & (0) & (21.49) & (28.88) \\
    Unweighted Sum of Retired Capacity Under 50 MW (0 to 100 km) & 0 & 0 & 10.74 & 12.58 \\
      & (0) & (0) & (23.09) & (29.73) \\
    Unweighted Sum of Retired Capacity Under 50 MW (0 to 50 km) & 0 & 0 & 2.905 & 3.357 \\
      & (0) & (0) & (10.51) & (12.07) \\
    Unweighted Sum of Retired Capacity Under 50 MW (0 to 25 km) & 0 & 0 & 1.018 & 0.843 \\
      & (0) & (0) & (5.897) & (5.674) \\
    Weighted Sum of Retired Capacity (0 to 100 km) & 0.0315 & 0.0467 & 3.216 & 1.505 \\
      & (0.480) & (0.441) & (6.524) & (4.358) \\
    Weighted Sum of Retired Capacity (0 to 50 km) & 0.0217 & 0.0310 & 1.686 & 0.702 \\
      & (0.464) & (0.414) & (5.858) & (3.812) \\
    Weighted Sum of Retired Capacity (0 to 25 km) & 0.0217 & 0.0255 & 0.844 & 0.337 \\
      & (0.464) & (0.396) & (5.071) & (3.585) \\
    Weighted Sum of Retired Capacity (50 to 100 km) & 0.00981 & 0.0158 & 1.531 & 0.802 \\
      & (0.123) & (0.156) & (2.843) & (1.939) \\
    Weighted Sum of Retired Capacity (25 to 100 km) & 0.00981 & 0.0213 & 2.372 & 1.168 \\
      & (0.123) & (0.197) & (4.098) & (2.569) \\
    Weighted Sum of Retired Capacity Under 50 MW (50 to 100 km) & 0 & 0 & 0.107 & 0.126 \\
      & (0) & (0) & (0.244) & (0.361) \\
    Weighted Sum of Retired Capacity Under 50 MW (25 to 100 km) & 0 & 0 & 0.154 & 0.188 \\
      & (0) & (0) & (0.355) & (0.467) \\
    Weighted Sum of Retired Capacity Under 50 MW (0 to 100 km) & 0 & 0 & 0.234 & 0.243 \\
      & (0) & (0) & (0.695) & (0.617) \\
    Weighted Sum of Retired Capacity Under 50 MW (0 to 50 km) & 0 & 0 & 0.131 & 0.120 \\
      & (0) & (0) & (0.619) & (0.466) \\
    Weighted Sum of Retired Capacity Under 50 MW (0 to 25 km) & 0 & 0 & 0.0813 & 0.0552 \\
      & (0) & (0) & (0.573) & (0.356) \\
    Yearly Desulphurization Capacity (10k ton) & 0 & 0 & 0.742 & 0.136 \\
      & (0) & (0) & (1.764) & (0.589) \\
    Sum of Operating Capacity Under Desulphurization & 0 & 0 & 170.8 & 63.12 \\
      & (0) & (0) & (396.3) & (277.1) \\
    Prim. GDP (CNY per capita) & 1629.2 & 1953.8 & 2523.9 & 3347.0 \\
      & (666.3) & (818.5) & (1051.5) & (1438.9) \\
    Sec. GDP (CNY per capita) & 3011.1 & 4703.1 & 7882.5 & 13448.7 \\
      & (3190.2) & (5329.2) & (9525.4) & (15179.7) \\
    Hospital Bed (per 10,000 Population) & 23.17 & 23.33 & 25.06 & 31.01 \\
      & (8.180) & (9.095) & (9.743) & (10.68) \\
    \bottomrule
    \bottomrule
    Standard Errors in Parenthesis &     &   &   &  \\
    \end{tabular}}%
  \label{tab:summary_ARCA1}%
  \\
  \begin{tablenotes}
      This table documents the mean and standard deviation of key variables of interest from 2001 to 2005 for the ARCZ subsample at county level. Variables documented are: under-5 mortality (outcome variable), $SO_2$ and PM$_{2.5}$ density (endogenous variable of interest), weighted and unweighted sums of capacities (coal plant phase-out policy instruments), and desulphurization capacities (desulphurization policy instruments).
  \end{tablenotes}
\end{table}%
\clearpage
\subsection{Sulphur Dioxide Pollution Control Zones Subsample}\label{scafig}
\begin{figure}[h]
    \centering
    \includegraphics[width=0.7\linewidth]{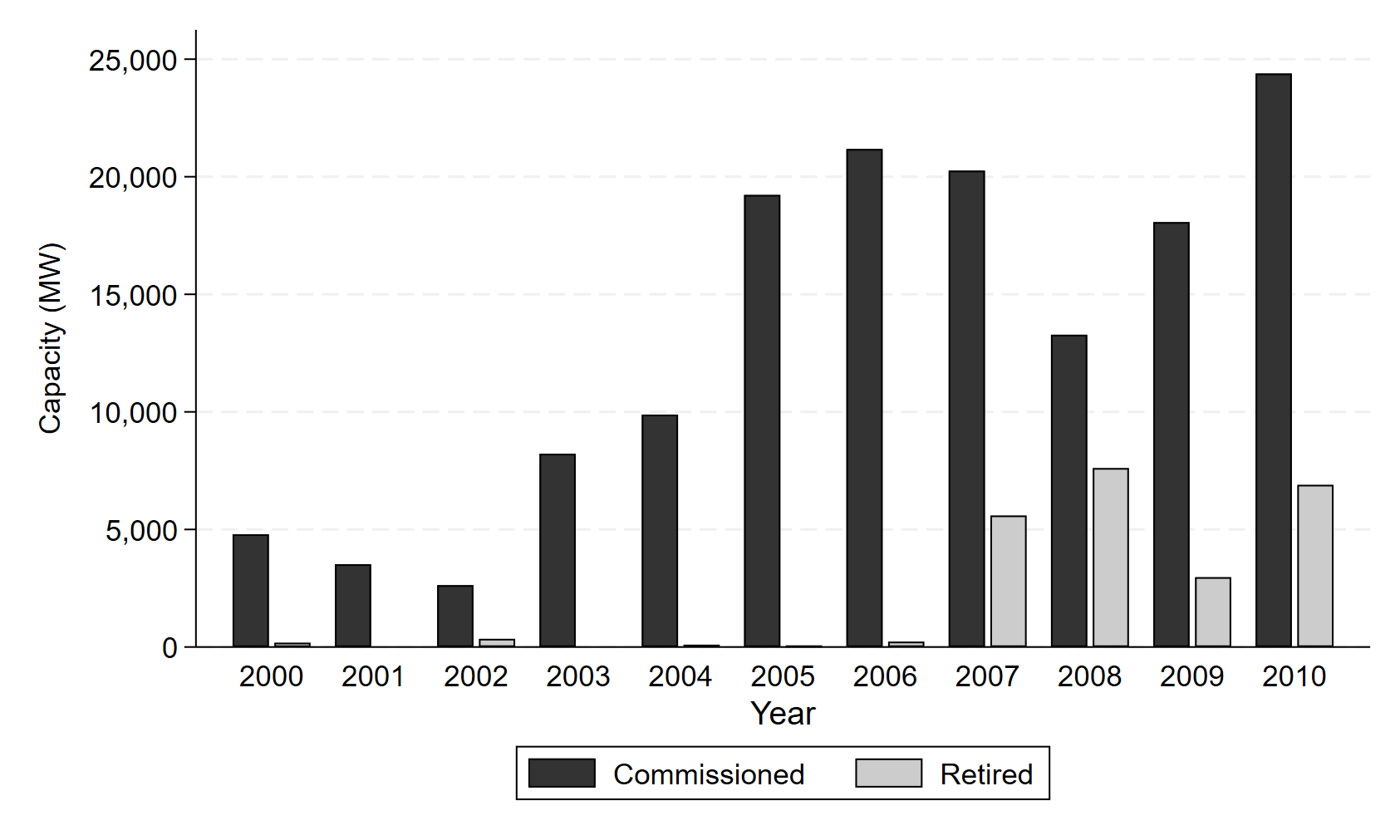}
    \caption{Commissioned and Retired Thermal Capacity Count from 2000 to 2010 (SO2CZ Subsample)}
    \label{fig:pollutantdensity_sca}
\end{figure}

\begin{figure}[htbp]
\centering
\begin{tabular}{cc}
\centering
    (a) $SO_2$ 2005 & (b) $SO_2$ 2010 \\
     \includegraphics[width=.5\textwidth]{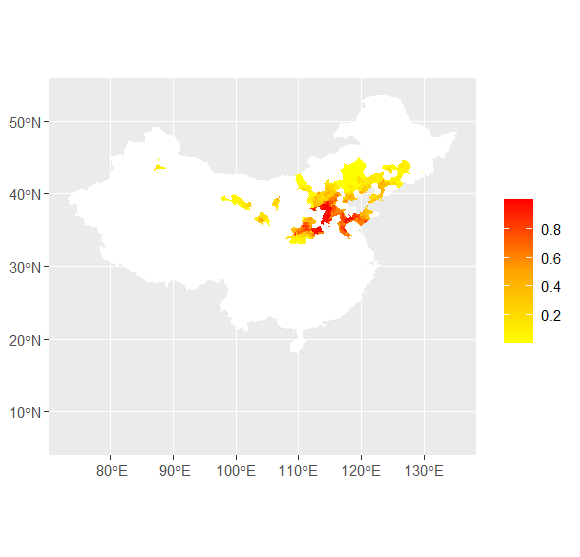}  &
     \includegraphics[width=.5\textwidth]{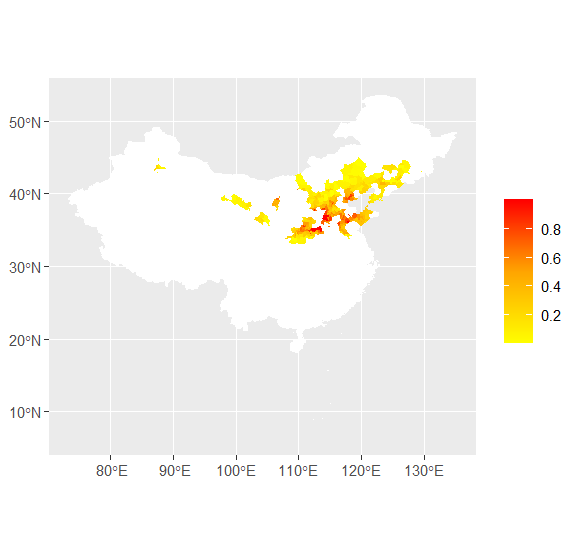} \\
     (c) $PM_{2.5}$ 2000 & (d)  $PM_{2.5}$ 2005 \\
     \includegraphics[width=.5\textwidth]{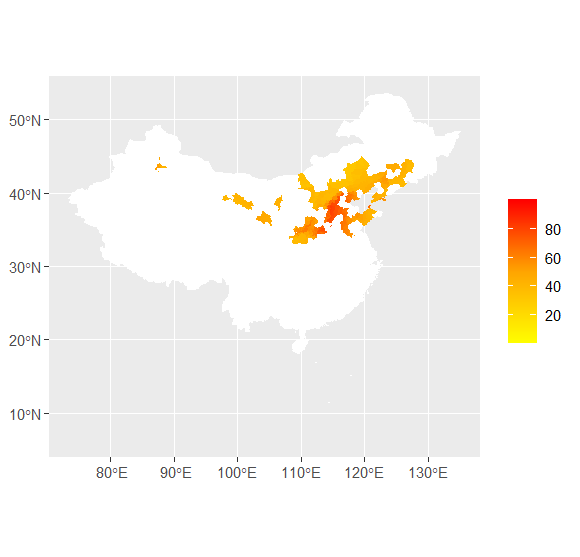}  &
     \includegraphics[width=.5\textwidth]{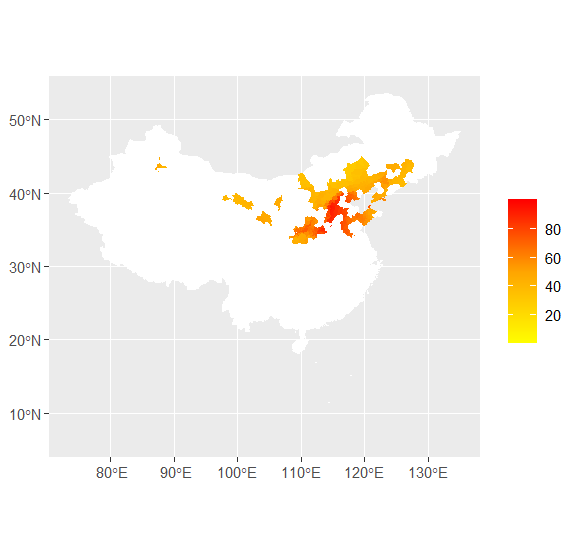} \\
     \multicolumn{2}{c}{(e) $PM_{2.5}$ 2010}\\
     \multicolumn{2}{c}{\includegraphics[width=.5\textwidth]{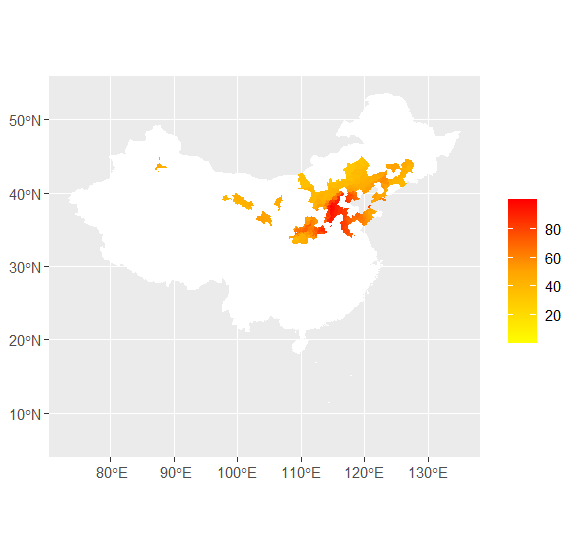}}
     
\end{tabular}
\caption{$SO_2$ and PM$_{2.5}$ Density Heat Maps (SO2CZ Subsample)}
\begin{figurenotes}
The figure shows the density of $SO_2$ (in DU) and $PM_{2.5}$ (in $\mu\text{g/m}^{3}$) in SO2CZ Subsample. Due to data availability, we document $SO_2$ density heatmaps in 2005 and 2010 while documenting PM$_{2.5}$ density heatmaps in 2000, 2005 and 2010.
\end{figurenotes}
\label{fig:sca_density}
\end{figure}

In this subsection we present the summary figures and the summary stats of key variables of the Sulphur Dioxide Pollution Control Zones policy subsample. The bar chart in Figure \ref{fig:pollutantdensity_sca} presents the sum of yearly retired and commissioned capacities while breaking down the total yearly capacity retired into capacity closed under 50 MW (the smaller-unit plant units forced to shut down by the phase-out policy). We present the density heat maps of the SO2CZ Policy subsample in Figure \ref{fig:sca_density}. Table \ref{tab:summary_sca1} documents the mean and standard deviations of our key variables of interest in 2001, 2004, 2007, and 2010, respectively for the SO2CZ subsample.

\begin{table}[htbp]
  \centering
  \caption{Summary Stats of Key Variables (SO2CZ Subsample)}
    \resizebox{\textwidth}{!}{\begin{tabular}{lcccc}
    \toprule
    \toprule
      & (1) & (2) & (3) & (4) \\
    Year  & 2001 & 2004 & 2007 & 2010 \\
    \midrule
    Under-5 Mortality (per 1000) & 30.19 & 24.06 & 19.33 & 15.68 \\
      & (10.75) & (8.967) & (7.168) & (5.997) \\
    $SO_2$ (DU) & 0 & 0 & 0.522 & 0.349 \\
      & (0) & (0) & (0.381) & (0.265) \\
    PM$_{2.5}$ ($\mu\text{g/m}^{3}$) & 58.01 & 62.06 & 70.95 & 64.23 \\
      & (14.87) & (17.68) & (21.30) & (19.00) \\
    Unweighted Sum of Retired Capacity (0 to 100 km) & 0 & 5.605 & 255.4 & 276.9 \\
      & (0) & (23.04) & (283.8) & (329.9) \\
    Unweighted Sum of Retired Capacity (0 to 50 km) & 0 & 2.065 & 68.02 & 73.95 \\
      & (0) & (14.24) & (139.9) & (141.6) \\
    Unweighted Sum of Retired Capacity (0 to 25 km) & 0 & 0.590 & 19.18 & 19.09 \\
      & (0) & (7.670) & (72.95) & (72.57) \\
    Unweighted Sum of Retired Capacity (50 to 100 km) & 0 & 3.540 & 187.4 & 203.0 \\
      & (0) & (18.51) & (237.3) & (257.9) \\
    Unweighted Sum of Retired Capacity (25 to 100 km) & 0 & 5.015 & 236.3 & 257.8 \\
      & (0) & (21.86) & (276.7) & (314.4) \\
    Unweighted Sum of Retired Capacity Under 50 MW (50 to 100 km) & 0 & 0 & 18.59 & 50.72 \\
      & (0) & (0) & (33.01) & (62.43) \\
    Unweighted Sum of Retired Capacity Under 50 MW (25 to 100 km) & 0 & 0 & 23.76 & 64.64 \\
      & (0) & (0) & (37.79) & (73.56) \\
    Unweighted Sum of Retired Capacity Under 50 MW (0 to 100 km) & 0 & 0 & 25.54 & 69.60 \\
      & (0) & (0) & (40.08) & (79.17) \\
    Unweighted Sum of Retired Capacity Under 50 MW (0 to 50 km) & 0 & 0 & 7.214 & 19.02 \\
      & (0) & (0) & (18.87) & (37.20) \\
    Unweighted Sum of Retired Capacity Under 50 MW (0 to 25 km) & 0 & 0 & 1.853 & 5.027 \\
      & (0) & (0) & (9.194) & (17.13) \\
    Weighted Sum of Retired Capacity (0 to 100 km) & 0 & 0.141 & 5.543 & 5.613 \\
      & (0) & (0.876) & (9.389) & (7.931) \\
    Weighted Sum of Retired Capacity (0 to 50 km) & 0 & 0.0954 & 3.020 & 2.872 \\
      & (0) & (0.846) & (8.316) & (6.287) \\
    Weighted Sum of Retired Capacity (0 to 25 km) & 0 & 0.0597 & 1.758 & 1.369 \\
      & (0) & (0.795) & (7.617) & (5.303) \\
    Weighted Sum of Retired Capacity (50 to 100 km) & 0 & 0.0457 & 2.523 & 2.741 \\
      & (0) & (0.245) & (3.257) & (3.581) \\
    Weighted Sum of Retired Capacity (25 to 100 km) & 0 & 0.0814 & 3.785 & 4.244 \\
      & (0) & (0.380) & (4.741) & (5.583) \\
    Weighted Sum of Retired Capacity Under 50 MW (50 to 100 km) & 0 & 0 & 0.248 & 0.678 \\
      & (0) & (0) & (0.445) & (0.831) \\
    Weighted Sum of Retired Capacity Under 50 MW (25 to 100 km) & 0 & 0 & 0.388 & 1.043 \\
      & (0) & (0) & (0.641) & (1.251) \\
    Weighted Sum of Retired Capacity Under 50 MW (0 to 100 km) & 0 & 0 & 0.596 & 1.468 \\
      & (0) & (0) & (1.952) & (2.567) \\
    Weighted Sum of Retired Capacity Under 50 MW (0 to 50 km) & 0 & 0 & 0.353 & 0.792 \\
      & (0) & (0) & (1.874) & (2.300) \\
    Weighted Sum of Retired Capacity Under 50 MW (0 to 25 km) & 0 & 0 & 0.210 & 0.426 \\
      & (0) & (0) & (1.782) & (2.033) \\
    Yearly Desulphurization Capacity (10k ton) & 0 & 0 & 0.838 & 0.475 \\
      & (0) & (0) & (1.369) & (1.168) \\
    Sum of Operating Capacity Under Desulphurization & 0 & 0 & 192.0 & 124.7 \\
      & (0) & (0) & (327.6) & (281.4) \\
    Prim. GDP (CNY per capita) & 1326.5 & 1944.5 & 2577.3 & 3755.1 \\
      & (778.8) & (970.7) & (1283.1) & (1809.4) \\
    Sec. GDP (CNY per capita) & 2597.5 & 5171.4 & 9204.4 & 14503.7 \\
      & (1972.2) & (4059.4) & (8114.7) & (12436.6) \\
    Hospital Bed (per 10,000 Population) & 29.50 & 27.92 & 30.32 & 36.40 \\
      & (10.42) & (10.13) & (8.827) & (9.397) \\
    \bottomrule
    \bottomrule
    Standard Errors in Parenthesis   &   &   &   &  \\
    \end{tabular}}%
  \label{tab:summary_sca1}%
  \\
  \begin{tablenotes}
      This table documents the mean and standard deviation of key variables of interest from 2001 to 2005 for the SO2CZ subsample at county level. Variables documented are: under-5 mortality (outcome variable), $SO_2$ and PM$_{2.5}$ density (endogenous variable of interest), weighted and unweighted sums of capacities (coal plant phase-out policy instruments), and desulphurization capacities (desulphurization policy instruments).
  \end{tablenotes}
\end{table}%
\clearpage

\subsection{Non-policy Area Subsample}\label{scafig}
\begin{figure}[h]
    \centering
    \includegraphics[width=0.7\linewidth]{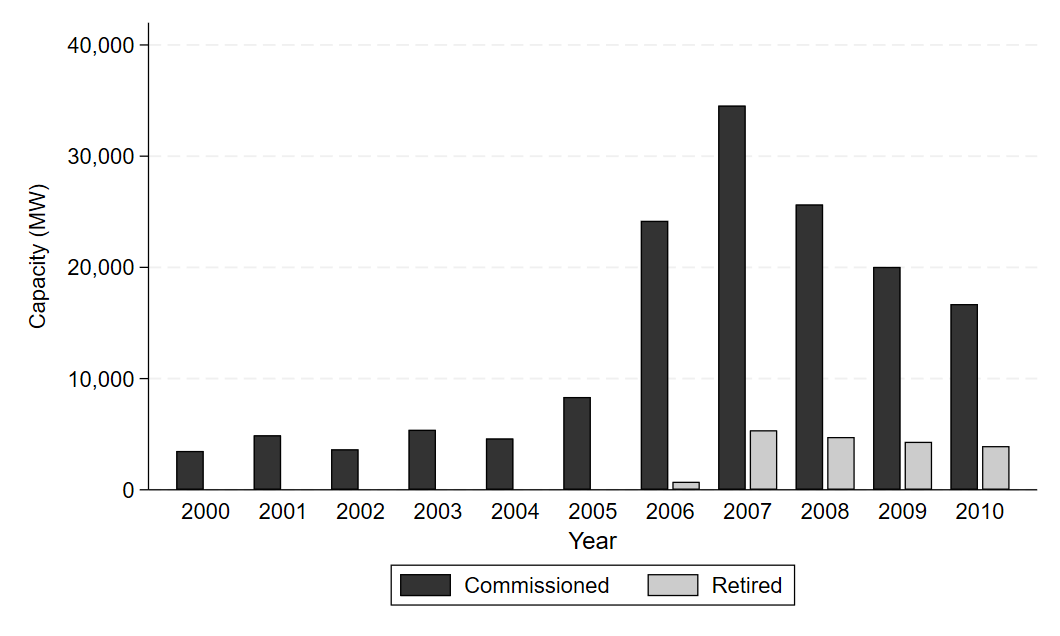}
    \caption{Commissioned and Retired Thermal Capacity Count from 2000 to 2010 (Non-policy Subsample)}
    \label{fig:pollutantdensity_none}
\end{figure}

\begin{figure}[htbp]
\centering
\begin{tabular}{cc}
\centering
    (a) $SO_2$ 2005 & (b) $SO_2$ 2010 \\
     \includegraphics[width=.5\textwidth]{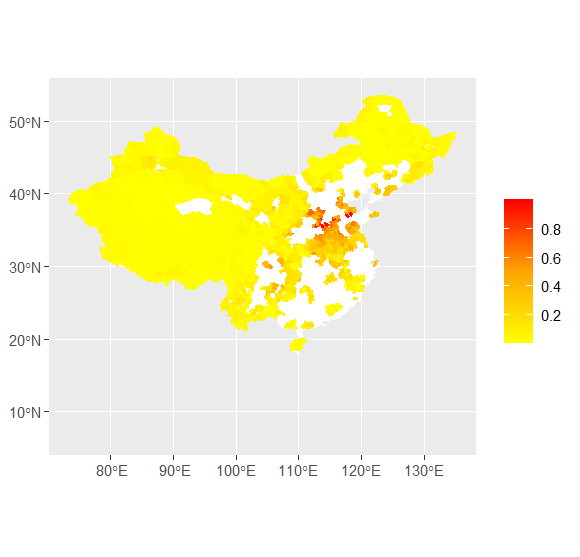}  &
     \includegraphics[width=.5\textwidth]{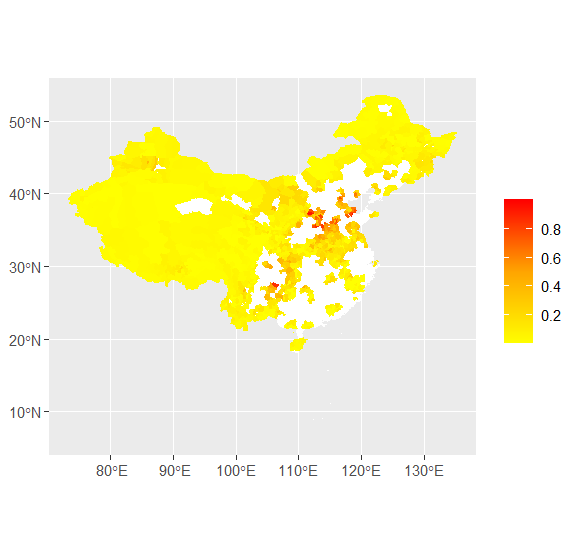} \\
     (c) $PM_{2.5}$ 2000 & (d)  $PM_{2.5}$ 2005 \\
     \includegraphics[width=.5\textwidth]{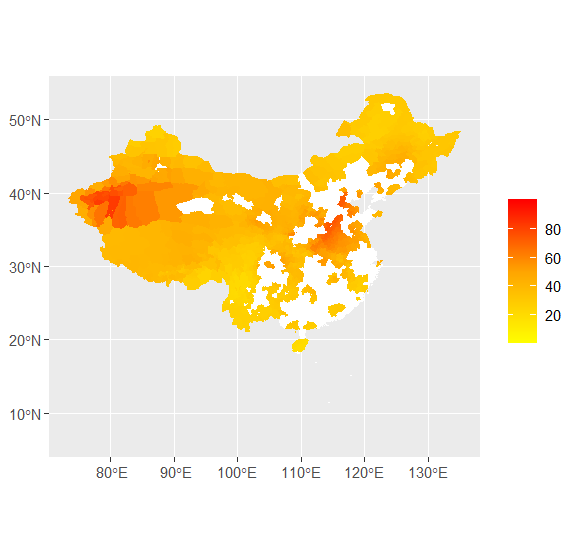}  &
     \includegraphics[width=.5\textwidth]{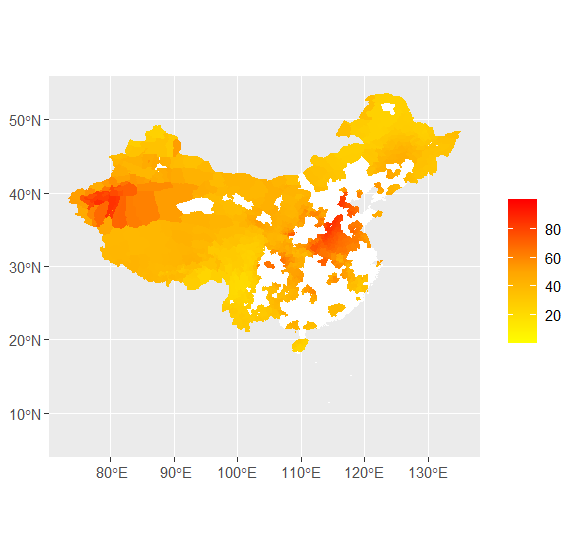} \\
     \multicolumn{2}{c}{(e) $PM_{2.5}$ 2010}\\
     \multicolumn{2}{c}{\includegraphics[width=.5\textwidth]{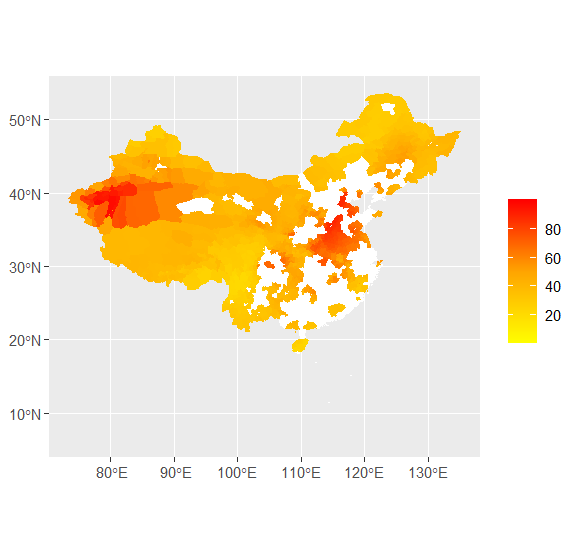}}
     
\end{tabular}
\caption{$SO_2$ and $PM_{2.5}$ Density Heat Maps (Non-policy Subsample)}
\begin{figurenotes}
The figure shows the density of $SO_2$ (in Dobson Unit (DU)) and $PM_{2.5}$ (in $\mu\text{g/m}^{3}$) in Non-policy Subsample. Due to data availability, we document $SO_2$ density heatmaps in 2005 and 2010 while documenting $PM_{2.5}$ density heatmaps in 2000, 2005 and 2010.
\end{figurenotes}
\label{fig:none_density}
\end{figure}

In this subsection we present the summary figures and the summary stats of key variables of the subsample outside the two aforementioned subsamples. The bar chart in Figure \ref{fig:pollutantdensity_none} presents the sum of yearly retired and commissioned capacities while breaking down the total yearly capacity retired into capacity closed under 50 MW (the smaller-unit plant units forced to shut down by the phase-out policy). We present the density heat maps of the Non-policy subsample in Figure \ref{fig:none_density}. Table \ref{tab:summary_none1} documents the mean and standard deviations of our key variables of interest in 2001, 2004, 2007, and 2010, respectively for the Non-policy subsample. 
\vspace{0cm}

\begin{table}[htbp]
  \centering
  \caption{Summary Stats of Key Variables (Non-policy Subsample)}
    \resizebox{\textwidth}{!}{\begin{tabular}{lcccc}
    \toprule
    \toprule
      & (1) & (2) & (3) & (4) \\
    Year  & 2001 & 2004 & 2007 & 2010 \\
    \midrule
    Under-5 Mortality (per 1000) & 34.86 & 27.05 & 21.73 & 17.34 \\
      & (14.06) & (10.84) & (8.455) & (6.993) \\
    $SO_2$ (DU) & 0 & 0 & 0.256 & 0.168 \\
      & (0) & (0) & (0.269) & (0.197) \\
    PM$_{2.5}$ ($\mu\text{g/m}^{3}$) & 48.92 & 54.25 & 58.23 & 54.70 \\
      & (13.60) & (15.38) & (18.76) & (16.44) \\
    Unweighted Sum of Retired Capacity (0 to 100 km) & 0.399 & 0.181 & 130.9 & 111.1 \\
      & (6.312) & (4.256) & (204.3) & (256.0) \\
    Unweighted Sum of Retired Capacity (0 to 50 km) & 0 & 0 & 27.35 & 24.46 \\
      & (0) & (0) & (79.61) & (89.00) \\
    Unweighted Sum of Retired Capacity (0 to 25 km) & 0 & 0 & 3.976 & 4.263 \\
      & (0) & (0) & (27.28) & (32.64) \\
    Unweighted Sum of Retired Capacity (50 to 100 km) & 0.399 & 0.181 & 103.5 & 86.60 \\
      & (6.312) & (4.256) & (180.1) & (205.2) \\
    Unweighted Sum of Retired Capacity (25 to 100 km) & 0.399 & 0.181 & 126.9 & 106.8 \\
      & (6.312) & (4.256) & (199.4) & (244.5) \\
    Unweighted Sum of Retired Capacity Under 50 MW (50 to 100 km) & 0 & 0 & 6.871 & 13.86 \\
      & (0) & (0) & (17.54) & (35.60) \\
    Unweighted Sum of Retired Capacity Under 50 MW (25 to 100 km) & 0 & 0 & 8.468 & 16.81 \\
      & (0) & (0) & (20.75) & (41.10) \\
    Unweighted Sum of Retired Capacity Under 50 MW (0 to 100 km) & 0 & 0 & 8.906 & 17.72 \\
      & (0) & (0) & (21.43) & (42.45) \\
    Unweighted Sum of Retired Capacity Under 50 MW (0 to 50 km) & 0 & 0 & 2.040 & 3.856 \\
      & (0) & (0) & (9.046) & (14.11) \\
    Unweighted Sum of Retired Capacity Under 50 MW (0 to 25 km) & 0 & 0 & 0.437 & 0.914 \\
      & (0) & (0) & (3.548) & (5.677) \\
    Weighted Sum of Retired Capacity (0 to 100 km) & 0.00513 & 0.00201 & 2.347 & 1.949 \\
      & (0.0814) & (0.0471) & (4.175) & (4.940) \\
    Weighted Sum of Retired Capacity (0 to 50 km) & 0 & 0 & 0.954 & 0.812 \\
      & (0) & (0) & (3.130) & (3.184) \\
    Weighted Sum of Retired Capacity (0 to 25 km) & 0 & 0 & 0.321 & 0.279 \\
      & (0) & (0) & (2.374) & (2.111) \\
    Weighted Sum of Retired Capacity (50 to 100 km) & 0.00513 & 0.00201 & 1.393 & 1.136 \\
      & (0.0814) & (0.0471) & (2.469) & (2.720) \\
    Weighted Sum of Retired Capacity (25 to 100 km) & 0.00513 & 0.00201 & 2.026 & 1.670 \\
      & (0.0814) & (0.0471) & (3.309) & (3.964) \\
    Weighted Sum of Retired Capacity Under 50 MW (50 to 100 km) & 0 & 0 & 0.0923 & 0.179 \\
      & (0) & (0) & (0.245) & (0.458) \\
    Weighted Sum of Retired Capacity Under 50 MW (25 to 100 km) & 0 & 0 & 0.133 & 0.257 \\
      & (0) & (0) & (0.352) & (0.640) \\
    Weighted Sum of Retired Capacity Under 50 MW (0 to 100 km) & 0 & 0 & 0.182 & 0.314 \\
      & (0) & (0) & (0.641) & (0.805) \\
    Weighted Sum of Retired Capacity Under 50 MW (0 to 50 km) & 0 & 0 & 0.0896 & 0.135 \\
      & (0) & (0) & (0.562) & (0.547) \\
    Weighted Sum of Retired Capacity Under 50 MW (0 to 25 km) & 0 & 0 & 0.0487 & 0.0566 \\
      & (0) & (0) & (0.500) & (0.385) \\
    Yearly Desulphurization Capacity (10k ton) & 0 & 0 & 0.111 & 0.0983 \\
      & (0) & (0) & (0.434) & (0.389) \\
    Sum of Operating Capacity Under Desulphurization & 0 & 0 & 37.74 & 53.65 \\
      & (0) & (0) & (144.1) & (184.4) \\
    Prim. GDP (CNY per capita) & 1425.5 & 1884.3 & 2477.8 & 3616.4 \\
      & (649.2) & (873.1) & (1191.5) & (1906.3) \\
    Sec. GDP (CNY per capita) & 1752.6 & 3189.9 & 5845.5 & 10444.6 \\
      & (1475.0) & (3188.0) & (7687.2) & (12283.6) \\
    Hospital Bed (per 10,000 Population) & 22.29 & 22.24 & 24.48 & 30.06 \\
      & (8.475) & (8.043) & (9.383) & (9.650) \\
    \bottomrule
    \bottomrule
    Standard Errors in Parenthesis  &   &   &   &  \\
    \end{tabular}}%
  \label{tab:summary_none1}%
  \\
  \begin{tablenotes}
      This table documents the mean and standard deviation of key variables of interest from 2001 to 2005 for the Non-policy subsample at county level. Variables documented are: under-5 mortality (outcome variable), $SO_2$ and PM$_{2.5}$ density (endogenous variable of interest), weighted and unweighted sums of capacities (coal plant phase-out policy instruments), and desulphurization capacities (desulphurization policy instruments).
  \end{tablenotes}
\end{table}%

\clearpage

\end{appendices}
\end{document}